\documentclass[a4paper,11pt]{article}

\usepackage{graphicx}
\usepackage{amssymb,amsmath}
\usepackage{rotating}

\usepackage{caption}
\usepackage{subcaption}
\usepackage[toc,page]{appendix}
\usepackage{bbold}

\usepackage{slashed}
\usepackage{bbm}
\usepackage{tikz}
\allowdisplaybreaks

\usepackage{multirow}

\makeatletter
 
\def\paragraph{\@startsection{paragraph}{4}{\z@}{+2.00ex plus
 +1ex minus +.2ex}{1.5ex plus .2ex}{\it\normalsize}}
 
\def\section{\@startsection {section}{1}{\z@}{+3.0ex plus +1ex minus
  +.2ex}{2.3ex plus .2ex}{\normalsize\bf}}
\def\subsection{\@startsection{subsection}{2}{\z@}{+2.5ex plus +1ex
minus +.2ex}{1.5ex plus .2ex}{\normalsize\bf}}
\def\subsubsection{\@startsection{subsubsection}{3}{\z@}{+3.25ex plus
 +1ex minus +.2ex}{1.5ex plus .2ex}{\normalsize\bf}}

\@addtoreset{equation}{section}

\def\appendix{\par
 \setcounter{section}{0} 
 \setcounter{subsection}{0}
 \setcounter{equation}{0}
 \def\thesection{\Alph{section}}}
 
\expandafter\ifx\csname mathrm\endcsname\relax\def\mathrm#1{{\rm #1}}\fi
 
\makeatletter

\newcount\@tempcntc
\def\@citex[#1]#2{\if@filesw\immediate\write\@auxout{\string\citation{#2}}\fi
  \@tempcnta\z@\@tempcntb\m@ne\def\@citea{}\@cite{\@for\@citeb:=#2\do
    {\@ifundefined
       {b@\@citeb}{\@citeo\@tempcntb\m@ne\@citea
        \def\@citea{,\penalty\@m\ }{\bf ?}\@warning
       {Citation `\@citeb' on page \thepage \space undefined}}%
    {\setbox\z@\hbox{\global\@tempcntc0\csname
b@\@citeb\endcsname\relax}%
     \ifnum\@tempcntc=\z@ \@citeo\@tempcntb\m@ne
       \@citea\def\@citea{,\penalty\@m}
       \hbox{\csname b@\@citeb\endcsname}%
     \else
      \advance\@tempcntb\@ne
      \ifnum\@tempcntb=\@tempcntc
      \else\advance\@tempcntb\m@ne\@citeo
      \@tempcnta\@tempcntc\@tempcntb\@tempcntc\fi\fi}}\@citeo}{#1}}

\def\@citeo{\ifnum\@tempcnta>\@tempcntb\else\@citea
  \def\@citea{,\penalty\@m}%
  \ifnum\@tempcnta=\@tempcntb\the\@tempcnta\else
   {\advance\@tempcnta\@ne\ifnum\@tempcnta=\@tempcntb \else
\def\@citea{--}\fi
    \advance\@tempcnta\m@ne\the\@tempcnta\@citea\the\@tempcntb}\fi\fi}

\def\asymp#1%
{\mathrel{\raisebox{-.4em}{$\widetilde{\scriptstyle #1}$}}}

\def\Nequal#1%
{\mathrel{\raisebox{-.5em}{$\stackrel{=}{\scriptstyle\rm#1}$}}}
\def\dsl{\mathpalette\make@slash}
\def\make@slash#1#2{\setbox\z@\hbox{$#1#2$}%
  \hbox to 0pt{\hss$#1/$\hss\kern-\wd0}\box0}

\def\beq#1\eeq{\begin{equation}#1\end{equation}}
\def\beqar{\begin{eqnarray}}
\def\eeqar{\end{eqnarray}}
\def\barr#1{\begin{array}{#1}}
\def\earr{\end{array}}
\def\bfi{\begin{figure}}
\def\efi{\end{figure}}
\def\btab{\begin{table}}
\def\etab{\end{table}}
\def\bce{\begin{center}}
\def\ece{\end{center}}

\newcommand{\TeV}{\unskip\,\mathrm{TeV}}
\newcommand{\GeV}{\unskip\,\mathrm{GeV}}

\def\mathswitchr#1{\relax\ifmmode{\mathrm{#1}}\else$\mathrm{#1}$\fi}

\def\mathswitch#1{\relax\ifmmode#1\else$#1$\fi}

\newcommand{\MSbar}{{\overline{\mathrm{MS}}}}

\renewcommand{\max}{\mathrm{max}}
\newcommand{\z}{\setbox0\hbox{+}\hbox to \wd0{\hss0\hss}}

\def\limfunc#1{\mathop{\rm #1}}

\def\tr{\limfunc{Tr}}

\def\slash#1{{\setbox0=\hbox{$#1$}
  \rlap{\ifdim\wd0>.7em\kern.22\wd0\else\kern.1\wd0\fi /}#1}}

\def\braket#1#2{\left\langle #1\vphantom{#2}
  \right. \kern-2.5pt\left| #2\vphantom{#1}\right\rangle }

\def\draftdate{\relax}
\def\mda{\relax}
\def\mua{\relax}
\def\mla{\relax}
\def\Mda{\relax}
\def\Mua{\relax}
\def\Mla{\relax}
\def\draft{
\def\thtystars{******************************}
\def\sixtystars{\thtystars\thtystars}
\typeout{}
\typeout{\sixtystars**}
\typeout{* Draft mode!
         For final version remove \protect\draft\space in source file *}
\typeout{\sixtystars**}
\typeout{}
\def\draftdate{\today}
\def\mua{\marginpar[\boldmath\hfil$\uparrow$]%
                   {\boldmath$\uparrow$\hfil}%
                    \typeout{marginpar: $\uparrow$}\ignorespaces}
\def\mda{\marginpar[\boldmath\hfil$\downarrow$]%
                   {\boldmath$\downarrow$\hfil}%
                    \typeout{marginpar: $\downarrow$}\ignorespaces}
\def\mla{\marginpar[\boldmath\hfil$\rightarrow$]%
                   {\boldmath$\leftarrow $\hfil}%
                    \typeout{marginpar: $\leftrightarrow$}\ignorespaces}
\def\Mua{\marginpar[\boldmath\hfil$\Uparrow$]%
                   {\boldmath$\Uparrow$\hfil}%
                    \typeout{marginpar: $\uparrow$}\ignorespaces}
\def\Mda{\marginpar[\boldmath\hfil$\Downarrow$]%
                   {\boldmath$\Downarrow$\hfil}%
                    \typeout{marginpar: $\downarrow$}\ignorespaces}
\def\Mla{\marginpar[\boldmath\hfil$\Rightarrow$]%
                   {\boldmath$\Leftarrow $\hfil}%
                    \typeout{marginpar: $\leftrightarrow$}\ignorespaces}
\overfullrule 5pt
\oddsidemargin -15mm
\marginparwidth 29mm
}

\begin{document}
\mbox{}\hfill FR-PHENO-2023-07
\thispagestyle{empty}
\def\thefootnote{\fnsymbol{footnote}}
\setcounter{footnote}{1}
\null
\strut\hfill 
\vskip 0cm
\vfill
\begin{center}
{\large \bf 
  \boldmath{Renormalization of a Standard Model Extension with a Dark Abelian Sector\\and Predictions for the W-Boson Mass}
\par} \vskip 2.5em
{\large
  {\sc Stefan Dittmaier$^1$, Jonas Rehberg$^1$
and Heidi Rzehak$^{2}$}\\[1ex]
{\normalsize 
\it 
$^1$ Albert-Ludwigs-Universit\"at Freiburg, 
Physikalisches Institut, \\
Hermann-Herder-Stra\ss{}e 3,
D-79104 Freiburg, Germany \\[.5em]
$^2$ Institute for Theoretical Physics, University of T\"ubingen, \\
Auf der Morgenstelle 14, 72076 T\"ubingen, Germany \\[.5em]
}
}

\par \vskip 1em
\end{center} \par
\vskip 2cm 
\noindent{\bf Abstract: }The described Dark Abelian Sector Model (DASM) extends the Standard Model (SM) by a ``dark'' sector containing a spontaneously broken $U(1)_\text{d}$ gauge group. Keeping this dark sector quite generic we only add one additional Higgs boson, one Dirac fermion, and right-handed SM-like neutrinos to the SM. Using the only two singlet operators of the SM with dimension less than 4 (the $U(1)_\text{Y}$ field-strength tensor and the SM Higgs mass operator $|\Phi|^2$) as well as the right-handed neutrino fields we open up three portals to the dark sector. Dark sectors, such as the one of the DASM, that introduce an additional Higgs boson $\text{H}$ as well as an additional $\text{Z}'$ gauge boson can have a large influence on the predictions for electroweak precision observables and even accommodate possible dark matter candidates. We consider one of the two Higgs bosons to be the known $125\,\text{GeV}$ Higgs boson and parameterize the extension of the scalar sector by the mass of the second Higgs boson, the Higgs mixing angle, and a Higgs self-coupling. We do not assume any mass hierarchy in the gauge sector and use the mass of the additional $\text{Z}'$ boson and a corresponding gauge-boson mixing angle to parameterize the extension of the gauge sector. The fermion sector is parameterized by the mass of the additional fermion and a fermion mixing angle. We describe an on-shell as well as an $\MSbar$ renormalization scheme for the DASM sectors and give explicit results for the renormalization constants at the 1-loop level, and, thus, prepare the ground for full NLO predictions for collider observables in the DASM. As a first example, we provide the DASM prediction for the W-boson mass derived from  muon decay.

\par
\vfill
\noindent 
\today \par
\vskip .5cm 
\null
\setcounter{page}{0}
\clearpage
\def\thefootnote{\arabic{footnote}}
\setcounter{footnote}{0}

\section{Introduction}
With the discovery of a SM-like Higgs boson \cite{CMS_Higgs,ATLAS_Higgs} at the Large Hadron Collider (LHC) in 2012, all particles described by the SM have been found.
Currently, measurements at collider experiments agree very precisely with predictions of the SM with only very few exceptions. Despite this success of the SM, there are several clear indications that it cannot be the ultimate theory describing Nature. Neither can the SM explain the observed matter--antimatter asymmetry in the visible universe, nor can it describe the origin of dark matter (DM) observed in the universe. Finally, a quantum-field-theoretical description of the gravitational force is missing within the SM, but most likely particle physics experiments cannot help to solve this problem. Among the measurements within particle physics, the tension \cite{g_2_Exp} between the high-precision SM prediction of the anomalous magnetic moment of the muon, $(g-2)_\mu$, and its measurements performed by the BNL and FNAL collaborations have somewhat tightened.
Further, the new result for the mass of the W boson published by the CDF collaboration shows a significant deviation of $7\sigma$ from the SM prediction \cite{CDFMW} (which, however, is at variance with the previous experimental world average \cite{PDG}). \\
There are several, quite fundamental SM extensions like supersymmetric models or grand unifying models that claim to solve some of the open problems mentioned above. However, at the moment there is no convincing experimental evidence hinting towards the realization of one of these models in Nature, and thus, it is not clear how the SM needs to be modified in order to match all observed phenomena. Since there are no new, truly elementary particles found by any experiment so far, a promising way to get hints towards the structure of possible new physics is to compare theoretical predictions within the SM and generic extensions at the highest possible precision with measurements. \\
In the search for physics beyond the SM (BSM), investigations of the structure of electroweak (EW) symmetry breaking are of great importance. In the past, analyses \cite{PDG,SESMana} probing this structure were able to limit the parameter space of various prominent extensions of the Higgs sector, like Singlet Extensions of the SM (SESM) \cite{SESM1,SESM2,SESM3,SchabWells}, substantially, but many of such extensions (including singlet Higgs extensions) are still viable. Recently the observed tensions in $(g-2)_\mu$ have drawn some attention towards non-standard models that enrich the gauge structure of the SM by an $U(1)$ or even more evolved gauge symmetries that might help to relax these tensions \cite{Zpg22,Zpg2,G2Zp}. There are several ideas on how these extensions of the gauge structure are added to the SM without introducing anomalies, to potentially solve some of the open questions like the origin of neutrino masses or dark matter. One way is to promote the global $B-L$ (baryon number minus lepton number) symmetry of the SM to a (possibly spontaneously broken) $U(1)_{B-L}$ gauge symmetry, which will introduce a massive neutral $Z'_{B-L}$ gauge boson to the theory coupling to the $B-L$ charge (see, e.g.,~Refs.~\cite{UnBroU1BL,UnBroU1BL2,NeutRev} and references therein). This opens a portal to possible BSM physics carrying $B-L$ charge. Another prominent class of $Z'$ or ``dark photon'' models make the basic assumption of a ``dark'' sector with a non-trivial gauge structure and at least one (possibly spontaneously broken) $U(1)_\text{d}$ gauge symmetry \cite{SchabWells,Zp1,Babu_ZZp_old,Wells,Hold,Tro,Pel,Bento}. Note that the two abelian gauge groups $U(1)_\text{d}$ of the dark sector and $U(1)_\text{Y}$ of the SM sector open the possibility of kinetic mixing between the (gauge-invariant) field-strength tensors of the respective gauge fields. As a result the two neutral massive gauge bosons $\text{Z}$ (which is SM like) and $\text{Z}'$ (the mass of which is only weakly constrained) connect the SM with the dark sector. In addition to the scalar and gauge sector extensions, the focus of BSM physics more recently also includes extensions of the fermion sector of the SM since they seem to provide promising DM candidates (see, e.g.,~Refs.~\cite{SterileNeutrinos2,SterileNeutrinos} and references therein).\\  
In this work we formulate a simple but quite generic model with a ``dark'' $U(1)_\text{d}$ group and call it the Dark Abelian Sector Model (DASM) in the following. The DASM is a simplified model featuring some phenomenological imprints of more comprehensive theories. More precisely, it introduces a dark sector with generic features that resembles a broad range of possible dark sectors embedded in more complete models. Similar to the proposals of Refs.~\cite{SchabWells,Babu_ZZp_old,Wells} the DASM introduces an additional neutral $U(1)_\text{d}$ gauge field, featuring kinetic mixing with the $U(1)_\text{Y}$ gauge field of the weak hypercharge, unlike models that exclude mixing by, e.g., postulating additional discrete symmetries in the dark sector, like in Ref.~\cite{Z2symZpfield}. The $U(1)_\text{d}$ gauge group is spontaneously broken by an additional Higgs field $\rho$ which is a singlet with respect to the SM, but carries dark charge and develops a non-vanishing vacuum expectation value (vev). This leads to an additional massive, neutral gauge boson as well as an additional Higgs boson. Finally, a generic Dirac fermion with dark charge as well as right-handed, SM-like neutrinos are introduced in order to allow for an additional portal to the dark sector of the DASM.\\ 
The SM itself is considered to be a singlet with respect to this additional $U(1)_{\text{d}}$ gauge symmetry. The full gauge group of the DASM is, thus, given by $SU(3)_{\text{C}} \times SU(2)_{\text{W}} \times U(1)_\text{Y}\times U(1)_{\text{d}}$. Via a scalar mixing in the Higgs potential and kinetic mixing between the two $U(1)$ gauge group factors of the theory, the DASM uses the only two gauge-invariant operators of the SM that can be employed to couple a dark (singlet) sector to SM particles, namely the mass operator $\Phi^\dagger\Phi$ for the complex Higgs doublet $\Phi$ and the field-strength tensor of the $U(1)_\text{Y}$. This opens two possible portals to the hypothetical dark sector with at least some $U(1)_{\text{d}}$ gauge symmetry. By adding right-handed neutrinos to the theory, a third portal to a possible dark fermion sector opens up via the mixing of the SM-like neutrino fields with the newly introduced fermion of the dark sector. Further, the DASM can accommodate effects of massive, SM-like neutrinos. The introduced model extensions can have large influence on the predictions of Higgs and EW precision observables, and on $(g-2)_\mu$. Moreover, for some regions of the parameter space the DASM can provide viable DM candidates. Recently, in Ref.~\cite{pheno_HAHM}, a comprehensive phenomenological study of the Hidden Abelian Higgs Model, a SM extension with extensions of the gauge and Higgs sectors similar to the DASM, was presented, further highlighting the relevance of abelian dark sector extensions in current searches for possible BSM physics.\\
In this paper, we present the full theoretical setup of the DASM, including a complete renormalization at NLO. We propose two alternative  renormalization schemes for the DASM: The first scheme is based on on-shell renormalization conditions for directly measurable quantities as far as possible and is, thus, particularly suited for a phenomenological study of EW and Higgs precision observables. The second alternative scheme is based on $\MSbar$ conditions for the new mixing angles. We are not aware of any comprehensive description of renormalization schemes for kinetic mixing in the gauge-boson sector existing in the literature\footnote{Recently a first, but very brief description of on-shell renormalization of the additional gauge-boson mixing angle was given in Ref.~\cite{Pel}, however, without giving the full details on the renormalization of the model.}. Our formulation may, thus, serve as a proposal for the renormalization of similar models as well.\\
The structure of this paper is as follows:
We introduce the DASM in detail in Sect.\ \ref{se:mod-setup} and define a particularly intuitive input parameter scheme that is well suited for a phenomenological study of collider observables. For the sector of neutrinos and the dark fermion, we formulate a simplified approximate  parameterization based on the smallness of the masses of the three known neutrinos. In the resulting approximation, which is sufficient for collider physics, only the dark fermion is massive. 
In Sect.\ \ref{se:renormierung} we define an on-shell renormalization scheme for the DASM and give the full NLO set of renormalization constants needed to perform precision calculations.
In Sect.~\ref{se:pheno} we give NLO predictions for the W-boson mass in the DASM (based on the measured muon decay width) as a first phenomenological application for the introduced renormalization schemes.
In Sect.\ \ref{se:conclusions} we give a summary and our conclusions.
In the appendix we display explicit expressions for lengthy, but important and interesting field-theoretical quantities of the DASM.

\section{Description of the DASM}
\label{se:mod-setup}
In this section we define the DASM and discuss some of its salient features. All parts of the Lagrangian that differ from their SM counterparts are discussed in detail. Moreover, we give a particularly intuitive set of input parameters for the DASM that is well suited for a phenomenological confrontation of theory and experiment.\\
The presence of an additional $U(1)_\text{d}$ gauge field $C^\mu$, an additional Higgs field $\rho$, and a newly introduced Dirac fermion field $f'_\text{d}$ leads to additional terms to the SM part of the Lagrangian. We will not give a detailed discussion of the SM parts of the theory that are not modified within the DASM.
They can be found, e.g., in Refs.\ \cite{Denner_Habil,Peskin_Schroed,Schwartz,BDJ,DEDI20201}. In detail, we adopt the notation and conventions for field-theoretical quantities from Ref.\ \cite{DEDI20201}.\\
The full Lagrangian of the DASM can be split up in the following way,
\begin{align}
\mathcal{L}_\text{DASM}=\mathcal{L}_\text{YM}+\mathcal{L}_\text{Fermion}+\mathcal{L}_\text{Higgs}+\mathcal{L}_\text{QCD}.
\end{align}
The individual parts of $\mathcal{L}_\text{DASM}$ will be explained in the following.
The QCD part $\mathcal{L}_\text{QCD}$ of the Lagrangian is not modified with respect to the SM and can, e.g., be found in Refs.~\cite{Peskin_Schroed,Schwartz,BDJ,DEDI20201}.

\subsection{Higgs sector}
\label{sec:higgs_pot}
The extension of the SM scalar sector with its complex Higgs doublet $\Phi$ by the additional complex Higgs field $\rho$ leads to 
\begin{align}
\mathcal{L}_{\text{Higgs}}= \left(D_\mu\Phi\right)^\dagger\left(D^\mu\Phi\right)+\left(D_{\text{d},\mu}\rho\right)^\dagger\left(D_\text{d}^\mu\rho\right) - V(\Phi,\rho),
\label{eq:lag-higgs}
\end{align} 
where the covariant derivatives are given by 
\begin{align}
D^\mu_\text{d}&=\partial^\mu+\text{i}\tilde{q}e_\text{d}C^\mu,\qquad \tilde{q}\rho=\tilde{q}_\rho \rho, \qquad \tilde{q}_\rho=1, \label{eq:cov-deriv-ym2}\\
D^\mu&=\partial^\mu-\frac{\text{i}g_2}{2}\tau^aW^{a,\mu}+\frac{\text{i}g_1}{2}B^\mu,
\label{eq:cov-deriv-ym}
\end{align}
with the charge operator $\tilde{q}$ and the coupling constant $e_\text{d}$ of the $U(1)_{\text{d}}$ gauge symmetry, and the quantities $\tau^a$ denote the Pauli matrices. The choice $\tilde{q}_\rho=1$ for the eigenvalue $\tilde{q}_\rho$ of $\tilde{q}$ simply provides the normalization of the $U(1)_\text{d}$ coupling strength $e_\text{d}$. The fields $W^{a,\mu}$, $B^\mu$ are the usual $SU(2)_\text{W}\times U(1)_\text{Y}$ gauge fields with respective gauge couplings $g_2$, $g_1$. The most general gauge-invariant, renormalizable potential $V(\Phi,\rho)$ allows for a mixing of the Higgs doublet $\Phi$ and the Higgs field $\rho$ and can be written as 
\begin{equation}
V(\Phi,\rho)=-\mu_2^2\Phi^\dagger\hspace{-2pt}\Phi -2\mu_1^2\rho^\dagger\rho+\frac{\lambda_2}{4}(\Phi^\dagger\hspace{-2pt}\Phi)^2+4\lambda_1(\rho^\dagger\rho)^2+2\lambda_{12}\Phi^\dagger\hspace{-2pt}\Phi\rho^\dagger\rho,
\label{eq:higgs-pot}
\end{equation}
with the five free real parameters $\mu_1^2, \mu_2^2,\lambda_1, \lambda_2, \lambda_{12}$.
The SM-like Higgs doublet $\Phi$ and the Higgs field $\rho$ can be parameterized according to 
\begin{equation}
\Phi= \begin{pmatrix}
\phi^+\\
\frac{1}{\sqrt{2}}(h_2+v_2+\text{i}\chi_2)\\
\end{pmatrix}\text{,}\qquad \rho=\frac{1}{\sqrt{2}}(h_1+v_1+\text{i}\chi_1),
\label{eq:higgs-fields}
\end{equation}
with $v_1$, $v_2$ representing constants quantifying the vacuum expectation values (vevs) of $\Phi$ and $\rho$, respectively. Without loss of generality the parameters $v_1$, $v_2$ can be taken real and positive by making use of global gauge transformations. Further, $\phi^+$ and $\phi^-=(\phi^+)^\dagger$ are SM-like charged would-be Goldstone-boson fields, $\chi_1$ and $\chi_2$ represent two neutral CP-odd would-be Goldstone-boson fields, and the two neutral CP-even Higgs fields $h_1$ and $h_2$ will eventually lead to two Higgs fields corresponding to two physical CP-even Higgs bosons. 
The vacuum stability conditions for the potential are given by 
\begin{equation}
\lambda_1>0\text{, }\hspace{10pt}\lambda_2>0\text{, }\hspace{10pt}\lambda_1\lambda_2-\lambda^2_{12}>0.
\label{eq:Higgs-pot-stab}
\end{equation}
The main features of the scalar potential of the DASM are reflected by most Higgs singlet extensions, see e.g.\ Refs.\ \cite{SESM1,SESM2,SESM3,BogDit}. Here we outline the main features adopting the parameterization given in Ref.\ \cite{BogDit}. \\
Expanding the Higgs potential \eqref{eq:higgs-pot} with the help of Eq.~\eqref{eq:higgs-fields} leads to
\begin{align}
V =& -t_{1}h_1-t_{2}h_2 \nonumber \\ &+ \frac{1}{2}(h_2,h_1)\mathrm{M}^2_{\text{Higgs}}\begin{pmatrix}
h_2\\
h_1\\
\end{pmatrix}+\frac{1}{2}(\chi_1,\chi_2)\mathrm{M}^2_{\chi}\begin{pmatrix}
\chi_1\\
\chi_2\\
\end{pmatrix}
+\mathrm{M}^2_{\phi^+\phi^-}
\phi^-\phi^+
\nonumber \\
&+ \text{interaction terms},
\end{align}
with the tadpole terms 
\begin{equation}
t_{1}=-v_1\left(4v^2_1\lambda_1+v_2^2\lambda_{12}-2\mu_1^2\right)\text{,}\qquad t_{2}=-v_2\left(\frac{v_2^2}{4}\lambda_2+v_1^2\lambda_{12}-\mu_2^2\right),
\label{eq:tadpols}
\end{equation}
the mass matrices
\begin{align}
\mathrm{M}^2_{\text{Higgs}}=\begin{pmatrix}
\frac{v_2^2}{2}\lambda_2-\frac{t_{2}}{v_2}& 2v_1v_2\lambda_{12}\\
2v_1v_2\lambda_{12} & 8v_1^2\lambda_1-\frac{t_{1}}{v_1}\\
\end{pmatrix}\text{, }\qquad
\mathrm{M}^2_\chi=\begin{pmatrix}
-\frac{t_{1}}{v_1}& 0\\
0 & -\frac{t_{2}}{v_2}\\
\end{pmatrix},
\end{align}
and the mass term of the charged would-be Goldstone-boson fields given by
\begin{align}
  \text{M}^2_{\phi^+\phi^-}=-\frac{t_{2}}{v_2}.
\end{align}
At leading order, the tadpole parameters $t_1$, $t_2$ are set to zero, in order to obtain the standard form for free propagation. For the renormalization procedure described below it is, however, convenient to keep the $t_1$, $t_2$ terms explicit here.
The Higgs fields $h$ and $H$ corresponding to mass eigenstates are obtained by a rotation by an angle $\alpha$, 
\begin{align}
\begin{pmatrix}
h\\
H\\
\end{pmatrix}=
\begin{pmatrix}
\cos\alpha\,&-\sin\alpha\\
\sin\alpha&\cos\alpha\\
\end{pmatrix}
\begin{pmatrix}
h_2\\
h_1\\
\end{pmatrix},
\end{align}
which diagonalizes the mass matrix $\mathrm{M}^2_{\text{Higgs}}$. Expressing the potential in terms of these fields one finds
\begin{align}
V =& -t_{{h}}h-t_{{H}}H \nonumber \\ &+ \frac{1}{2}(h,H)\begin{pmatrix}
M^2_{\text{h}}&M^2_{\text{hH}}\\
M^2_{\text{hH}}&M^2_{\text{H}}\\
\end{pmatrix}\begin{pmatrix}
h\\
H\\
\end{pmatrix}+\frac{1}{2}(\chi_1,\chi_2)\mathrm{M}^2_{\chi}\begin{pmatrix}
\chi_1\\
\chi_2\\
\end{pmatrix}
+M^2_{\phi^+\phi^-}
\phi^-\phi^+
\nonumber \\
&+ \text{interaction terms},
\end{align}
with the rotated tadpole terms 
\begin{align}
  t_{{h}}=c_\alpha t_2-s_\alpha t_1\text{,}\qquad t_{{H}}=s_\alpha t_2+c_\alpha t_1,
  \label{eq:tapo}
\end{align}
where the shorthands $c_\alpha=\cos\alpha$ and $s_\alpha=\sin\alpha$ are used.\\
For the potential to acquire a minimum at the vevs, i.e.\ for $h_i=\chi_i=\phi^+=0$ with $i=1,2$, one has to set the tadpole terms to zero in leading order. Using $t_h=t_H=0$ the diagonalization of the mass matrix $\mathrm{M}^2_{\text{Higgs}}$ fixes the mixing angle $\alpha$ at leading order to
\begin{equation}
t_{2\alpha}\equiv \tan (2\alpha)=\frac{8v_1v_2\lambda_{12}}{16v_1^2\lambda_1-v_2^2\lambda_2}.
\end{equation}
We enforce the mass hierarchy $M_\text{h}\leq M_\text{H}$ by allowing for  $\alpha\in(-\frac{\pi}{2},\frac{\pi}{2}]$ and choosing $\alpha$ such that $\lambda_{12}s_{2\alpha}\geq0$, leading to 
\begin{align}
s_{2\alpha}=\frac{8v_1v_2\lambda_{12}}{\sqrt{(8v_1v_2\lambda_{12})^2+(16v_1^2\lambda_1-v_2^2\lambda_2)^2}},\\
c_{2\alpha}=\frac{16v_1^2\lambda_1-v_2^2\lambda_2}{\sqrt{(8v_1v_2\lambda_{12})^2+(16v_1^2\lambda_1-v_2^2\lambda_2)^2}},
\label{eq:s-c-alpha}
\end{align}
with eigenvalues
\begin{align}
M^2_\text{h}&=\frac{1}{4}v_2^2\lambda_2+4v_1^2\lambda_1-\frac{1}{4}\sqrt{(8v_1v_2\lambda_{12})^2+(16v_1^2\lambda_1-v_2^2\lambda_2)^2},\\
M^2_\text{H}&=\frac{1}{4}v_2^2\lambda_2+4v_1^2\lambda_1+\frac{1}{4}\sqrt{(8v_1v_2\lambda_{12})^2+(16v_1^2\lambda_1-v_2^2\lambda_2)^2}.
\end{align}
Input parameters of a theory should always be intuitive and phenomenologically most easily accessible. We choose the masses $M_\text{h}$, $M_\text{H}$, the mixing angle $\alpha$, and the dimensionless coupling $\lambda_{12}$. The tadpole constants $t_{h}$ and $t_{H}$ are fixed by the definition of the EW vacuum, i.e.\ they do not count as input parameters, and are kept here only for later convenience. For the parameters introduced in the Higgs potential, expressed in terms of the new input parameters, we find
\begin{align}
\lambda_1&=\frac{1}{8 v_1^2}\left(c_\alpha^2M_\text{H}^2+s_\alpha^2M_\text{h}^2\right)+\frac{1}{8v_1^3}t_1,\\
\lambda_2&=\frac{2}{v_2^2}\left(c_\alpha^2 M_\text{h}^2+s_\alpha^2 M_\text{H}^2\right)+\frac{2}{v_2^3}t_2,\\
\mu_1^2&=\frac{1}{4}\left(s_\alpha^2 M_\text{h}^2+c_\alpha^2 M_\text{H}^2\right)+\frac{1}{2}v^2_2 \lambda_{12}+\frac{3}{4 v_1}t_1,\\
\mu_2^2&=v_1^2\lambda_{12}+\frac{1}{2}\left(c_\alpha^2 M_\text{h}^2+s_\alpha^2M_\text{H}^2\right)+\frac{3}{2 v_2} t_2,\\
v_1&=\frac{\left(M_\text{H}^2-M_\text{h}^2\right)s_{2\alpha}}{4v_2\lambda_{12}}.
\label{eq:par-rel-pot}
\end{align}
Using these relations it is easy to see that the stability conditions for the potential given in Eq.~\eqref{eq:Higgs-pot-stab} are automatically fulfilled for physical input values for $M_\text{h}$ and $M_\text{H}$. Note that the requirement of symmetry breaking, i.e.\ $\mu_1^2>0$ or $\mu_2^2>0$, is in the physical parameter space automatically fulfilled\footnote{To achieve the desired symmetry breaking only $\mu_1^2>0$ or  $\mu_2^2>0$ is needed. Therefore, the stated inequalities given in Eq.~(2.16) of Ref.~\cite{BogDit}, forcing $\mu_1^2>0$ \textit{and}  $\mu_2^2>0$, are not necessary.}.
Expanding the potential one finds scalar interaction terms of the form 
\begin{align}
V_{\text{int}}={}&c_\text{hhh}h^3+c_\text{hhH}h^2H+c_\text{hHH}hH^2+c_\text{HHH}H^3\nonumber\\&+
c_{\text{hhhh}}h^4+c_\text{hhhH}h^3H+c_\text{hhHH}h^2H^2+c_\text{hHHH}hH^3+c_\text{HHHH}H^4\nonumber\\&
+\frac{1}{2}(c_{\text{h}\phi\phi}h+c_{\text{H}\phi\phi}H+c_{\text{hh}\phi\phi}h^2+c_{\text{hH}\phi\phi}hH+c_{\text{HH}\phi\phi}H^2)(2\phi^+\phi^-+\chi_2^2)\nonumber\\&
+\frac{1}{2}(c_{\text{h}\chi\chi}h+c_{\text{H}\chi\chi}H+c_{\text{hh}\chi\chi}h^2+c_{\text{hH}\chi\chi}hH+c_{\text{HH}\chi\chi}H^2)\chi_1^2\nonumber\\&+\frac{\lambda_2}{16}(2\phi^+\phi^-+\chi_2^2)^2+\frac{\lambda_{12}}{2}\chi_1^2(\chi_2^2+2\phi^-\phi^+)+\lambda_1\chi_1^4,
\label{eq:scalar-self-coup}
\end{align}
with
\begin{alignat}{2}
  c_{\text{h}\chi\chi}&=2 v_2 c_\alpha \lambda_{12}-8v_1 s_\alpha \lambda_1,\qquad&  c_{\text{H}\chi\chi}&=8 v_1 c_\alpha \lambda_1+2 v_2 s_\alpha \lambda_{12},\\
  c_{\text{hh}\chi\chi}&=c_\alpha^2 \lambda_{12}+4 s_\alpha^2 \lambda_1,\qquad& c_{\text{hH}\chi\chi}&=2 s_\alpha c_\alpha \lambda_{12}-8s_\alpha c_\alpha\lambda_1 ,\\
  c_{\text{HH}\chi\chi}&=4 c_\alpha^2\lambda_1+s_\alpha^2 \lambda_{12},&&
\end{alignat}
and all other coupling constants of the Higgs potential are spelled out explicitly in Eq.~(2.18) of Ref.~\cite{BogDit}.

\subsection{Gauge part and physical gauge bosons}
\label{sec:gauge_part}
Different versions of generic $U(1)$ extensions of the SM have been already discussed qualitatively a long time ago in the literature, see e.g.\ Refs.~\cite{SchabWells,Zp1,Babu_ZZp_old,Wells,Hold}. To achieve a precision necessary for a meaningful confrontation of predictions with EW precision data all predictions should include at least NLO corrections. In the following, we present the DASM in $R_\xi$ gauge, which provides the most convenient and most common framework for calculating higher-order corrections.\\
Owing to the presence of the $U(1)_\text{d}$ gauge group, new terms in the Yang--Mills (YM) part of the Lagrangian are present in addition to the SM YM part,
\begin{align}
\mathcal{L}_\text{YM}=\mathcal{L}^\text{SM}_\text{YM}+\mathcal{L}_\text{YM}^\text{d}.
\end{align}
The new part
\begin{align}
\mathcal{L}_{\text{YM}}^{\text{d}}= -\frac{1}{4}C^{\mu\nu}C_{\mu\nu}-\frac{a}{2}  B^{\mu\nu}C_{\mu\nu}
\label{eq:lag-ym}
\end{align}
includes a mixing term of the gauge field $B^\mu$ of the SM $U(1)_\text{Y}$ group and the gauge field $C^\mu$ of the additional $U(1)_{\text{d}}$ group via the gauge-invariant field-strength tensors
\begin{align}
B_{\mu\nu}=\partial_\mu B_\nu-\partial_\nu B_\mu, \qquad C_{\mu\nu}=\partial_\mu C_\nu-\partial_\nu C_\mu
\end{align}
of the $U(1)_{\text{Y}}$ and the $U(1)_{\text{d}}$, respectively, with the parameter $a$ ruling the strength of this mixing. The kinetic terms are diagonalized by a redefinition of the fields \cite{Wells}
\begin{equation}
\begin{pmatrix}
C_\mu\\
B_\mu\\
\end{pmatrix}
=
\begin{pmatrix}
\frac{1}{\sqrt{1-a^2}} & 0 \\
-\frac{a}{\sqrt{1-a^2}} & 1 \\
\end{pmatrix}
\begin{pmatrix}
C'_\mu\\
B'_\mu\\
\end{pmatrix},
\label{eq:gauge-field-redef}
\end{equation}
leading to
\begin{align}
\mathcal{L}_{\text{YM}}&= -\frac{1}{4}C^{\mu\nu}C_{\mu\nu}-\frac{a}{2}B^{\mu\nu}C_{\mu\nu}-\frac{1}{4}  B^{\mu\nu}B_{\mu\nu}-\frac{1}{4}W^{b,\mu\nu}W^b_{\mu\nu}\nonumber\\&=-\frac{1}{4}C'^{\mu\nu}C'_{\mu\nu}-\frac{1}{4}B'^{\mu\nu}B'_{\mu\nu}-\frac{1}{4}W^{b,\mu\nu}W^b_{\mu\nu}.
\end{align}
Choosing $|a|>1$ would lead to a relative sign factor between the eigenvalues of the quadratic form of the fields $B_{\mu\nu}$ and $C_{\mu\nu}$ (or equivalently of $B'_{\mu\nu}$ and $C'_{\mu\nu}$) and thus to a wrong signature for one of the kinetic terms. Therefore, we constrain the parameter $a$ according to $|a|<1$ to maintain the self-consistency of the DASM.\\
Rewriting Eq.~\eqref{eq:cov-deriv-ym} with the help of Eq.~\eqref{eq:gauge-field-redef} and expanding the kinetic terms in Eq.~\eqref{eq:lag-higgs} with the decompositions \eqref{eq:higgs-fields} of the Higgs doublet and singlet fields leads to
\begin{align}
\mathcal{L}_{\text{M}_\text{V}}=\frac{1}{2}\left(B'_\mu,W^3_\mu,C'_\mu\right)\mathrm{M}^2_\text{V}\begin{pmatrix}
B'_\mu\\W^3_\mu\\C'_\mu
\end{pmatrix}
+M_\text{W}^2W^+W^-,
\end{align}
with the mass matrix for the neutral vector bosons
\begin{align}
\mathrm{M}^2_\text{V}=\begin{pmatrix}
\frac{s_\text{w}^2M_\text{W}^2 }{c_\text{w}^2} & \frac{s_\text{w} M_\text{W}^2}{c_\text{w}}  & -\frac{\eta s_\text{w}^2 M_\text{W}^2 }{c_\text{w}^2} \\
\frac{s_\text{w} M_\text{W}^2}{c_\text{w}}& M_\text{W}^2 & -\frac{\eta s_\text{w} M_\text{W}^2}{c_\text{w}} \\
-\frac{\eta s_\text{w}^2M_\text{W}^2}{c_\text{w}^2} & -\frac{\eta s_\text{w} M_\text{W}^2}{c_\text{w}} & \frac{\eta^2s_\text{w}^2 M_\text{W}^2}{c_\text{w}^2}+M_\text{C}^2
\\
\end{pmatrix},
\label{eq:mm-vb}
\end{align}
where
\begin{alignat}{3}
  \eta={}&\frac{a}{\sqrt{1-a^2}},\qquad&M_\text{C}={}&\tilde{e} v_1,&\qquad\tilde{e}={}&\frac{e_\text{d}}{\sqrt{1-a^2}},\nonumber\\
  s_\text{w}\equiv \sin\theta_\text{w}={}&\frac{g_1}{\sqrt{g_1^2+g_2^2}},&M_\text{W}={}&\frac{g_2 v_2}{2}.&&
\label{eq:v2_para_dep}
\end{alignat}
Diagonalizing the matrix $\mathrm{M}^2_\text{V}$ leads to the field basis for the mass eigenstates of the neutral vector bosons. Since $\text{rank}\left(\mathrm{M}^2_\text{V}\right)=2$, this can be achieved with a combination of two appropriate rotations of the fields in the form
\begin{align}
\begin{pmatrix}
B'_\mu\\W^3_\mu\\C'_\mu
\end{pmatrix}
=\mathrm{R}_\text{V}
\begin{pmatrix}
A_\mu\\Z_\mu\\Z'_\mu
\end{pmatrix},\qquad
\mathrm{R}_\text{V}
=
\begin{pmatrix}
c_\text{w}&s_\text{w}&\hspace{2pt}0\\
-s_\text{w}&c_\text{w}&\hspace{2pt}0\\
0 & 0 & \hspace{2pt}1\\
\end{pmatrix}
\begin{pmatrix}
  1 & \hspace{2pt}0 & 0 \\
  0 & \hspace{2pt}c_\gamma & -s_\gamma & \\
  0 & \hspace{2pt}s_\gamma & c_\gamma \\
\end{pmatrix},
\label{eq:gb_rot}
\end{align}
where $s_\gamma=\sin\gamma$, $c_\gamma=\cos\gamma$ for an appropriate mixing angle $\gamma$. For $s_\gamma=0$ the whole rotation \eqref{eq:gb_rot} reduces to a mere rotation by the mixing angle $\theta_\text{w}$ as in the SM. The rotation \eqref{eq:gb_rot} transforms $\mathrm{M}^2_\text{V}$ into
\begin{equation}
\mathrm{R}^\text{T}_\text{V}\mathrm{M}^2_\text{V}\mathrm{R}_\text{V}=\begin{pmatrix}
0 & 0 & 0 \\
 0 & M^2_\text{Z}&M^2_{\text{ZZ}'} \\
 0 & M^2_{\text{ZZ}'} & M^2_{\text{Z}'} 
\end{pmatrix},
\end{equation}
with 
\begin{align}
M^2_\text{Z}&=s_\gamma^2 M_\text{C}^2+\frac{M_\text{W}^2\left(c_\gamma- s_\gamma s_\text{w}\eta \right)^2}{c_\text{w}^2},\nonumber \\
M^2_{\text{Z}'}&=c_\gamma^2 M_\text{C}^2+\frac{ M_{\text{W}}^2\left(s_\gamma+c_\gamma s_\text{w} \eta\right)^2}{c_\text{w}^2},\nonumber \\
M^2_{\text{ZZ}'}&=s_\gamma c_\gamma M_\text{C}^2 +\frac{ M_{\text{W}}^2\left[s_{2\gamma}\left( s_\text{w}^2\eta^2 -1 \right)-2 s_\text{w} c_{2\gamma}\eta \right]}{2c_\text{w}^2}.
\label{eq:diag-cond-nc}
\end{align}
The diagonalization condition $M^2_{\text{ZZ}'}=0$ fixes the mixing angle $\gamma$ to 
\begin{equation}
t_{2\gamma}\equiv\tan2\gamma=\frac{-2 \eta s_\text{w}}{1-\eta^2s_\text{w}^2-c_\text{w}^2\frac{M_\text{C}^2}{M_{\text{W}}^2}}.
\label{eq:mixing-angle-gamma}
\end{equation}
The mass of the photon remains zero,
\begin{equation}
M^2_\text{A}=0,
\end{equation}
and requiring the photon--fermion couplings to reproduce its QED form, the electric unit charge $e$ is, as in the SM, directly related to the gauge couplings $g_1$, $g_2$ according to
\begin{align}
e=\frac{g_1g_2}{\sqrt{g_1^2+g_2^2}}.
\end{align}
Choosing $\gamma\in(-\frac{\pi}{4},\frac{\pi}{4}]$, Eq.~\eqref{eq:mixing-angle-gamma} leads to
\begin{align}
  s_{2\gamma}&=\frac{t_{2\gamma}}{\sqrt{1+t_{2\gamma}^2}}=-\frac{2 \eta s_\text{w} \text{sgn}\left\{1-\eta^2 s_\text{w}^2-c_\text{w}^2\frac{M_\text{C}^2}{M_{\text{W}}^2}\right\}}{\sqrt{\left(1-\eta^2 s_\text{w}^2-c_\text{w}^2\frac{M_\text{C}^2}{M_{\text{W}}^2}\right)^2+4\eta^2 s_\text{w}^2}},\\
c_{2\gamma}&=\frac{1}{\sqrt{1+t_{2\gamma}^2}}=\frac{\left|1-\eta^2 s_\text{w}^2-c_\text{w}^2\frac{M_\text{C}^2}{M_{\text{W}}^2}\right|}{\sqrt{\left(1-\eta^2 s_\text{w}^2-c_\text{w}^2\frac{M_\text{C}^2}{M_{\text{W}}^2}\right)^2+4\eta^2 s_\text{w}^2}},
\end{align}
and one finds the masses of the two massive, neutral gauge bosons $\text{Z}$ and $\text{Z}'$ to be  
\begin{align}
  M^2_\text{Z}={}&\frac{M_{\text{W}}^2}{c_\text{w}^2}\left(1- s_\text{w}t_{\gamma}\eta\right),\\
M^2_{\text{Z}'}={}&\frac{M_{\text{W}}^2}{c_\text{w}^2}\left(1+ \frac{s_\text{w}\eta}{t_\gamma}\right).
\end{align}
We do not impose a mass hierarchy between the $\text{Z}$ and $\text{Z}'$ bosons, i.e.~both $M_{\text{Z}'}\geq M_\text{Z}$ and $M_{\text{Z}'}<M_\text{Z}$ are possible.
In addition to the SM-like parameters $e$ and $M_\text{W}$, we choose $M_{\text{Z}'}$ and $\gamma$ as the two input parameters that fully fix the parameters of the gauge-boson sector of the DASM. Using Eqs.~\eqref{eq:diag-cond-nc} the dependent parameters are given by
\begin{align}
c_\text{w}=\frac{M_\text{W}}{\sqrt{c_\gamma^2 M_\text{Z}^2+s_\gamma^2M_{\text{Z}'}^2}},\qquad M^2_\text{C}=\frac{c_\text{w}^2M^2_\text{Z}M^2_{\text{Z}'}}{M_\text{W}^2},\qquad \eta=\frac{s_{2\gamma} c_\text{w}^2\left(M_{\text{Z}'}^2-M_\text{Z}^2\right)}{2s_\text{w}M_\text{W}^2},
\label{eq:par-rel-ym}
\end{align}
and further
\begin{align}
  \tilde{e}=\frac{c_\text{w}M_\text{Z}M_{\text{Z}'}}{v_1 M_\text{W}},\qquad e_\text{d}=\hspace{2pt}\frac{c_\text{w}M_\text{Z}M_{\text{Z}'}}{v_1 M_\text{W} \sqrt{1+\eta^2}},\qquad a=\frac{\eta}{\sqrt{1+\eta^2}}.
\end{align}
We have not inserted the full analytical dependence on the input parameters to keep the expressions compact. This includes in particular $v_1$ (see Eqs.~\eqref{eq:par-rel-pot} and \eqref{eq:v2_para_dep} for its explicit relation to input parameters). Without loss of generality we choose $e_\text{d}\geq 0$, i.e.\ we absorb the sign into the definition of $C_\mu$, which is possible since the sign of the parameter $a$, ruling the strength of the kinetic mixing in Eq.~\eqref{eq:lag-ym}, is not constrained.
Finally, we note that in contrast to the SM the W-boson mass $M_\text{W}$ is not equal to $c_\text{w}M_\text{Z}$, not even in lowest order, i.e.\ the $\rho$ parameter \cite{rhoparam,rhoparam2} is not equal to 1 in lowest order. This means that in order to describe EW precision data in the DASM, we have to pay the price of fine-tuning the mixing angle $\gamma$ to some small value.
By virtue of Eq.~\eqref{eq:par-rel-ym} the consistency of the theory further demands the additional restriction
  \begin{align}
    M_\text{W}^2<c_\gamma^2 M_\text{Z}^2+s_\gamma^2M_{\text{Z}'}^2
\end{align}
for the parameter space of the DASM.
\subsection{Gauge-fixing and ghost Lagrangian}
The kinetic terms of the Higgs fields contain mixing terms between the gauge-boson fields and the corresponding would-be Goldstone-boson fields. To avoid unpleasant non-diagonal propagators between gauge bosons and would-be Goldstone bosons at tree level we choose $R_\xi$ gauge-fixing conditions, i.e.~we introduce a gauge-fixing Lagrangian of the form
\begin{align}
  \mathcal{L}_\text{fix}=-\frac{1}{2\xi_A}\left(F^A\right)^2-\frac{1}{2\xi_Z}\left(F^Z\right)^2-\frac{1}{2\xi_{Z'}}\bigl(F^{Z'}\bigr)^2-\frac{1}{\xi_W}F^+F^-,
\end{align}
with the gauge-fixing functionals
\begin{align}
F^\pm&=\partial^\mu W^\pm_\mu\mp \text{i}\xi_W M_\text{W}\phi^\pm,\nonumber\\
F^Z&=\partial^\mu Z_\mu-\xi_Z\left[M_\text{C}s_\gamma\chi_1+\frac{ c_\text{w}c_\gamma M_\text{Z}^2}{M_\text{W}}\chi_2\right],\nonumber\\
F^{Z'}&=\partial^\mu Z'_\mu-\xi_{Z'}\left[M_\text{C}c_\gamma\chi_1-\frac{c_\text{w}s_\gamma M_{\text{Z}'}^2}{M_\text{W}}\chi_2\right],\nonumber\\
F^A&=\partial^\mu A_\mu.
\label{eq:gauge-fixing-funct}
\end{align}
Introducing a gauge-fixing of this form leads to a non-diagonal mass matrix $\mathrm{M}_\chi^2$ for the neutral would-be Goldstone fields. This mass matrix can be diagonalized by an appropriate rotation of the would-be Goldstone fields,
 \begin{align}
\begin{pmatrix}
\chi'\\
\chi\\
\end{pmatrix}=
\mathrm{R}^\text{T}_\chi
\begin{pmatrix}
\chi_1\\
\chi_2\\
\end{pmatrix},\qquad \mathrm{R}^\text{T}_\chi=  \begin{pmatrix}
c_x&-s_x\\
s_x&c_x\\
  \end{pmatrix},
\end{align}
with $s_x=\sin\theta_x$, $c_x=\cos\theta_x$. Choosing a common gauge parameter $\xi_i=\xi_\text{V}$, for the neutral gauge bosons \mbox{$i= Z, Z'$}, to avoid unnecessarily complicated expressions, leads to 
\begin{equation}
\mathrm{R}^\text{T}_\chi\mathrm{M}^2_{\chi}\mathrm{R}_\chi=\xi_\text{V}\begin{pmatrix}
M^2_{\chi'\chi'}&M^2_{\chi\chi'} \\
M^2_{\chi\chi'} & M^2_{\chi\chi} 
\end{pmatrix},
\end{equation}
with
\begin{align}
M^2_{\chi'\chi'}&=M_\text{C}^2c_x^2+\frac{M_{\text{W}}^2s_x^2+\eta s_\text{w} s_x M_{\text{W}} \left(2M_\text{C}c_xc_\text{w}+M_{\text{W}}\eta s_\text{w} s_x\right)}{c^2_\text{w}}-\frac{s_x^2 v_1 t_2+c_x^2 v_2 t_1}{v_1 v_2 \xi_\text{V}},\nonumber\\
M^2_{\chi\chi}&=M_\text{C}^2s_x^2+\frac{M_{\text{W}}^2c_x^2+\eta s_\text{w} c_xM_{\text{W}}\left(M_{\text{W}}\eta s_\text{w} c_x-2M_\text{C}c_\text{w}s_x\right)}{c_\text{w}^2}-\frac{c_x^2 v_1 t_2+s_x^2 v_2 t_1}{v_1v_2 \xi_\text{V}},\nonumber\\
M^2_{\chi\chi'}&=\frac{s_{2x}}{2}\left(M_\text{C}^2-\frac{M_{\text{W}}^2}{c_\text{w}^2}\right) -\frac{M_{\text{W}}\eta s_\text{w}\left(2M_\text{C} c_\text{w} c_{2x}+M_{\text{W}}\eta s_\text{w} s_{2x}\right)}{2c_\text{w}^2}+\frac{s_x c_x \left(v_1 t_2-v_2 t_1\right)}{v_1 v_2 \xi_\text{V}}.
\end{align}
Requiring $M^2_{\chi\chi'}=0$ and setting tadpole terms to zero suggests the definition
\begin{align}
  t_{2x}\equiv\tan (2\theta_x)=\frac{-2\eta M_\text{C}s_\text{w}c_\text{w}}{M_{\text{W}}\biggl(1+\eta^2s_\text{w}^2-c_\text{w}^2\frac{M^2_\text{C}}{M^2_{\text{W}}}\biggr)}=\frac{c_\text{w} M_\text{C} t_{2\gamma} }{M_\text{W}\left(1-\eta s_\text{w} t_{2\gamma}\right)},
  \end{align}
which eliminates mixing terms between $\chi$ and $\chi'$ in lowest order.
We choose
\begin{align}
  &s_{2x}=\frac{c_\text{w}s_{2\gamma}M_\text{C}}{M_\text{W}},\qquad c_{2x}=c_{2\gamma}-s_\text{w}\eta s_{2\gamma},
\label{eq:sin-cos-thetax}
\end{align}
to align the masses of the two would-be Goldstone bosons  
\begin{align}
M_{\chi}=\sqrt{\xi_\text{V}}M_Z,\qquad M_{\chi'}=\sqrt{\xi_\text{V}}M_{Z'}
\end{align}
with the masses of the respective gauge bosons (for $t_1,t_2\to 0$).
Moreover, we find
\begin{align}
s_x=\frac{c_\text{w} s_\gamma M_{\text{Z}'}}{M_\text{W}},\qquad c_x=\frac{c_\text{w} c_\gamma M_\text{Z}}{M_\text{W}},\qquad t_x=\frac{M_{\text{Z}'}}{M_\text{Z}}t_\gamma.
\end{align}
With these simplifications the gauge-fixing functionals given by Eq.~\eqref{eq:gauge-fixing-funct} reduce to
\begin{align}
F^\pm&=\partial^\mu W^\pm_\mu\mp \text{i}\xi_\text{W}M_\text{W}\phi^\pm,\nonumber\\
F^Z&=\partial^\mu Z_\mu-\xi_\text{V}M_\text{Z}\chi,\nonumber\\
F^{Z'}&=\partial^\mu Z'_\mu-\xi_\text{V}M_{\text{Z}'}\chi',\nonumber\\
F^A&=\partial^\mu A_\mu.
\label{eq:gauge-fixing-funct-FtH}
\end{align}
Following the Faddeev--Popov procedure, unphysical anti-commuting scalar ghost fields $u^a$, $\bar{u}^a$, ($a=\pm,Z,Z',A$), needed for the consistency of the gauge-fixing procedure in the functional integral, are introduced.
The ghost Lagrangian is given by 
\begin{align}
\mathcal{L}_{\text{FP}}(x)=-\int d^4y\,\bar{u}^a(x)\left(\frac{\delta F^a(x)}{\delta \theta^b(y)}\right)u^b(y),
\label{eq:lag-fp}
\end{align}
with $a,b=\pm,Z,Z',A$. In the chosen $R_\xi$ gauge all masses of the ghost fields coincide with the masses of the corresponding would-be Goldstone bosons. The gauge transformations of the fields, which are needed for the evaluation of Eq.~\eqref{eq:lag-fp}, as well as the explicit form of $\mathcal{L}_{\text{FP}}$ can be found in Appendix \ref{app:ghost-part}.

\subsection{Fermion sector}

The DASM extends the fermion sector of the SM by right-handed neutrino fields $\nu'^{\text{R}}_j$, \mbox{$j=e,\mu,\tau$,} corresponding to the left-handed SM-like neutrino fields $\nu'^{\text{L}}_j$, and an additional non-chiral Dirac fermion $f'_{\text{d}}$ of the hidden sector. The primes on the fields indicate the use of the gauge-interaction eigenbasis. The field $f'_\text{d}$ is assumed to carry only the charge $\tilde{q}_{\text{f}}e_\text{d}$ of the $U(1)_\text{d}$ gauge group, but neither weak isospin, nor weak hypercharge, nor colour. Since $f'_\text{d}$ is non-chiral, no anomalies are introduced in the DASM. For the relative coupling strength we choose
\begin{align}
  \tilde{q}_{\text{f}}=\tilde{q}_{\rho}=1,
\end{align}
where $q_{\text{d},\rho}e_\text{d}$ is the $U(1)_\text{d}$ charge of the Higgs singlet $\rho$. While $\tilde{q}_{\rho}=1$, as already mentioned above, simply provides the normalization of the $U(1)_\text{d}$ coupling $e_\text{d}$, the choice of $\tilde{q}_{\text{f}}=1$ allows for a Yukawa coupling in the Lagrangian connecting the Higgs field $\rho$, the new fermion field of the dark sector $f'_{\text{d}}$, and the right-handed neutrinos $\nu'^{\text{R}}_{j}$, opening another portal between the SM and the dark sector. 
The terms that need to be introduced in the Lagrangian in addition to the SM-like fermionic terms $\mathcal{L}^\text{SM}_\text{Fermion}$ (with massless left-handed neutrinos) read 
\begin{align}
\mathcal{L}_\text{Fermion}={}&\mathcal{L}^\text{SM}_\text{Fermion}+\bar{f}'_{\text{d}}\left(\text{i}\slashed{D}_\text{d}-m_{\text{f}_{\text{d}}}\right)f'_{\text{d}}+\sum_{j=e,\mu,\tau}\left[\bar\nu'^{\text{R}}_{j}\text{i}\slashed{\partial}\nu'^{\text{R}}_{j}-\left(y_{\rho,j}\rho\bar{f}'^{\text{L}}_{\text{d}}\nu'^{\text{R}}_{j}+\text{h.c.}\right)\right]\nonumber\\
  &-\sum_{k,l=\text{e},\mu,\tau} \left(\bar{L}'^{\text{L}}_{k} G'^\nu_{kl}\nu'^{\text{R}}_{l} \Phi^\text{C}+ \text{h.c.}\right),
\label{eq:lag-ferm-exten}
\end{align}
with the covariant derivative $D_\text{d}^\mu$ of the dark sector given in Eq.~\eqref{eq:cov-deriv-ym2}. The first term in Eq.~\eqref{eq:lag-ferm-exten} contains the gauge interactions of $f'_{\text{d}}$ as well as its Dirac mass term with the mass parameter $m_{\text{f}_\text{d}}$. We can assume $m_{\text{f}_\text{d}}$ as real and non-negative, $m_{\text{f}_\text{d}}\geq 0$, upon adjusting a chiral phase of $f'_\text{d}$ appropriately. The sum over $j$ contains the kinetic terms of the right-handed neutrinos as well as the Yukawa terms including $f'_{\text{d}}$, the right-handed neutrino fields $\nu'^{\text{R}}_{j}$, and the Higgs singlet $\rho$. The sum over $k$, $l$ introduces Yukawa terms involving the SM-like neutrinos and the charge-conjugate \mbox{$\Phi^\text{C}=\text{i} \tau^2 \Phi^*$} of the SM-like Higgs doublet. Here $G'^\nu_{kl}$ denote the respective Yukawa coupling constants. We do not consider the possibility of Majorana mass terms for the right-handed neutrinos. Using Eq.~\eqref{eq:higgs-fields} the mass terms in $\mathcal{L}_\text{Fermion}$ for the neutral fermion fields $\nu'^{\text{L}}_{k}$, $\nu'^{\text{R}}_{k}$, and $f'_{\text{d}}= f'^{\text{L}}_{\text{d}}+f'^{\text{R}}_{\text{d}}$ (with $f'^{\text{L/R}}_{\text{d}}=\omega_{\text{L/R}}f'_{\text{d}}$ and the chiral projectors $\omega_{\text{L/R}}=\frac{1}{2}(1\mp \gamma_5)$) become   
\begin{align}
\mathcal{L}_{f'_{\text{d}}\nu}^\text{m}=-
\biggl(\bar{\nu}'^{\text{L}}_{\text{e}},\bar{\nu}'^{\text{L}}_{\mu},\bar{\nu}'^{\text{L}}_{\tau},\bar{f}'^{\text{L}}_{\text{d}}\biggr){\mathrm{M}}'_{\text{f}_\text{d}}\begin{pmatrix}
\nu'^{\text{R}}_{\text{e}}\\
\nu'^{\text{R}}_{\mu}\\
\nu'^{\text{R}}_{\tau}\\
f'^{\text{R}}_{\text{d}}
\end{pmatrix} + \text{h.c. ,}
\end{align}
with
\begin{align}
{\mathrm{M}}'_{\text{f}_\text{d}}=
\begin{pmatrix}
 m_{11} & m_{12} &  m_{13} & 0 \\
 m_{21} & m_{22} &  m_{23} & 0 \\
 m_{31} & m_{32} &  m_{33} & 0 \\
 \tilde{y}_\text{e} & \tilde{y}_\mu & \tilde{y}_\tau & m_{\text{f}_\text{d}} \\
 \end{pmatrix}, \hspace{0.5cm} \tilde{y}_i=\frac{v_1 y_{\rho,i}}{\sqrt{2}}, \hspace{0.5cm} m_{ij}=\frac{v_2 G^\nu_{ij}}{\sqrt{2}}.
\label{eq:mass-matrix-ferm}
\end{align}
In general $\mathrm{M}'_{\text{f}_\text{d}}$ is a complex, non-symmetric matrix. A bi-unitary transformation
\begin{align}
\begin{pmatrix}
\nu'^{\text{R}}_{\text{e}}\\
\nu'^{\text{R}}_{\mu}\\
\nu'^{\text{R}}_{\tau}\\
f'^{\text{R}}_{\text{d}}
\end{pmatrix}=\mathrm{U}_\text{R}\begin{pmatrix}
\nu^{\text{R}}_{1}\\
\nu^{\text{R}}_{2}\\
\nu^{\text{R}}_{3}\\
\nu^{\text{R}}_{4}
\end{pmatrix},\qquad\begin{pmatrix}
\nu'^{\text{L}}_{\text{e}}\\
\nu'^{\text{L}}_{\mu}\\
\nu'^{\text{L}}_{\tau}\\
f'^{\text{L}}_{\text{d}}
\end{pmatrix}=\mathrm{U}_\text{L}\begin{pmatrix}
\nu^{\text{L}}_{1}\\
\nu^{\text{L}}_{2}\\
\nu^{\text{L}}_{3}\\
\nu^{\text{L}}_{4}
\end{pmatrix},
\end{align}
can be used to transform the fields into the mass eigenbasis of the resulting four neutrino-like states,
\begin{align}
{\mathrm{U}}^\dagger_\text{L}{\mathrm{M}}'_{\text{f}_\text{d}}{\mathrm{U}}_\text{R}= {\mathrm{M}}_{\nu}, \hspace{1cm} {\mathrm{M}}_{\nu,\alpha\beta}= m_{\nu,\alpha}\delta_{\alpha\beta}, \hspace{1cm} m_{\nu,\alpha}\geq 0,\qquad \alpha,\beta=1,2,3,4.
\label{eq:bi-uni-trafo}
\end{align}
\subsubsection{Qualitative discussion of the neutrino mixing}
\label{sec:simplified-case}
In this section we consider the main features of the diagonalization process of the mass matrix of the neutral fermions in the limit of small SM-like neutrino masses. The results from this consideration are then used in the next section to define an approximation to the complete fermion sector of the DASM that is sufficient for applications to collider phenomenology.\\
We assume all entries of the SM-like mass matrix $(m_{kl})$ to be of the scale $m_\nu$ representing the generic scale of the SM-like neutrino masses, $m_{kl}=\mathcal{O}(m_\nu)$. Inspired by the experimental evidence for very small neutrino masses we consider the case
\begin{align}
  m_\nu \ll \widetilde{m}\equiv \max\{\tilde{y},m_{\text{f}_\text{d}}\},\qquad \tilde{y}^2\equiv|\tilde{y}_\text{e}|^2+|\tilde{y}_\mu|^2+|\tilde{y}_\tau|^2,\qquad \tilde{y}\geq 0.
\end{align}
The four neutrino masses are obtained as the square roots of the (non-negative) eigenvalues of the matrices $\text{M}'^\dagger_{\text{f}_\text{d}}\text{M}'_{\text{f}_\text{d}}$ or $\text{M}'_{\text{f}_\text{d}}\text{M}'^\dagger_{\text{f}_\text{d}}$. The latter has the structure
\begin{align}
\text{M}'_{\text{f}_\text{d}}\text{M}'^\dagger_{\text{f}_\text{d}}= \begin{pmatrix}
 \begin{matrix}\mbox{\text{\Large $\mathcal{O}(m_\nu^2)$}}\end{matrix} &\begin{matrix} \mathcal{O}\left(m_\nu \widetilde{m}\right) \\   \mathcal{O}\left(m_\nu \widetilde{m}\right) \\  \mathcal{O}\left(m_\nu \widetilde{m}\right) \end{matrix}\\
\begin{matrix} \mathcal{O}\left(m_\nu \widetilde{m}\right) & \mathcal{O}\left(m_\nu \widetilde{m}\right) & \mathcal{O}\left(m_\nu \widetilde{m}\right) \end{matrix} &\begin{matrix} m^2_{\text{f}_\text{d}} + \tilde{y}^2 \end{matrix}  \\
\end{pmatrix}.
\label{eq:mat_mmd_order}
\end{align}
Note that in the notation used in Eq.~\eqref{eq:mat_mmd_order} the symbol $\mathcal{O}(m_\nu^2)$ should be read as a $3\times 3$ matrix with elements of order $\mathcal{O}(m_\nu^2)$.
From this form, we read that the neutrino masses obey the hierarchy
\begin{align}
  m_{\nu_k}=\mathcal{O}(m_\nu),\quad k=1,2,3, \qquad m_{\nu_4}=\sqrt{m^2_{\text{f}_\text{d}} + \tilde{y}^2}+\mathcal{O}(m_\nu),
\end{align}
and that the mixing matrix $\text{U}_\text{L}$, which is the unitary matrix diagonalizing $\text{M}'_{\text{f}_\text{d}}\text{M}'^\dagger_{\text{f}_\text{d}}$,
\begin{align}
\text{U}^\dagger_{\text{L}}\text{M}'_{\text{f}_\text{d}}\text{M}'^\dagger_{\text{f}_\text{d}}\text{U}_\text{L}= \text{diag}(m^2_{\nu_1},m^2_{\nu_2},m^2_{\nu_3},m^2_{\nu_4}),
\end{align}
has the form
\begin{align}
  \mathrm{U}_\text{L}=
    \begin{pmatrix}
\begin{array}{ *{4}{c} }
    & & & 0 \\
    & & & 0 \\
    \multicolumn{3}{c}
      {\,\raisebox{\dimexpr\normalbaselineskip+.7\ht\strutbox-.43\height}[0pt][0pt]
        {\scalebox{1.3}{$\hat{\text{U}}_\text{L}$}}} & 0 \\
   0 \,& 0\, & 0\, & 1
\end{array}
    \end{pmatrix}+\mathcal{O}\left(\frac{m_\nu}{\widetilde{m}}\right),
    \qquad \hat{\text{U}}_\text{L}=\mathcal{O}(1),
    \label{eq:ULHat_neut}
\end{align}
where $\hat{\text{U}}_\text{L}$ is a unitary $3\times3$ matrix ruling the mixing in the sector of the light SM-like neutrinos.
The unitary matrix $\text{U}_\text{R}$, which diagonalizes $\text{M}'^\dagger_{\text{f}_\text{d}}\text{M}'_{\text{f}_\text{d}}$,
\begin{align}
\text{U}^\dagger_{\text{R}}\text{M}'^\dagger_{\text{f}_\text{d}}\text{M}'_{\text{f}_\text{d}}\text{U}_\text{R}= \text{diag}(m^2_{\nu_1},m^2_{\nu_2},m^2_{\nu_3},m^2_{\nu_4}),
\end{align}
is more complicated and leads to mixing between the right-handed SM-like neutrino fields and $f'^{\text{R}}_{\text{d}}$ which is, in contrast to the respective case for the left-handed fields, not suppressed by $m_\nu$. Motivated by the hierarchy of the coefficients in $\text{M}'_{\text{f}_\text{d}}$,
\begin{align}
 \text{M}'_{\text{f}_\text{d}}=
  \begin{pmatrix}
    \begin{array}{ *{4}{c} }
      & & & 0 \\
      & & & 0 \\
      \multicolumn{3}{c}
      {\,\raisebox{\dimexpr\normalbaselineskip+.7\ht\strutbox-.5\height}[0pt][0pt]
        {\scalebox{1.15}{$\mathcal{O}(m_\nu)$}}} & 0 \\
    \tilde{y}_\text{e} & \tilde{y}_\mu & \tilde{y}_\tau & m_{\text{f}_\text{d}}  \\
    \end{array}
  \end{pmatrix},
\end{align}
we decompose $\text{U}_\text{R}$ into two factors, $\text{U}_\text{R}=\text{U}_{\text{R},1}\text{U}_{\text{R},2}$, where $\text{U}_{\text{R},1}$ aligns the vector $(\tilde{y}_\text{e}, \tilde{y}_\mu,\tilde{y}_\tau)$ along the $(0,0,1)$ direction and $\text{U}_{\text{R},2}$ takes care of the remaining rotation in the $\nu_3$-$\nu_4$ plane. In detail, we define
\begin{align}
\text{U}_{\text{R},1}=  \begin{pmatrix}
  \begin{array}{ *{4}{c} }
    \multirow{3}{*}{\text{\Large$\mathbf{e'}$}}& \multirow{3}{*}{\text{\Large$\mathbf{e''}$}} & \multirow{3}{*}{\text{\Large$\hspace{-0.5pt}\mathbf{e\textcolor{white}{'}}$}}&\, 0 \\ && &\,0\\ &&&\,0\\
    0 & \hspace{-1pt}0 &\hspace{-3pt} 0 &\,1\\
    \end{array}
  \end{pmatrix},\qquad (\tilde{y}_\text{e}, \tilde{y}_\mu,\tilde{y}_\tau)=\tilde{y}\, \mathbf{e}^\dagger,\qquad \tilde{y}\geq 0,
\label{eq:rot_UR1}
\end{align}
with \{$\mathbf{e},\mathbf{e'},\mathbf{e''}$\} forming an orthonormal system of 3-vectors, i.e.~$|\mathbf{e}|=|\mathbf{e'}|=|\mathbf{e''}|=1$, $\mathbf{e}^\dagger \mathbf{e'}=0$, etc., so that
\begin{align}
 \text{M}'_{\text{f}_\text{d}}\text{U}_{\text{R},1}=
  \begin{pmatrix}
    \begin{array}{ *{4}{c} }
      & & & \hspace{-2pt}0 \\
      & & & \hspace{-2pt}0 \\
      \multicolumn{3}{c}
      {\,\raisebox{\dimexpr\normalbaselineskip+.7\ht\strutbox-.5\height}[0pt][0pt]
        {\scalebox{1.15}{$\mathcal{O}(m_\nu)$}}} & \hspace{-2pt}0 \\
    0\, & \,0 & \hspace{-2pt}\tilde{y} & \hspace{-2pt}m_{\text{f}_\text{d}}  \\
    \end{array}
 \hspace{-2pt} \end{pmatrix}.
\end{align}
The second matrix $\text{U}_{\text{R},2}$ rotates the 2-vector $(\tilde{y},m_{\text{f}_\text{d}})$ in the $\nu_3$-$\nu_4$ plane into $(0,m_{\nu_4})$,
\begin{align}
  \text{M}'_{\text{f}_\text{d}}\text{U}_{\text{R},1}\text{U}_{\text{R},2}=\text{M}'_{\text{f}_\text{d}}= \begin{pmatrix}
    \begin{array}{ *{4}{c} }
      & & & \hspace{-2pt}0 \\
      & & & \hspace{-2pt}0 \\
      \multicolumn{3}{c}
      {\,\raisebox{\dimexpr\normalbaselineskip+.7\ht\strutbox-.5\height}[0pt][0pt]
        {\scalebox{1.15}{$\mathcal{O}(m_\nu)$}}} & \hspace{-2pt}\mathcal{O}\left(m_\nu\right) \\
    0\, & \,0 & 0 & \hspace{-2pt}m_{\nu_4} \\
    \end{array}
    \hspace{-2pt} \end{pmatrix}.
\end{align}
Thus, while $\text{U}_{\text{R},1}$ simply aligns the SM-like right-handed neutrino fields, the rotation $\text{U}_{\text{R},2}$ dictates the mixing of the right-handed SM-like neutrino fields with the fermion from the dark sector.

\subsubsection{Simplified fermion sector of the DASM}
\label{sec:simp_ferm_DASM}
The results of the previous section illustrate that it is possible to restrict any analysis of observables that are not sensitive to SM-like neutrino masses to the case $m_{ij}=0$ ($i,j=1,2,3$). After aligning the right-handed SM-like neutrino fields with the help of \eqref{eq:rot_UR1} the Lagrangian \eqref{eq:lag-ferm-exten} simplifies to
\begin{align}
\mathcal{L}_{\text{Fermion}}^\text{add}={}&\bar{f}'_{\text{d}}\left(\text{i}\slashed{D}_\text{d}-m_{{\text{f}_\text{d}}}\right)f'_{\text{d}}-\left(y_{\rho}\rho\bar{f}'^{\text{L}}_{\text{d}}\nu'^{\text{R}}_{3}+\text{h.c.}\right)+\sum_{j=1,2,3}\left[\bar\nu'^{\text{R}}_{j}\text{i}\slashed{\partial}\nu'^{\text{R}}_{j}\right],
\label{eq:L_ferm_simp_align}
\end{align}
with $y_\rho=\sqrt{2}\tilde{y}/v_1$.
The resulting mass matrix for the SM-like neutrinos and the dark fermion $f'_\text{d}$ reads
\begin{align}
\mathrm{M}'_{\text{f}_\text{d}}=
\begin{pmatrix}
 0 & 0 & 0 & 0 \\
 0 & 0 & 0 & 0 \\
 0 & 0 & 0 & 0 \\
 0 & 0 & \tilde{y} & m_{\text{f}_\text{d}} \\
\end{pmatrix}.
\end{align}
Note that both $\tilde{y}$ and $m_{\text{f}_\text{d}}$ can be assumed real and non-negative as became clear in the discussion of the previous section.
The very simple form of $\mathrm{M}'_{\text{f}_\text{d}}$ allows for the ansatz
\begin{align}
\mathrm{U}_\text{L}=\begin{pmatrix}
 1 & 0 & 0 & 0 \\
 0 & 1 & 0 & 0 \\
 0 & 0 & \cos \theta_\text{l} & \sin \theta_\text{l}  \\
 0 & 0 & - \sin \theta_\text{l}  & \cos \theta_\text{l}  \\
 \end{pmatrix}, \qquad \mathrm{U}_{\text{R},2}=\begin{pmatrix}
 1 & 0 & 0 & 0 \\
 0 & 1 & 0 & 0 \\
 0 & 0 & \cos \theta_\text{r} & \sin \theta_\text{r}  \\
 0 & 0 & - \sin \theta_\text{r}  & \cos \theta_\text{r}  \\
\end{pmatrix},
\label{eq:unit-matices-ferm}
\end{align}
for the rotation matrices leading to the fields $\nu^{\text{L/R}}_{i}$, $i=1,2,3,4$, that correspond to mass eigenstates. The diagonalization conditions then read
\begin{align}
0= \sin\theta_\text{l}\left(\tilde{y}\sin\theta_\text{r}+ m_{\text{f}_\text{d}} \cos \theta_\text{r}\right),\qquad
  0= \cos\theta_\text{l}\left(\tilde{y}\cos\theta_\text{r}- m_{\text{f}_\text{d}} \sin \theta_\text{r}\right).
\label{eq:diag-cond-ferm}
\end{align}
Equations~\eqref{eq:diag-cond-ferm} have the solution
\begin{align}
  \sin\theta_\text{l}= 0,\qquad
  \tan\theta_\text{r}=\frac{\tilde{y}}{m_{\text{f}_\text{d}}},
  \label{eq:theta-l-r}
\end{align}
so that we can choose
\begin{align}
  \theta_\text{l}=0,\qquad  s_{\theta_\text{r}}\equiv \sin\theta_\text{r}= \frac{\tilde{y}}{\sqrt{\tilde{y}^2+m^2_{\text{f}_\text{d}}}},\qquad  c_{\theta_\text{r}}\equiv\cos\theta_\text{r}= \frac{m_{\text{f}_\text{d}}}{\sqrt{\tilde{y}^2+m^2_{\text{f}_\text{d}}}}.
\end{align}
In particular, we have $\mathrm{U}_\text{L}= \mathbbm{1}_4$. This means that in the approximation of neglecting $m_{\nu_1,\nu_2,\nu_3}$ the flavour eigenstates for each of the left-handed fields are already aligned to the mass eigenstates, but the right-handed flavour eigenstates need to be rotated to correspond to mass eigenstates. The masses are given by
\begin{align}
m_{\nu_1}={}m_{\nu_2}=m_{\nu_3}=0,\qquad m_{\nu_4}={}\sqrt{\tilde{y}^2+m^2_{\text{f}_\text{d}}}.
\label{eq:m-nue-fd}
\end{align}
Moreover, it should be noted that for $m_{\nu_1,\nu_2,\nu_3}=0$ any further unitarity transformation of the SM-like neutrino fields (provided by some $\text{U}_{\text{L}}$ of the form \eqref{eq:ULHat_neut} with $m_\nu=0$) is unobservable as in the SM, so that there is no CKM-like mixing in the charged-current interaction of the leptons.
As an intuitive set of input parameters we choose the mass $m_{\nu_4}$ of the heavy neutrino $\nu_4$ and the mixing angle $\theta_\text{r}\in[0,\frac{\pi}{2}]$ for the right-handed fields. Using Eqs.~\eqref{eq:theta-l-r} and \eqref{eq:m-nue-fd} we find
\begin{align}
  \tilde{y}= s_{\theta_\text{r}} m_{\nu_4},\qquad
  m_{\text{f}_\text{d}}=c_{\theta_\text{r}}m_{\nu_4}
\label{eq:para-rel-ferm-sec}
\end{align}
for the additional parameters introduced in the fermion part of the Lagrangian.\\
Rewriting Eq.~\eqref{eq:L_ferm_simp_align} in terms of $\nu_i$ ($i=1,2,3,4$) leads to
\begin{align}
  \mathcal{L}^\text{add}_\text{Fermion}={}&\bar{\nu}_{4}\left(\text{i}\slashed{\partial}-m_{\nu_4}\right)\nu_{4}+\sum_{j=1,2,3}\bar\nu^{\text{R}}_{j}\text{i}\slashed{\partial}\nu^{\text{R}}_{j}\nonumber\\
  &+\frac{s_{\theta_\text{r}}m_{\nu_4}}{v_1}\left(s_{\theta_\text{r}}s_\alpha h \bar{\nu}^{\text{L}}_{4}\nu^{\text{R}}_{4}+c_{\theta_\text{r}}s_\alpha h \bar{\nu}^{\text{L}}_{4}\nu^{\text{R}}_{3}-s_{\theta_\text{r}} c_\alpha H \bar{\nu}^{\text{L}}_{4}\nu^{\text{R}}_{4}-c_{\theta_\text{r}}c_\alpha H \bar{\nu}^{\text{L}}_{4}\nu^{\text{R}}_{3}+\text{h.c.}\right)\nonumber\\
  & -\tilde{e}\left(s_\gamma Z_\mu+c_\gamma Z'_\mu\right)\bigl(\bar{\nu}^{\text{L}}_{4}\gamma^\mu \nu^{\text{L}}_{4}+c^2_{\theta_\text{r}}\bar{\nu}^{\text{R}}_{4}\gamma^\mu \nu^{\text{R}}_{4}+s^2_{\theta_\text{r}}\bar{\nu}^{\text{R}}_{3}\gamma^\mu \nu^{\text{R}}_{3}\nonumber\\
&  -\left[s_{\theta_\text{r}}c_{\theta_\text{r}} \bar{\nu}^{\text{R}}_{4}\gamma^\mu \nu^{\text{R}}_{3}+\text{h.c.}\right]\bigl),
\end{align}
from which the couplings of the neutrinos $\nu_i$ ($i=1,2,3,4$) can be easily read off.

\subsection{Input parameters}
For the input parameters originating from the SM part of the DASM we use the precisely measured masses of the gauge bosons, $M_\text{W}$, $M_\text{Z}$, the electromagnetic coupling $\alpha_\text{em}=e^2/(4\pi)$, the fermion masses, the CKM matrix elements $V_{ij}$, and the mass of the SM-like Higgs boson, which can be $\text{h}$ or $\text{H}$. The various extensions introduced above lead to additional free parameters: 
\begin{itemize}
\item There are two additional free parameters in the gauge sector. Using Eq.~\eqref{eq:par-rel-ym} we express $a$ and $e_\text{d}$ in terms of $M_{\text{Z}'}$ and $\gamma$ and use the latter two as input parameters.
\item There are three extra parameters, $\lambda_1$, $\lambda_{12}$, $\mu_1^2$, introduced in the Higgs sector in addition to the SM Higgs parameters. Using Eq.~\eqref{eq:par-rel-pot} we express the additional parameters coming from the Higgs potential in terms of $M_{\text{h}'}$, $\alpha$, and $\lambda_{12}$, which are used as input parameters, where $M_{\text{h}'}$ is the non-SM-like Higgs boson of $\text{h}, \text{H}$. Note that the mixing angle $\alpha$ is most directly connected to the ``signal strength'' of the SM-like Higgs boson as measured by LHC collaborations.
\item The new parameters $\tilde{y}$ and $m_{\text{f}_\text{d}}$ of the fermion sector are expressed in terms of the input quantities $\theta_\text{r}$ and $m_{\nu_4}$ using Eq.~\eqref{eq:para-rel-ferm-sec}.
\end{itemize}
In total we use 
\begin{equation}
\{M_\text{W},M_\text{Z},M_{\text{Z}'},M_\text{H},M_\text{h},\alpha_\text{em},\gamma,\alpha,\theta_\text{r},\lambda_{12},m_{f,i},m_{\nu_4},V_{ij}\}
\label{eq:parameterset}
\end{equation}
as a particularly intuitive set of input parameters for the DASM.

\section{Renormalization of the DASM}
\label{se:renormierung}
In order to match the accuracy of EW precision data from the LHC and previous colliders, including the $(g-2)_\mu$ result, at least NLO predictions for precision observables are needed. Thus, the theory needs to be renormalized.
In this section we introduce two types of renormalization schemes for the DASM and give the resulting expressions for the various renormalization constants at NLO. The first scheme is based on on-shell (OS) renormalization conditions as far as possible, the second employs $\MSbar$ conditions for some of the parameters of the non-SM sectors.\\

\subsection{Renormalization transformation}
\label{se:reno_trafos}
We choose a multiplicative renormalization procedure for the input parameters and the fields. To this end, at NLO we define the following renormalization transformations,   
\begin{align}
&M^2_{\text{h},0}=M^2_\text{h}+\delta M^2_\text{h},&{}&M^2_{\text{H},0}=M^2_\text{H}+\delta M^2_\text{H},\nonumber\\
&M^2_{\text{Z},0}=M^2_{\text{Z}}+\delta M^2_{\text{Z}},&{}&M^2_{{\text{Z}'},0}=M^2_{\text{Z}'}+\delta M^2_{\text{Z}'},\nonumber\\
&M^2_{\text{W},0}=M^2_{\text{W}}+\delta M^2_{\text{W}},&{}&e_0=\left(1+\delta Z_e\right)e=e+\delta e,\nonumber\\
&\gamma_0=\gamma+\delta\gamma,&{}&\alpha_0=\alpha+\delta\alpha,\nonumber\\
&\lambda_{12,0}=\lambda_{12}+\delta\lambda_{12},&{}&m_{f,i,0}=m_{f,i}+\delta m_{f,i},\nonumber\\
&V_{ij,0}=V_{ij}+\delta V_{ij},&{}&\theta_{\text{r},0}= \theta_\text{r}+\delta\theta_\text{r},&
\label{eq:ren_trafos_para}
\end{align}
for the input parameters of the SM and dark parts of the model. Here we denote the bare parameters and fields with a subscript ``0'' to distinguish them from their renormalized counterparts without this subscript; the quantities $\delta p$ are the renormalization constants for the respective parameters $p$. After the rotation of the fields into a basis corresponding to mass eigenstates, the renormalization transformation for the fields reads
\begin{align}
&&\begin{pmatrix}h_{0}\\H_{0}\\\end{pmatrix}&=\left(\mathbbm{1}_2 +\frac{1}{2} \delta \mathrm{Z}_\text{S}\right)\begin{pmatrix}h\\H\\\end{pmatrix},\qquad &\delta\mathrm{Z}_\text{S}&=\begin{pmatrix}\delta Z_{hh}&&\delta Z_{hH}\\ \delta Z_{Hh}&&\delta Z_{HH}\\\end{pmatrix},\nonumber\\
&&\begin{pmatrix}A_{0}\\Z_{0}\\Z'_0\\\end{pmatrix}&={}\left(\mathbbm{1}_3 +\frac{1}{2}\delta\mathrm{Z}_\text{V}\right)\begin{pmatrix}A\\Z\\Z'\\\end{pmatrix},\qquad &\delta\mathrm{Z}_\text{V}&=\begin{pmatrix} \delta Z_{AA}&\delta Z_{AZ}&\delta Z_{AZ'}\\\delta Z_{ZA}&\delta Z_{ZZ}&\delta Z_{ZZ'}\\\delta Z_{Z'A}&\delta Z_{Z'Z}&\delta Z_{Z'Z'}\\\end{pmatrix},\nonumber\\
      &&W^\pm_0&={}\left(1+\frac{1}{2}\delta Z_W\right)W^\pm,&\nonumber\\
      &&f^\text{L}_{i,0}&={}\left(1+\frac{1}{2}\delta Z_{i}^{f,\text{L}}\right)f_i^\text{L},
      &f^\text{R}_{i,0}&={}\left(\delta_{ij}+\frac{1}{2}\delta Z_{ij}^{f,\text{R}}\right)f_j^\text{R},
\label{eq:field_RT}
\end{align}
where the matrix structures for the Higgs fields $h$ and $H$, the neutral gauge-boson fields $A, Z, Z'$, and the right-handed fermions $f^\text{R}_i$ arise due to their possible mixing. For charged leptons, up-type and down-type quarks, the indices $i$, $j$ label the three generations. Since the new fermion $f'_\text{d}$ mixes with right-handed neutrinos, it has to be taken into account in the renormalization of the neutrino fields and, thus, for $f=\nu$ the indices $i$, $j$ run over the three SM-like neutrino generations as well as the non-standard fermion $\nu_4$. Note that we do not introduce a matrix structure in the quark sector, since we set the CKM matrix in the following to the unit matrix\footnote{A generalization to a non-diagonal CKM matrix proceeds exactly as in the SM (see, e.g., Refs.~\cite{Denner_Habil,DEDI20201} and references therein).}. For our purposes this is an adequate approximation as we will not consider flavour-sensitive observables.\\
Additionally, we introduce renormalization constants for the tadpole parameters, which were introduced in Eqs.~\eqref{eq:tadpols} and \eqref{eq:tapo}, according to
\begin{align}
  t_{h,0}=t_h+\delta t_h, \hspace{0.5cm} t_{H,0}=t_H+\delta t_H.
  \label{eq:tad_rt}
\end{align}
Note that the relation \eqref{eq:tapo} is true for the sets of the bare and renormalized quantities, \linebreak$\{t_{1,0},t_{2,0},t_{\text{h},0},t_{\text{H},0}\}$ and $\{t_{1},t_{2},t_{\text{h}},t_{\text{H}}\}$, respectively. However, the parameter relations obtained from Eq.~\eqref{eq:tadpols} only hold for the bare quantities $t_{h,0}$ and $t_{H,0}$, but not for $t_h$ and $t_H$.\\
The renormalization of the unphysical sector, i.e.~the renormalization of the would-be Goldstone-boson fields, ghost fields, and gauge parameters, is not relevant for the calculation of S-matrix elements and we choose to not renormalize them at all. In particular, this means that the gauge-fixing functionals do not generate any counterterms at all (for more details on possible renormalization schemes of the unphysical sector in the SM see, e.g., Refs.~\cite{DEDI20201} and references therein).

\subsection{Renormalization conditions}
First we set up a renormalization scheme that, as far as possible, adopts OS renormalization conditions that lead to intuitive and direct relations between the input parameters and physical observables. The only exception is the coupling constant $\lambda_{12}$ where no OS renormalization condition is phenomenologically appropriate as long as the non-SM-like Higgs boson is not found. Thus, $\delta\lambda_{12}$ will be fixed by an $\MSbar$ renormalization condition. The dimensionless coupling $\lambda_{12}$ is not related to masses, and  therefore, its renormalization does not depend on the details of the tadpole renormalization (see Section \ref{sec:tadpol}). As an alternative to the OS renormalization conditions for the non-SM parameters, we describe $\MSbar$ renormalization of the corresponding parameters. Once OS renormalization is established, the transition to $\MSbar$ renormalization is provided by simply keeping only the standard UV divergences in the renormalization constants of the respective parameters.\\
In the following we present the various renormalization conditions and the explicit form of the renormalization constants. 

\subsubsection{Tadpole renormalization}
\label{sec:tadpol}
In EW renormalization---in particular if $\MSbar$ conditions are involved---the treatment of tadpoles plays an important role. Historically, the two most popular tadpole schemes are the \textit{Parameter Renormalized Tadpole Scheme} (PRTS) \cite{Denner_Habil,PRTS} and the \textit{Fleischer--Jegerlehner Tadpole Scheme} (FJTS) \cite{FLJETad}. In addition, recently the \textit{Gauge-Invariant Vacuum expectation value Scheme} (GIVS) was proposed in Refs.~\cite{https://doi.org/10.48550/arxiv.2203.07236,SDTad2} as an alternative to the former two. Here we briefly sketch the renormalization procedure of the tadpoles, for a more detailed and comprehensive description we refer, e.g.,~to Refs.~\cite{DEDI20201,https://doi.org/10.48550/arxiv.2203.07236,SDTad2}.\\
In spontaneously broken gauge theories, like the DASM, explicit 1-loop contributions originating from tadpole diagrams, i.e.~Feynman diagrams of the form
\begin{equation}
  \setlength{\unitlength}{1pt}
 \text{i}\Gamma^h=\text{i} T^h=\quad\raisebox{-13pt}{\includegraphics{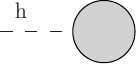}}\text{ },\qquad
\text{i} \Gamma^H=\text{i} T^H=\quad\raisebox{-13pt}{\includegraphics{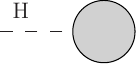}}\text{ },
\label{eq:tadpole_diags}
\end{equation}
where the blobs represent any 1-loop subdiagrams, arise if the effective Higgs potential does not acquire its minimum at the renormalized vevs $v_1,v_2$ (see also Appendix C of Ref.~\cite{SDMixingAngles}). Technically it is desirable to avoid the appearance of these tadpole contributions in calculations as far as possible. This can be done by an appropriate choice of parameter and field definitions in the course of the renormalization programme. At LO, tadpole contributions can be eliminated by demanding $t_{h,0}=t_{H,0}=0$ (see Sect.~\ref{sec:higgs_pot}). At NLO, explicit tadpole contributions can be canceled by generating tadpole counterterms $\delta t_h h$ and $\delta t_H H$ in the Lagrangian and demanding for the renormalized 1-point functions $\Gamma^h_{\text{R}}$ and $\Gamma^H_{\text{R}}$ of the physical Higgs fields to vanish,
\begin{alignat}{2}
  \Gamma^h_{\text{R}}&=T^h +\delta t_h=0\qquad& &\Rightarrow\qquad \delta t_h=-T^h,\nonumber\\
  \Gamma^H_{\text{R}}&=T^H+\delta t_H=0& &\Rightarrow\qquad \delta t_H=-T^H.
\label{eq:tad-rc}
\end{alignat}
There are different ways to introduce the tadpole counterterms $\delta t_h h$ and $\delta t_H H$ in the Lagrangian.
\newline \\ In the \textit{Fleischer--Jegerlehner Tadpole Scheme} (FJTS) \cite{FLJETad} the bare tadpoles $t_{h,0}$ and $t_{H,0}$ are consistently set to zero, and the tadpole counterterms are introduced via shifts of the Higgs fields
\begin{align}
  h&\to h+\Delta v_h^\text{FJTS},\qquad \Delta v_h^\text{FJTS}=-\frac{\delta t_h^\text{FJTS}}{M_\text{h}^2},\nonumber\\
  H&\to H+\Delta v_H^\text{FJTS},\qquad \Delta v_H^\text{FJTS}=-\frac{\delta t_H^\text{FJTS}}{M_\text{H}^2}.
  \label{eq:FJTS_shifts}
\end{align}
Even though $\Delta v_h^\text{FJTS}$ and $\Delta v_H^\text{FJTS}$ turn out to be gauge dependent, the FJTS does not destroy the gauge independence in predictions for observables in OS or $\MSbar$ renormalization schemes since the field redefinitions \eqref{eq:FJTS_shifts} can be viewed as mere shifts in the integration variables in the functional integral used to quantize the theory. However, the FJTS tends to lead to large tadpole contributions to renormalization constants of mass parameters in $\MSbar$ renormalization schemes, often jeopardizing the perturbative stability of predictions (see e.g.~Refs.~\cite{https://doi.org/10.48550/arxiv.2203.07236,SDTad2,SDMixingAngles,Lukas_Alt,RECOLA2,Sym_MixingAngles}). \newline \\In contrast to the FJTS the \textit{Parameter Renormalized Tadpole Scheme} (PRTS) \cite{Denner_Habil,PRTS} treats the tadpoles similar to model parameters and introduces the desired tadpole counterterms simply via the parameter renormalization transformation given in Eq.~\eqref{eq:tad_rt}. For the renormalized effective Higgs potential to have a minimum at $v_1,v_2$, the renormalized tadpoles are set to zero in the PRTS, $t_h=t_H=0$, so that $\delta t_h= t_{h,0}$ and $\delta t_H= t_{H,0}$. Starting from
\begin{align}
  \delta t_{1}&=-v_{1,0}\left(4v^2_{1,0}\lambda_{1,0}+v_{2,0}^2\lambda_{12,0}-2\mu_{1,0}^2\right)\text{,}\qquad   \delta t_{2}=-v_{2,0}\left(\frac{v_{2,0}^2}{4}\lambda_{2,0}+v_{1,0}^2\lambda_{12,0}-\mu_{2,0}^2\right),\nonumber\\
\lambda_{1,0}&=\frac{v_{1,0}\left(c_{\alpha,0}^2 M_{\text{H},0}^2+s_{\alpha,0}^2 M_{\text{h},0}^2\right)+c_{\alpha,0} \delta t_H-s_{\alpha,0} \delta t_h}{8 v_{1,0}^3},\nonumber\\
\lambda_{2,0}&=\frac{2\left[v_{2,0}\left(c_{\alpha,0}^2 M_{\text{h},0}^2+s_{\alpha,0}^2 M_{\text{H},0}^2\right)+c_{\alpha,0} \delta t_{h}+s_{\alpha,0} \delta t_{H}\right]}{v_{2,0}^3},
\end{align}
and using Eq.~\eqref{eq:par-rel-pot} for the bare ($v_{1,0}$) and renormalized ($v_1$) vev of the Higgs field $\rho$, one finds the simple replacements\footnote{These replacement rules, due to the similar construction of the Higgs sectors, are equivalent to the ones given in Ref.~\cite{SDTad2} for the pure Higgs singlet extension.}
\begin{alignat}{2}
  \mu_{1,0}^2&\to \mu_{1,0}^2 +\frac{3 \left(c_\alpha \delta t_H^\text{PRTS}-s_\alpha \delta t_h^\text{PRTS}\right)}{4 v_1},\qquad &\mu_{2,0}^2\to\mu_{2,0}^2+\frac{3\left(c_\alpha \delta t_h^\text{PRTS}+s_\alpha  \delta t_H^\text{PRTS}\right)}{2 v_2},\nonumber\\
  \lambda_{1,0}&\to\lambda_{1,0}+\frac{c_\alpha\delta t_H^\text{PRTS}-s_\alpha \delta t_h^\text{PRTS}}{8 v_1^3},\qquad &\lambda_{2,0}\to \lambda_{2,0} +\frac{2\left(c_\alpha\delta t_h^\text{PRTS}+s_\alpha \delta t_H^\text{PRTS}\right)}{v_2^3},\nonumber\\
  \lambda_{12,0}&\to\lambda_{12,0},
  \label{eq:PRTS_reps}
\end{alignat}
for the bare parameters of the Higgs potential to generate all tadpole counterterms from the bare Lagrangian in which $t_{h,0}$ and $t_{H,0}$ are set to zero. The PRTS leads to small corrections to renormalized mass parameters, but in general introduces a gauge dependence in the parameterization of predictions for observables in $\MSbar$ schemes (there is no such dependence introduced in OS schemes). Thus, to produce consistent predictions using the PRTS in combination with an $\MSbar$ scheme, either a gauge has to be fixed once and for all or the values of the input parameters have to be converted between different gauge choices.\\ In both tadpole schemes the tadpole counterterms are chosen in such a way that they cancel the respective contributions from explicit tadpole diagrams in Green functions. Nevertheless, the tadpole renormalization constants $\delta t_h$ and $\delta t_H$ enter the Lagrangian in different places, and therefore, lead to a dependence of counterterms on the chosen tadpole renormalization scheme.\newline\\
Combining the benefits of the FJTS and the PRTS---the gauge independence of the FJTS and the perturbative stability of the PRTS---recently a hybrid version of the two schemes, the GIVS, was proposed in Refs.~\cite{https://doi.org/10.48550/arxiv.2203.07236,SDTad2}. The GIVS constructs the tadpole renormalization constant $\delta t_S$ ($S=h,H$) from two contributions, $\delta t^\text{GIVS}_{S,1}$ and $\delta t^\text{GIVS}_{S,2}$, where $\delta t^\text{GIVS}_{S,1}$ is introduced as in the PRTS and $\delta t^\text{GIVS}_{S,2}$ as in the FJTS. The GIVS switches to the non-linear Higgs representation for the calculation of the $\delta t^\text{GIVS}_{S,1}$. In the non-linear representation the scalar components corresponding to CP-even neutral physical Higgs fields are gauge invariant, rendering the corresponding tadpole functions gauge independent. These tadpole renormalization constants are given by
  \begin{align}
    \delta t_{h,1}^\text{GIVS}=- T_\text{nl}^h,\qquad \delta t_{H,1}^\text{GIVS}=-T_\text{nl}^H,
  \end{align}
  where the explicit expressions for the NLO Higgs 1-point functions in the non-linear Higgs representation (see Appendix \ref{app:non_lin_Higgs}) are given by
  \begin{align}
    T_\text{nl}^S={}&\sum_{V=Z,Z',W}\lambda_{SV} \left[3A_0(M_\text{V}^2)-2 M_\text{V}^2\right]+\sum_{S'=h,H}\left[\lambda_{SS'}A_0(M_{S'}^2)\right]\nonumber\\&+\lambda_{SF}\sum_{f=l,u,d}\left[N_{C,f} m_f^2 A_0(m_f^2)\right]+\lambda_{S\nu_4}m_{\nu_4}^2 A_0(m_{\nu_4}^2), \qquad S=h,H,
  \end{align}
  with the respective color factor $N_{C,f}$ of a fermion $f$ and
  \begin{alignat}{4}
    \lambda_{hZ}&=\frac{c_\text{w}^2 M_\text{Z}^2\left(v_1 c_\alpha c_\gamma^2M_\text{Z}^2-v_2s_\alpha s_\gamma^2M_{\text{Z}'}^2\right)}{16 \pi^2 v_1v_2 M_\text{W}^2},\qquad& \lambda_{HZ}&={}\frac{c_\text{w}^2 M_\text{Z}^2\left(v_1 s_\alpha c_\gamma^2M_\text{Z}^2+v_2c_\alpha s_\gamma^2M_{\text{Z}'}^2\right)}{16 \pi^2v_1v_2 M_\text{W}^2},\nonumber\\
    \lambda_{hZ'}&=\frac{c_\text{w}^2 M_{\text{Z}'}^2\left(v_1 c_\alpha s_\gamma^2 M_{\text{Z}'}^2-v_2 s_\alpha c_\gamma^2 M_{\text{Z}}^2\right)}{16 \pi^2 v_1 v_2 M_\text{W}^2},\qquad& \lambda_{HZ'}&={}\frac{c_\text{w}^2 M_{\text{Z}'}^2\left(v_1 s_\alpha s_\gamma^2 M_{\text{Z}'}^2+v_2 c_\alpha c_\gamma^2 M_{\text{Z}}^2\right)}{16 \pi^2 v_1 v_2 M_\text{W}^2},\nonumber\\
    \lambda_{hW}&=\frac{c_\alpha M_\text{W}^2}{8 \pi^2 v_2},\qquad& \lambda_{HW}&={}\frac{s_\alpha M_\text{W}^2}{8 \pi^2 v_2},\nonumber\\
    \lambda_{hh}&=\frac{3c_\text{hhh}}{16 \pi^2},\qquad& \lambda_{Hh}&={}\frac{c_\text{hhH}}{16\pi^2},\nonumber\\
    \lambda_{hH}&=\frac{c_\text{hHH}}{16\pi^2},\qquad& \lambda_{HH}&={}\frac{3c_\text{HHH}}{16\pi^2},\nonumber\\
    \lambda_{hF}&=-\frac{c_\alpha}{4 \pi^{2}v_2},\qquad& \lambda_{HF}&={}{-}\frac{s_\alpha}{4 \pi^{2}v_2},\nonumber\\
    \lambda_{h\nu_4}&=\frac{s_\alpha s_{\theta_\text{r}}^2}{4 \pi^{2}v_1},\qquad& \lambda_{H\nu_4}&={-}\frac{c_\alpha s_{\theta_\text{r}}^2}{4 \pi^{2}v_1}.
  \end{alignat}
The explicit expressions for the Higgs self-couplings $c_{S_1S_2S_2}$ ($S_1,S_2=\text{h},\text{H}$) can be found in Eq.~(2.18) of Ref.~\cite{BogDit}. The function $A_0(m^2)$ is the standard 1-loop 1-point integral in dimensional regularization using the conventions of Refs.~\cite{Denner_Habil,BDJ,DEDI20201}.
 The 1-point functions $T_\text{nl}^S$ are then used in tadpole counterterms that are generated in the course of parameter renormalization as in the PRTS, i.e.,
\begin{alignat}{2}
  \mu_{1,0}^2&\to \mu_{1,0}^2 +\frac{3 \left(c_\alpha \delta t_{H,1}^\text{GIVS}-s_\alpha \delta t_{h,1}^\text{GIVS}\right)}{4 v_1},\qquad &\mu_{2,0}^2\to\mu_{2,0}^2+\frac{3 \left(c_\alpha \delta t_{h,1}^\text{GIVS}+s_\alpha  \delta t_{H,1}^\text{GIVS}\right)}{2 v_2},\nonumber\\
  \lambda_{1,0}&\to\lambda_{1,0}+\frac{c_\alpha\delta t_{H,1}^\text{GIVS}-s_\alpha \delta t_{h,1}^\text{GIVS}}{8 v_1^3},\qquad &\lambda_{2,0}\to \lambda_{2,0} +\frac{2\left(c_\alpha\delta t_{h,1}^\text{GIVS}+s_\alpha \delta t_{H,1}^\text{GIVS}\right)}{v_2^3},\nonumber\\
  \lambda_{12,0}&\to\lambda_{12,0}.
\end{alignat}
Apart from the calculation of $\delta t^\text{GIVS}_{S,1}$, the whole loop calculation proceeds in the linear Higgs representation as usual. Therefore, the mere use of $\delta t^\text{GIVS}_{S,1}$ as tadpole renormalization constants would not lead to a full cancellation of tadpole loops in the loop calculation. The ``missing'' parts in the tadpole renormalization constants, $\delta t_{h,2}^\text{GIVS}$ and $\delta t_{H,2}^\text{GIVS}$ (which are gauge dependent), are introduced by field shifts as in the FJTS,
\begin{align}
  h&\to h+\Delta v_h^\text{GIVS},\qquad \Delta v_h^\text{GIVS}=-\frac{\delta t_{h,2}^\text{GIVS}}{M_\text{h}^2},\nonumber\\
  H&\to H+\Delta v_H^\text{GIVS},\qquad \Delta v_H^\text{GIVS}=-\frac{\delta t_{H,2}^\text{GIVS}}{M_\text{H}^2},
\end{align}
with
\begin{align}
\Delta v_S^\text{GIVS}= \frac{T^S-T^S_\text{nl}}{M_S^2}=\lambda_{S\chi}A_0(\xi_\text{V} M_\text{Z})+\lambda_{S\chi'}A_0(\xi_\text{V} M_{\text{Z}'})+\lambda_{S\phi}A_0(\xi_\text{W} M_\text{W}),\qquad S=h,H,
\end{align}
and
\begin{alignat}{3}
  \lambda_{h\chi}&=\frac{v_1 c_\alpha c_x^2-v_2 s_\alpha s_x^2}{32\pi^2 v_1v_2 },&\qquad  \lambda_{H\chi}&={}\frac{v_1 s_\alpha c_x^2+v_2 c_\alpha s_x^2}{32\pi^2 v_1 v_2},\nonumber \\
  \lambda_{h\chi '}&=\frac{v_1 c_\alpha s_x^2-v_2 s_\alpha c_x^2}{32 \pi^2 v_1 v_2 },&\qquad  \lambda_{H\chi '}&={}\frac{v_1 s_\alpha s_x^2+v_2 c_\alpha c_x^2}{32 \pi^2 v_1 v_2},\nonumber\\
  \lambda_{h\phi}&=\frac{c_\alpha}{16 \pi^2 v_2},&\qquad  \lambda_{H\phi}&={}\frac{s_\alpha}{16 \pi^2 v_2}.
\end{alignat}
The full tadpole renormalization constants $\delta t_S^\text{GIVS}$, thus, consist of two parts,
\begin{align}
  \delta t_{h}^\text{GIVS}=\delta t_{h,1}^\text{GIVS}+\delta t_{h,2}^\text{GIVS}=-T^h,\qquad \delta t_{H}^\text{GIVS}=\delta t_{H,1}^\text{GIVS}+\delta t_{H,2}^\text{GIVS}=-T^H.
\end{align}
The GIVS is designed in such a way that the vevs $v_1, v_2$ of the Higgs fields are tied to the minimum of the renormalized Higgs potential, rendering the GIVS a perturbatively stable, gauge-independent tadpole renormalization scheme. \\
In the following we give explicit results for OS and $\MSbar$ renormalization constants for all parameters of the DASM that are related to masses. The OS scheme leads to a systematic cancellation between tadpole corrections in mass renormalization constants and self-energies in calculations of observables in all three tadpole schemes so that predictions for observables in OS schemes do not depend on the tadpole scheme. However, for $\MSbar$ renormalization the choice of the tadpole scheme matters. By default, we choose the (most commonly used) PRTS for the renormalization of the DASM (see Sect.~\ref{se:reno_trafos}), but the use of each of the three tadpole schemes described above is straightforward in the DASM. In the following we give substitution rules for the translation of the PRTS results to the respective FJTS and GIVS results for all mixing angles of the DASM.

\subsubsection{Mass and field renormalization}
We adopt OS renormalization conditions for the masses and fields as, e.g., formulated in Refs.~\cite{Denner_Habil,DEDI20201,BogDit,SDMixingAngles} and follow Ref.~\cite{DEDI20201} in the notation and conventions for field-theoretical quantities. This means that the renormalized mass parameters of physical particles correspond to the locations of the zeroes in the real parts of the respective inverse propagators, that on-shell fields do not mix with other fields, and that the renormalized fields are canonically normalized. In terms of renormalized two-point vertex functions $\Gamma_\text{R}^{ab}$ of fields $a$, $b$ the renormalization conditions are given by 
\begin{align}
  &\text{Re}\,\Gamma^{V^\dagger V'}_{\text{R},\mu\nu} (-k,k)\varepsilon^\nu(k)|_{k^2=M_V^2}=0,\qquad V,V'=A,Z,Z',W^\pm,\\
  &\lim_{k^2\rightarrow M_V^2}\frac{1}{k^2-M_V^2}\,\text{Re}\,\Gamma^{V^\dagger V}_{\text{R},\mu\nu} (-k,k)\varepsilon^\nu(k)=-\varepsilon_\mu(k),\\
&\text{Re}\,\Gamma^{SS'}_{\text{R}} (-k,k)|_{k^2=M_S^2}=0,\qquad S,S'=h,H,\label{eq:scalar_field_rcon}\\
&\lim_{k^2\rightarrow M_S^2}\frac{1}{k^2-M_S^2}\,\text{Re}\,\Gamma^{SS}_{\text{R}} (-k,k)=1,\label{eq:scalar_mass_rcon}\\
&\text{Re}\,\Gamma^{\bar{f}f}_{\text{R},ij}(-p,p)u_j(p)\biggr|_{p^2=m_{f,j}^2}=0,\label{eq:ferm_field_rcon}\\
&\lim_{p^2\rightarrow m_{f,i}^2}\frac{\slashed{p}+m_{f,i}}{p^2-m_{f,i}^2}\,\text{Re}\,\Gamma^{\bar{f}f}_{\text{R},ii}(-p,p)u_i(p)=u_i(p),
\label{eq:ren-cond-gene}
\end{align}
where $\varepsilon^\mu(k)$ and $u(k)$ represent the polarization vectors and spinors for on-shell external gauge bosons and fermions with momentum $k$, respectively.
Using the covariant decomposition of the two-point functions for gauge bosons in 't Hooft--Feynman gauge, 
\begin{align}
\Gamma^{V^\dagger V'}_{\text{R},\mu\nu}(-k,k)=-g_{\mu\nu}(k^2-M_V^2)\delta_{VV'}-\left(g_{\mu\nu}-\frac{k_\mu k_\nu}{k^2}\right)\Sigma_{\text{R,T}}^{V^\dagger V'}(k^2)-\frac{k_\mu k_\nu}{k^2}\Sigma_\text{R,L}^{V^\dagger V'}(k^2), 
\end{align}
where $\Sigma_{\text{R,T}}^{V^\dagger V'}(k^2)$ and $\Sigma_\text{R,L}^{V^\dagger V'}(k^2)$ are the renormalized transversal and longitudinal self-energies, we find the well-known results
\begin{alignat}{3}
  \delta M^2_V&=\text{Re}\,\Sigma_\text{T}^{V^\dagger V}(M^2_V),\quad & \delta Z_{V^\dagger V}&=-\text{Re}\,\frac{\partial \Sigma^{V^\dagger V}_\text{T}(k^2)}{\partial k^2}\biggr|_{k^2=M_{\text{V}}^2},\quad & V&= A, Z, Z',W^\pm,\nonumber\\\delta Z_{VV'}&=-2\text{Re}\,\frac{\Sigma^{V^\dagger V'}_\text{T}(M_{V'}^2)}{M_{V'}^2-M_{V}^2},\quad & VV'&= AZ,ZA,AZ',Z'A,ZZ',Z'Z,
\label{eq:def_RC_Vboson}
\end{alignat}
for the renormalization constants in the gauge sector. Note that in our convention all self-energies include one-particle irreducible (1PI) contributions $\Sigma_{1\text{PI}}\left(k^2\right)$ as well as 1- and 2-point tadpole counterterms $\Sigma_{\delta t}$ and reducible tadpole loop corrections $\Sigma_\text{tad}$ (for more details see Section 3.1.7 of Ref.~\cite{DEDI20201}),
\begin{align}
\Sigma\left(k^2\right)=\Sigma_{1\text{PI}}\left(k^2\right) + \Sigma_{\delta t}+\Sigma_\text{tad}.
\end{align}
In contrast to the SM, the Higgs sector of the DASM contains two physical Higgs bosons. Higher-order effects lead to mixing between the fields of these two scalar bosons. 
Therefore, similarly to the sector of the neutral gauge bosons, the renormalized two-point function with two external scalar fields $S, S'=h, H$ is not diagonal and might be decomposed according to
\begin{alignat}{2}
\Gamma_\text{R}^{SS'}(-k,k)= \left(k^2-M^2_S\right)\delta_{SS'}+\Sigma^{SS'}_\text{R}(k^2),
\end{alignat} 
with the renormalized self-energies
\begin{align}
\Sigma^{SS'}_\text{R}(k^2)=\Sigma^{SS'}(k^2)+\frac{1}{2}(k^2-M_S^2)\delta Z_{SS'}+\frac{1}{2}(k^2-M_{S'}^2)\delta Z_{S'S}-\delta_{SS'} \delta M_{S}^2.
\end{align}
Using Eqs.~\eqref{eq:scalar_field_rcon} and \eqref{eq:scalar_mass_rcon} one finds for the mass and field renormalization constants of the Higgs sector,
\begin{alignat}{2}
\delta M^2_{S}&=\text{Re}\,\Sigma^{SS}(M^2_{S}),\qquad \delta Z_{SS}&=&-\text{Re}\,\frac{\partial \Sigma^{SS}(k^2)}{\partial k^2}\biggr|_{k^2=M_{S}^2},\nonumber\\
\delta Z_{SS'}&=-2\text{Re}\,\frac{\Sigma^{S'S}(M_{S'}^2)}{M_{S'}^2-M_{S}^2},\qquad S&\neq&S'.
\end{alignat}
The renormalization procedure of the fermion part of the DASM can be carried out in analogy to the SM procedure. We parameterize the renormalized two-point function for fermions by covariants as follows,
\begin{align}
\Gamma_{\text{R},ij}^{f\bar{f}}(-p,p)=\slashed{p}\omega_\text{L}\Gamma_{\text{R},ij}^{f,\text{L}}(p^2)+\slashed{p}\omega_\text{R}\Gamma_{\text{R},ij}^{f,\text{R}}(p^2)+\omega_\text{L}\Gamma_{\text{R},ij}^{f,\text{l}}(p^2) + \omega_\text{R} \Gamma_{\text{R},ij}^{f,\text{r}}(p^2),
\end{align}
where $\omega_\text{L/R}=\frac{1}{2}\left(1\mp\gamma_5\right)$ denote the chiral projectors and $\Gamma_{\text{R},ij}^{f,\text{L}}(p^2)$, $\Gamma_{\text{R},ij}^{f,\text{R}}(p^2)$, $\Gamma_{\text{R},ij}^{f,\text{l}}(p^2)$, and  $\Gamma_{\text{R},ij}^{f,\text{r}}(p^2)$ are the renormalized left- and right-handed vector and scalar parts of the fermion two-point vertex function. Adopting this notation for the covariants of the fermion self-energy we find
  \begin{align}
    \Gamma_{\text{R},ij}^{f,\text{L}}(p^2)&=\delta_{ij}+\Sigma_{ij}^{f\text{,L}}\left(p^2\right)+\frac{1}{2}\left(\delta Z_{ij}^{f\text{,L}}+\delta Z_{ij}^{f\text{,L}\dagger}\right),\\
    \Gamma_{\text{R},ij}^{f,\text{R}}(p^2)&=\delta_{ij}+\Sigma_{ij}^{f\text{,R}}\left(p^2\right)+\frac{1}{2}\left(\delta Z_{ij}^{f\text{,R}}+\delta Z_{ij}^{f\text{,R}\dagger}\right),\\
    \Gamma_{\text{R},ij}^{f,\text{l}}(p^2)&=- m_{f,i}\delta_{ij}+\Sigma_{ij}^{f,l}\left(p^2\right)-\frac{1}{2}\left(m_{f,i}\delta Z_{ij}^{f\text{,L}}+m_{f,j}\delta Z_{ij}^{f\text{,R}\dagger}\right)-\delta_{ij}\delta m_{f,i},\\
    \Gamma_{\text{R},ij}^{f,\text{r}}(p^2)&=- m_{f,i}\delta_{ij}+\Sigma_{ij}^{f,r}\left(p^2\right)-\frac{1}{2}\left(m_{f,i}\delta Z_{ij}^{f\text{,R}}+m_{f,j}\delta Z_{ij}^{f\text{,L}\dagger}\right)-\delta_{ij}\delta m_{f,i},
  \end{align}
  where $\delta Z_{ij}^\dagger=\delta Z_{ji}^*$.
In combination with the renormalization conditions \eqref{eq:ferm_field_rcon} and \eqref{eq:ren-cond-gene} one finds the same form as in the SM \cite{Denner_Habil,BDJ,DEDI20201},
\begin{align}
\delta m_{f,i}={}&\frac{1}{2}\text{Re}\,\Bigl[m_{f,i}\Bigl(\Sigma_{ii}^{f,\text{L}}(m_{f,i}^2)+\Sigma_{ii}^{f,\text{R}}(m_{f,i}^2)\Bigr)+\Sigma_{ii}^{f,\text{l}}(m_{f,i}^2)+\Sigma_{ii}^{f,\text{r}}(m_{f,i}^2)\Bigr],\nonumber\\
\delta Z_{ii}^{f,\text{L}}={}&-\text{Re}\,\Sigma_{ii}^{f,\text{L}}(m_{f,i}^2)-m_{f,i}\frac{\partial}{\partial p^2}\text{Re}\,\Bigl[{}m_{f,i}\Bigl(\Sigma_{ii}^{f,\text{L}}(p^2)+\Sigma_{ii}^{f,\text{R}}(p^2)\Bigr)\nonumber\\
    &+\Sigma_{ii}^{f,\text{l}}(p^2)+\Sigma_{ii}^{f,\text{r}}(p^2)\Bigr]\Bigr|_{p^2=m_{f,i}^2},\nonumber\\
  \delta Z_{ii}^{f,\text{R}}={}&-\text{Re}\,\Sigma_{ii}^{f,\text{R}}(m_{f,i}^2)-m_{f,i}\frac{\partial}{\partial p^2}\text{Re}\,\Bigl[{}m_{f,i}\Bigl(\Sigma_{ii}^{f,\text{L}}(p^2)+\Sigma_{ii}^{f,\text{R}}(p^2)\Bigr)\nonumber\\
    &+\Sigma_{ii}^{f,\text{l}}(p^2)+\Sigma_{ii}^{f,\text{r}}(p^2)\Bigr]\Bigr|_{p^2=m_{f,i}^2},\nonumber\\
  \delta Z_{ij}^{f,\text{R}}={}&\frac{2}{m_{f,i}^2-m^2_{f,j}}\text{Re}\Bigl[{}m_{f,j}^2\Sigma_{ij}^{f,\text{R}}(m_{f,j}^2)+m_{f,i}m_{f,j}\Sigma_{ij}^{f,\text{L}}(m_{f,j}^2)+m_{f,j}\Sigma_{ij}^{f,\text{l}}(m_{f,j}^2)\nonumber\\
    {}& +m_{f,i}\Sigma_{ij}^{f,\text{r}}(m_{f,j}^2)\Bigr],\hspace{1cm} i\neq j,
  \end{align}
 for the mass and field renormalization constants of the fermion sector.

\subsubsection{Renormalization of the mixing angle {\boldmath$\gamma$}}
\label{sec:dgamma}
For the renormalization of the mixing angle $\gamma$ we follow the OS renormalization approach that was introduced in Ref.~\cite{SDMixingAngles} for mixing angles in scalar sectors. To this end, we introduce an extra ``fake fermion'' field $\omega_\text{d}$ with appropriate infinitesimal couplings which, thus, does not change any predictions for observables, but still can be used in the formulation of renormalization conditions. The additional fermionic field $\omega_\text{d}$ is a singlet under the gauge group of the SM, but is charged under the additional $U(1)_\text{d}$ gauge group of the DASM. The relative $U(1)_\text{d}$ charge of $\omega_\text{d}$ is called $\tilde{q}_\omega$ in the following. The respective Lagrangian for $\omega_\text{d}$ is given by
\begin{align}
\mathcal{L}_{\omega_\text{d}}=\bar{\omega}_\text{d}\left(\text{i}\slashed{D}_\text{d}- m_{\omega_\text{d}}\right){\omega}_\text{d}=\bar{\omega}_\text{d}\left[\text{i}\slashed{\partial}-\tilde{e}\tilde{q}_{\omega}\left(s_\gamma \slashed{Z}+c_\gamma \slashed{Z}'\right)- m_{\omega_\text{d}}\right]{\omega}_\text{d}.
\label{eq:L-artif-ferm}
\end{align}
The fake fermion $\omega_\text{d}$ is non-chiral, so that we can attribute to it a Dirac mass term with mass parameter $m_{\omega_\text{d}}$. Moreover, no anomalies are introduced to the theory because of the non-chirality of $\omega_\text{d}$. The Feynman rules for the two new vertices introduced by $\mathcal{L}_{\omega_\text{d}}$ are given by
\begin{equation}
\setlength{\unitlength}{1pt}
\raisebox{-18pt}{\includegraphics{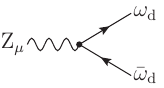}} 
= -\text{i}\tilde{q}_{\omega}\tilde{e}s_\gamma\gamma_\mu ,\hspace{0.8cm}
  \raisebox{-18pt}{\includegraphics{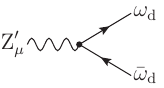}} 
= -\text{i}\tilde{q}_{\omega}\tilde{e}c_\gamma\gamma_\mu.
\end{equation}
Obviously, the original theory is recovered by taking $\tilde{q}_{\omega}\rightarrow 0$.
Following Ref.~\cite{SDMixingAngles} we consider OS formfactors $\mathcal{F}^{V\bar{\omega}_\text{d}{\omega_\text{d}}}$, ($V=Z,Z'$), defined by the decay matrix elements
\begin{align}
  \mathcal{M}^{\text{Z}\rightarrow\bar{\omega}_\text{d}{\omega_\text{d}}}= [\bar{u}_{\omega_\text{d}}\slashed{\varepsilon}v_{\omega_\text{d}}]_{Z} \mathcal{F}^{Z\bar{\omega}_\text{d}{\omega_\text{d}}},\qquad\mathcal{M}^{\text{Z}'\rightarrow\bar{\omega}_\text{d}{\omega_\text{d}}}= [\bar{u}_{\omega_\text{d}}\slashed{\varepsilon}v_{\omega_\text{d}}]_{Z'} \mathcal{F}^{Z'\bar{\omega}_\text{d}{\omega_\text{d}}},
  \label{eq:Form_fac_psi}
\end{align}
where $\bar{u}_{\omega_\text{d}}$ and $v_{\omega_\text{d}}$ are the spinors of the final-state fermions, $\varepsilon_\mu$ denotes the polarization vector of the respective gauge boson, and the $[\dots]_{Z/Z'}$ indicate the respective decay kinematics.
To fix $\delta \gamma$ we demand that higher-order corrections to the ratio of the two formfactors defined in Eq.~\eqref{eq:Form_fac_psi} vanish,
\begin{align}
\lim_{\tilde{q}_{\omega}\rightarrow 0}\frac{\mathcal{F}^{Z\bar{\omega}_\text{d}{\omega_\text{d}}}}{\mathcal{F}^{Z'\bar{\omega}_\text{d}{\omega_\text{d}}}}\stackrel{!}{=}\frac{\mathcal{F}_\text{LO}^{Z\bar{\omega}_\text{d}{\omega_\text{d}}}}{\mathcal{F}_\text{LO}^{Z'\bar{\omega}_\text{d}{\omega_\text{d}}}} = \frac{s_\gamma}{c_\gamma}.
\label{eq:gamma-rc}
\end{align}
Using the OS renormalization scheme we find for the NLO-corrected formfactors $\mathcal{F}^{Z/Z'\bar{\omega}_\text{d}\omega_\text{d}}$
\begin{align}
&\mathcal{F}_\text{NLO}^{Z\bar{\omega}_\text{d}{\omega_\text{d}}}=\mathcal{F}_\text{LO}^{Z\bar{\omega}_\text{d}{\omega_\text{d}}}\left(1+\frac{\delta \tilde{e}}{\tilde{e}}+\frac{\delta s_\gamma}{s_\gamma}+\delta Z_{\omega_\text{d}}+\frac{1}{2} \delta Z_{ZZ}+\frac{c_\gamma}{2s_\gamma} \delta Z_{Z'Z}+\delta^{Z\bar{\omega}_\text{d}{\omega_\text{d}}}_{\text{loop}}\right),\nonumber\\
  &\mathcal{F}_\text{NLO}^{Z'\bar{\omega}_\text{d}{\omega_\text{d}}}=\mathcal{F}_\text{LO}^{Z'\bar{\omega}_\text{d}{\omega_\text{d}}}\hspace{-4pt}\left(\hspace{-2pt}1+\frac{\delta \tilde{e}}{\tilde{e}}+\frac{\delta c_\gamma}{c_\gamma}+\delta Z_{\omega_\text{d}}+\frac{1}{2} \delta Z_{Z'Z'}+\frac{s_\gamma}{2c_\gamma} \delta Z_{ZZ'}+\delta^{Z'\bar{\omega}_\text{d}{\omega_\text{d}}}_{\text{loop}}\right),
\end{align}
where $\delta^{Z/Z'\bar{\omega}_\text{d}{\omega_\text{d}}}_{\text{loop}}$ represent the unrenormalized relative 1-loop corrections to the $Z/Z'\bar{\omega_\text{d}}\omega_\text{d}$ vertices and $\delta Z_{\omega_\text{d}}$ is the field renormalization constant of the non-chiral field $\omega_\text{d}$.
At NLO this leads to 
\begin{align}
\frac{\mathcal{F}_\text{NLO}^{Z\bar{\omega}_\text{d}{\omega_\text{d}}}}{\mathcal{F}_\text{NLO}^{Z'\bar{\omega}_\text{d}{\omega_\text{d}}}}={}\frac{\mathcal{F}_\text{LO}^{Z\bar{\omega}_\text{d}{\omega_\text{d}}}}{\mathcal{F}_\text{LO}^{Z' \bar{\omega}_\text{d}{\omega_\text{d}}}}\biggl[&1+\frac{\delta s_\gamma}{s_\gamma}-\frac{\delta c_\gamma}{c_\gamma}+\frac{1}{2}\left( \delta Z_{ZZ}- \delta Z_{Z'Z'}+\frac{c_\gamma}{s_\gamma} \delta Z_{Z'Z}-\frac{s_\gamma}{c_\gamma} \delta Z_{ZZ'}\right)\nonumber\\
&+\delta^{Z\bar{\omega}_\text{d}{\omega_\text{d}}}_{\text{loop}}-\delta^{Z'\bar{\omega}_\text{d}{\omega_\text{d}}}_{\text{loop}}\biggr].
\end{align}
The 1-loop vertex corrections are induced by the diagrams 
\begin{equation}
  \raisebox{-19pt}{\includegraphics{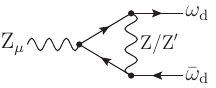}}
  \hspace{1cm}
  \raisebox{-19pt}{\includegraphics{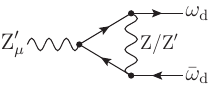}}
\end{equation}
and thus the terms $\delta^{Z/Z'\bar{\omega}_\text{d}{\omega_\text{d}}}_{\text{loop}}$ are of $\mathcal{O}(\tilde{q}^2_{\omega})$ for $\tilde{q}_{\omega}\rightarrow 0$. Similarly we obtain $\delta Z_{\omega_\text{d}}= \mathcal{O}\left(\tilde{q}_\omega^2\right)$. Evaluating the NLO renormalization condition \eqref{eq:gamma-rc} in the limit $\tilde{q}_\omega\rightarrow 0$ and using \mbox{$\delta s_\gamma= c_\gamma \delta \gamma$}, thus, leads to
\begin{align}
  \delta\gamma_\text{OS}=\frac{1}{2}s_\gamma c_\gamma\left(\delta Z_{Z'Z'}-\delta Z_{ZZ}\right)+\frac{1}{2}\left(s_\gamma^2\delta Z_{ZZ'}-c_\gamma^2\delta Z_{Z'Z}\right).
  \label{eq:dg_rc}
\end{align}
Note that the OS renormalization condition \eqref{eq:gamma-rc} has several of the desirable properties that were discussed in Refs.~\cite{SDMixingAngles,RCProperties}:
\begin{itemize}
\item[(i)] The OS renormalization of $\gamma$ is symmetric in the fields $Z$ and $Z'$ of the neutral gauge bosons.
\item[(ii)] The renormalization constant of the mixing angle $\gamma$ is given by a gauge-independent combination of field renormalization constants.
\item[(iii)] The renormalization does not depend on a specific physical process.
\item[(iv)] $\delta\gamma_\text{OS}$ is well behaved for exceptional values of $\gamma$, i.e.\ it has smooth limits for $s_\gamma\rightarrow 0$ and $c_\gamma\rightarrow 0$.
\item[(v)] Combining Eq.~\eqref{eq:gb_rot} for the bare fields with Eq.~\eqref{eq:field_RT} and using the NLO expansion
\begin{align}
\mathrm{R}_\text{V}(\gamma_0;c_{\text{w},0})=\mathrm{R}_\text{V}(\gamma,c_\text{w})+\delta\mathrm{R}_\text{V}(\gamma,c_\text{w},\delta \gamma,\delta c_\text{w}),
\label{eq:mix_mat_NLO}
\end{align}
we find to 1-loop accuracy
\begin{align}
\begin{pmatrix}
B'_\mu\\W^3_\mu\\C'_\mu
\end{pmatrix}
=\mathrm{R}_\text{V}(\gamma,c_\text{w})\left(1+\mathrm{R}^\text{T}_\text{V}(\gamma,c_\text{w})\delta \mathrm{R}_\text{V}(\gamma,c_\text{w},\delta\gamma,\delta c_\text{w})+\frac{1}{2}\delta \mathrm{Z}_\text{V} \right)
\begin{pmatrix}
A_\mu\\Z_\mu\\Z'_\mu
\end{pmatrix}.
\end{align}
This field redefinition leads to explicit $\delta \gamma$ terms originating from $\gamma$ in $\mathrm{R}_\text{V}$, i.e.~from $\delta \mathrm{R}_V$ in Eq.~\eqref{eq:mix_mat_NLO}. Similar to the situation for scalar mixing (see Ref.~\cite{SDMixingAngles}) these explicit $\delta \gamma$ terms are always introduced via the combinations
\begin{align}
  -\delta\gamma+\frac{1}{2}\delta Z_{ZZ'},\qquad \delta\gamma +\frac{1}{2}\delta Z_{Z'Z}.
  \label{eq:dg_combinations}
\end{align}
The dependence of \eqref{eq:dg_combinations} on $\delta Z_{Z'Z}$ and $\delta Z_{ZZ'}$ in the chosen OS renormalization \eqref{eq:dg_rc} is, in either case, given by the combination
\begin{align}
\delta Z_{Z'Z}+\delta Z_{ZZ'}=2\text{Re}\frac{\Sigma^{ZZ'}_\text{T}(M^2_\text{Z})-\Sigma^{Z'Z}_\text{T}(M^2_{\text{Z}'})}{M_{\text{Z}'}^2-M_\text{Z}^2}.
\label{eq:gamma_RC_comb}
\end{align}
This combination of renormalization constants is numerically stable for degenerate masses, i.e.~for $M_\text{Z}\approx M_{\text{Z}'}$. Further, all appearances of $\gamma$ coming from rewriting the original parameters of the Lagrangian in terms of the chosen input parameters include a prefactor of $M_\text{Z}^2-M_{\text{Z}'}^2$. This shows that no artifacts occur in the degeneracy limit $M_{\text{Z}'}\to M_\text{Z}$, i.e.~the OS scheme is perturbatively stable in this limit.
In the OS scheme all renormalization conditions for parameters related to masses and mixing are based on S-matrix elements. Thus, even though $\delta\gamma_\text{OS}$ itself depends on the tadpole scheme, all predictions of observables made within this OS scheme are independent of the chosen tadpole scheme.\\
\end{itemize}
As an alternative to the OS condition \eqref{eq:gamma-rc}, the mixing angle $\gamma$ could be renormalized with an $\MSbar$ prescription. Using Eq.~\eqref{eq:dg_rc}, the $\MSbar$ version of $\delta\gamma$ can be obtained according to 
\begin{align}
  \delta\gamma_\MSbar = \delta\gamma_\text{OS}\Bigl|_\text{UV},
  \label{eq:gammaMSbOS}
\end{align}
where the subscript UV indicates that only the UV-divergent parts of $\delta\gamma_\text{OS}$ that are proportional to the standard UV divergence $\Delta_\text{UV}$, defined in Eq.~\eqref{eq:stand_UV}, are kept.
The explicit expression for $\delta\gamma_\MSbar$ in the PRTS is given in Appendix~\ref{sec:exp_MSbar_CTs}. We recall that even though Eq.~\eqref{eq:gammaMSbOS} is true in any tadpole scheme, NLO predictions based on $\gamma_\MSbar$ are, in contrast to predictions based on $\gamma_\text{OS}$, in general dependent on the tadpole renormalization scheme (see Sect.~\ref{sec:tadpol}). The tadpole contributions to $\delta \gamma_\MSbar$ in the tadpole schemes described in Sect.~\ref{sec:tadpol} explicitly read
\begin{align}
  \delta \gamma^{\text{PRTS}}_{\MSbar,\text{tad}}&=0,\label{eq:gammaMSbarPRTS}\\
  \delta\gamma^{\text{FJTS}}_{\MSbar,\text{tad}}&=-\frac{ \left(c_{\text{ZZ}'\text{h}}\Delta v^\text{FJTS}_\text{h}+c_{\text{ZZ}'\text{H}}\Delta v^\text{FJTS}_\text{H}\right)}{ M^2_\text{Z}-M_{\text{Z}'}^2}\biggr|_\text{UV},\label{eq:gammaMSbarFJTS}\\
  \delta\gamma^{\text{GIVS}}_{\MSbar,\text{tad}}&=-\frac{ \left(c_{\text{ZZ}'\text{h}}\Delta v^\text{GIVS}_\text{h}+c_{\text{ZZ}'\text{H}}\Delta v^\text{GIVS}_\text{H}\right)}{M^2_\text{Z}-M_{\text{Z}'}^2}\biggr|_\text{UV},\label{eq:gammaMSbarGIVS}
\end{align}
with the shorthands
\begin{align}
  c_{\text{ZZ}'\text{h}}=-s_{2\gamma}\left(c_\alpha v_1+s_\alpha v_2\right)\frac{M_\text{C}^2}{v_1v_2},\qquad  c_{\text{ZZ}'\text{H}}=s_{2 \gamma}\left(c_\alpha v_2-s_\alpha v_1\right)\frac{M_\text{C}^2}{v_1 v_2}
\end{align}
for the couplings $c_{\text{ZZ}'\text{h}}h Z'_\mu Z^\mu+c_{\text{ZZ}'\text{H}}H Z'_\mu Z^\mu$ in the Lagrangian. In contrast to the FJTS, in the PRTS and GIVS tadpole terms explicitly appear in the relations between the original bare parameters $\lambda_{1,0},\lambda_{2,0}, \lambda_{12,0},\mu_{1,0}^2,\mu_{2,0}^2$ of the Higgs potential and the bare masses, vevs, and mixing angles. Thus, the latter depend on the tadpole scheme. The bare mixing angles $\gamma_0^\text{PRTS}$ and $\gamma_0^\text{FJTS}$ are related via
\begin{align}
  \gamma_0^\text{PRTS}=\gamma_0^\text{FJTS}-\frac{1}{M^2_\text{Z}-M_{\text{Z}'}^2} \left(c_{\text{ZZ}'\text{h}}\frac{\delta t^\text{FJTS}_\text{h}}{M_\text{h}^2}+c_{\text{ZZ}'\text{H}}\frac{\delta t^\text{FJTS}_\text{H}}{M_\text{H}^2}\right).
\end{align}
For the $\MSbar$-renormalized mixing angle we find the (gauge-dependent) shift
\begin{align}
  \gamma_\MSbar^\text{PRTS}-\gamma_\MSbar^\text{FJTS}&=\gamma_0^\text{PRTS}-\gamma_0^\text{FJTS}-\left(\delta\gamma_\MSbar^\text{PRTS}-\delta\gamma_\MSbar^\text{FJTS}\right)\nonumber\\
  &=\frac{1}{M^2_\text{Z}-M_{\text{Z}'}^2}\left(c_{\text{ZZ}'\text{h}}\frac{T^h}{M_\text{h}^2}+c_{\text{ZZ}'\text{H}}\frac{T^H}{M_\text{H}^2}\right)\biggr|_\text{finite}
\end{align}
between the PRTS and the FJTS values, where the subscript ``finite'' indicates that all UV-dependent parts proportional to $\Delta_\text{UV}$ are dropped. Similarly, we find the (gauge-independent) shift
\begin{align}
\gamma_\MSbar^\text{GIVS}-\gamma_\MSbar^\text{FJTS}=\frac{1}{M^2_\text{Z}-M_{\text{Z}'}^2}\left(c_{\text{ZZ}'\text{h}}\frac{T^h_\text{nl}}{M_\text{h}^2}+c_{\text{ZZ}'\text{H}}\frac{T^H_\text{nl}}{M_\text{H}^2}\right)\biggr|_\text{finite}
\end{align}
for the conversion between the GIVS and the FJTS.

\subsubsection{Renormalization of the mixing angle {\boldmath$\alpha$}}
Similar to our procedure for the renormalization of $\gamma$, we apply OS or alternatively $\MSbar$ renormalization for the mixing angle $\alpha$. Further possible schemes based on symmetry-inspired conditions are described in Refs.~\cite{SDMixingAngles,Sym_MixingAngles,Sym_MixingAngles2}.\\
If the mass hierarchy $M_\text{h}>2m_{\nu_4}$ holds (recall $M_\text{h}\leq M_\text{H}$), the OS renormalization condition for the Higgs mixing angle $\alpha$ can be formulated for the OS formfactors of the decays $\text{h/H}\to \bar{\nu}_4\nu_4$, defined by
\begin{align}
  \mathcal{M}^{\text{h}\rightarrow \bar{\nu}_4\nu_4}=[\bar{u}_{\nu_4}v_{\nu_4}]_{h}\hspace{1pt}\mathcal{F}^{h \bar{\nu}_4\nu_4},\qquad
 \mathcal{M}^{\text{H}\rightarrow \bar{\nu}_4\nu_4}=[\bar{u}_{\nu_4}v_{\nu_4}]_{H}\hspace{1pt}\mathcal{F}^{H \bar{\nu}_4\nu_4},
  \label{eq:Form_fac_nu4}
\end{align}
where again the fermion spinors of the final state are denoted by $\bar{u}_{\nu_4}$ and $v_{\nu_4}$ and the decay kinematics are indicated by $[\dots]_{h/H}$. As renormalization condition we demand the higher-order corrections to the ratio of the real parts of the two formfactors defined in Eq.~\eqref{eq:Form_fac_nu4} to vanish\footnote{Since we choose an OS renormalization scheme, i.e.~$\alpha$ to be real, the absorptive parts  on the left side of Eq.~\eqref{eq:Form_fac_nu4} are not taken into account in the renormalization condition.},
  \begin{align}
\frac{\text{Re}\,\mathcal{F}^{h\bar{\nu}_4\nu_4}}{\text{Re}\,\mathcal{F}^{H\bar{\nu}_4\nu_4}}\stackrel{!}{=}\frac{\mathcal{F}_\text{LO}^{h\bar{\nu}_4\nu_4}}{\mathcal{F}_\text{LO}^{H\bar{\nu}_4\nu_4}}=-\frac{s_\alpha}{c_\alpha}.
\label{eq:rc-alpha}
\end{align}
At NLO the two formfactors are given by
\begin{align}
\mathcal{F}_\text{NLO}^{h\bar{\nu}_4\nu_4}=\mathcal{F}^{h\bar{\nu}_4\nu_4}_\text{LO}\biggl[&1+\frac{\delta s_\alpha}{s_\alpha}+\frac{\delta s_{\theta_\text{r}}}{s_{\theta_\text{r}}}+\frac{\delta\tilde{y}}{\tilde{y}}-\frac{\delta v_1}{v_1}&\nonumber\\& + \frac{1}{2}\left(2\delta Z^{\nu}_{44}+\delta Z_{hh}-\frac{c_\alpha}{s_\alpha}\delta Z_{Hh}+\frac{c_{\theta_\text{r}}}{s_{\theta_\text{r}}}\delta Z^{\nu,\text{R}}_{34}\right)+\delta_\text{loop}^{h\bar{\nu}_4\nu_4}\biggr],\nonumber\\
\mathcal{F}_\text{NLO}^{H\bar{\nu}_4\nu_4}=\mathcal{F}^{H\bar{\nu}_4\nu_4}_\text{LO}\biggl[&1+\frac{\delta c_\alpha}{c_\alpha}+\frac{\delta s_{\theta_\text{r}}}{s_{\theta_\text{r}}}+\frac{\delta\tilde{y}}{\tilde{y}}-\frac{\delta v_1}{v_1}\nonumber\\
  &+ \frac{1}{2}\left(2\delta Z^{\nu}_{44}+\delta Z_{HH}-\frac{s_\alpha}{c_\alpha}\delta Z_{hH}+\frac{c_{\theta_\text{r}}}{s_{\theta_\text{r}}}\delta Z^{\nu,\text{R}}_{34}\right)+\delta_\text{loop}^{H\bar{\nu}_4\nu_4}\biggr],
\label{eq:ffhH44}
\end{align}
with $\delta_\text{loop}^{h\bar{\nu}_4\nu_4}$ and $\delta_\text{loop}^{H\bar{\nu}_4\nu_4}$ denoting the relative unrenormalized 1-loop corrections to the decays, respectively, and
\begin{align}
  \delta Z^{\nu}_{44}=\frac{1}{2}\left(\delta Z^{\nu,\text{L}}_{44}+\delta Z^{\nu,\text{R}}_{44}\right).
\label{eq:ferm_RC_Hff}
\end{align}
The ratio of the real parts of the two formfactors at NLO is given by
\begin{align}
\frac{\text{Re}\,\mathcal{F}_\text{NLO}^{h\bar{\nu}_4\nu_4}}{\text{Re}\,\mathcal{F}_\text{NLO}^{H\bar{\nu}_4\nu_4}}={}-\frac{s_\alpha}{c_\alpha}\biggl\{&1+\frac{\delta s_\alpha}{s_\alpha}-\frac{\delta c_\alpha}{c_\alpha}+\frac{1}{2}\left[\delta Z_{hh}-\delta Z_{HH}+\frac{s_\alpha}{c_\alpha}\delta Z_{hH} -\frac{c_\alpha}{s_\alpha}\delta Z_{Hh}\right] \nonumber\\
&+\text{Re}\left[\delta_\text{loop}^{h\bar{\nu}_4\nu_4}-\delta_\text{loop}^{H\bar{\nu}_4\nu_4}\right]\biggr\}.
\end{align}
Using $\delta s_\alpha = c_\alpha \delta \alpha$, the renormalization condition \eqref{eq:rc-alpha} then finally leads to
\begin{align}
\delta \alpha_{\text{OS}1}=\frac{1}{2}c_\alpha s_\alpha\left(\delta Z_{HH}-\delta Z_{hh}\right)+\frac{1}{2}\left(c_\alpha^2\delta Z_{Hh} -s_\alpha^2\delta Z_{hH} \right)+c_\alpha s_\alpha\text{Re}\left[\delta_\text{loop}^{H\bar{\nu}_4\nu_4}-\delta_\text{loop}^{h\bar{\nu}_4\nu_4}\right],
\label{eq:alpha-RC}
\end{align}
where OS1 stands for the particular OS condition \eqref{eq:rc-alpha} based on the assumed mass hierarchy $M_\text{h}>2 m_{\nu_4}$. To relax this mass hierarchy, for $m_{\nu_3}\ll M_\text{h},M_\text{H}, m_{\nu_4}$ we can formulate an OS renormalization condition based on the OS formfactors of the $\text{h}/\text{H}\to \bar{\nu}_3\nu_4$ decays or $\bar{\nu}_4\to \text{h}/\text{H} \bar{\nu}_3$ depending on the mass hierarchy between $\text{h}, \text{H}$, and $\nu_4$\footnote{In the ``collider approximation'' for the neutrino sector outlined in Section \ref{sec:simp_ferm_DASM}, where $m_{\nu_1}=m_{\nu_2}=m_{\nu_3}=0$, the state $\nu_3$, which is specifically aligned to couple to the dark sector maximally, corresponds to a mass eigenstate, so that OS renormalization conditions can make use of $\nu_3$ directly. For non-vanishing neutrino masses $m_{\nu_j}$ \mbox{($j=1,2,3$)}, the renormalization of the full neutrino sector becomes way more complicated, an issue that is, however, not relevant for collider physics and, thus, beyond the scope of this paper.}. For $M_\text{h}>m_{\nu_4}$, e.g., we have 
\begin{align}
  \mathcal{M}^{\text{h}\rightarrow \bar{\nu}_3\nu_4}=[\bar{u}_{\nu_4}\omega_\text{R}v_{\nu_3}]_{h}\hspace{1pt}\mathcal{F}^{h \bar{\nu}_4\nu_3},\qquad
 \mathcal{M}^{\text{H}\rightarrow \bar{\nu}_3\nu_4}=[\bar{u}_{\nu_4}\omega_\text{R}v_{\nu_3}]_{H}\hspace{1pt}\mathcal{F}^{H \bar{\nu}_4\nu_3},
 \label{eq:Form_fac_nu4_nu3}
\end{align}
while for $M_\text{H}<m_{\nu_4}$ we can write
\begin{align}
  \mathcal{M}^{\bar{\nu}_4\rightarrow \text{h}\bar{\nu}_3}=[\bar{v}_{\nu_4}\omega_\text{R}v_{\nu_3}]_{h}\hspace{1pt}\mathcal{F}^{h \bar{\nu}_4\nu_3},\qquad
 \mathcal{M}^{\bar{\nu}_4\rightarrow \text{H}\bar{\nu}_3}=[\bar{v}_{\nu_4}\omega_\text{R}v_{\nu_3}]_{H}\hspace{1pt}\mathcal{F}^{H \bar{\nu}_4\nu_3},
\end{align}
with the same formfactors\footnote{Note that in our notation the labels of the formfactors always denote incoming fields.} $\mathcal{F}^{S\bar{\nu}_4\nu_3}$ and \mbox{$\omega_\text{R}=\frac{1}{2}(1+\gamma_5)$} denoting the right-handed chiral projection operator.
Here the spinors of the final-state particles are denoted by $\bar{u}_{\nu_4},v_{\nu_3}$ and $\bar{v}_{\nu_4},v_{\nu_3}$, respectively, and the decay kinematics are indicated by $[\dots]_{h/H}$. As renormalization condition we now demand the higher-order corrections to the ratio of the real parts of these two formfactors to vanish,
  \begin{align}
\frac{\text{Re}\,\mathcal{F}^{h\bar{\nu}_4\nu_3}}{\text{Re}\,\mathcal{F}^{H\bar{\nu}_4\nu_3}}\stackrel{!}{=}\frac{\mathcal{F}_\text{LO}^{h\bar{\nu}_4\nu_3}}{\mathcal{F}_\text{LO}^{H\bar{\nu}_4\nu_3}}=-\frac{s_\alpha}{c_\alpha}.
\label{eq:rc-alpha_nu4_nu3}
\end{align}
The renormalized NLO formfactors are given by
\begin{align}
  \mathcal{F}^{h\bar{\nu}_4\nu_3}_\text{NLO}=\mathcal{F}^{h\bar{\nu}_4\nu_3}_\text{LO}\biggl[&1+\frac{\delta s_\alpha}{s_\alpha}+\frac{\delta c_{\theta_\text{r}}}{c_{\theta_\text{r}}}+\frac{\delta\tilde{y}}{\tilde{y}}-\frac{\delta v_1}{v_1}&\nonumber\\& + \frac{1}{2}\left(\delta Z^{\nu,\text{R}}_{33}+\delta Z^{\nu,\text{L}}_{44}+\delta Z_{hh}-\frac{c_\alpha}{s_\alpha}\delta Z_{Hh}+\frac{s_{\theta_\text{r}}}{c_{\theta_\text{r}}}\delta Z^{\nu,\text{R}}_{43}\right)+\delta_\text{loop}^{h\bar{\nu}_4\nu_3}\biggr],\nonumber\\
\mathcal{F}^{H\bar{\nu}_4\nu_3}_\text{NLO}=\mathcal{F}^{H\bar{\nu}_4\nu_3}_\text{LO}\biggl[&1+\frac{\delta c_\alpha}{c_\alpha}+\frac{\delta c_{\theta_\text{r}}}{c_{\theta_\text{r}}}+\frac{\delta\tilde{y}}{\tilde{y}}-\frac{\delta v_1}{v_1}\nonumber\\
  &+ \frac{1}{2}\left(\delta Z^{\nu,\text{L}}_{44}+\delta Z^{\nu,\text{R}}_{33}+\delta Z_{HH}-\frac{s_\alpha}{c_\alpha}\delta Z_{hH}+\frac{s_{\theta_\text{r}}}{c_{\theta_\text{R}}}\delta Z^{\nu,\text{R}}_{43}\right)+\delta_\text{loop}^{H\bar{\nu}_4\nu_3}\biggr],
\label{eq:ffhH43}
\end{align}
where the $\delta_\text{loop}^{h/H\bar{\nu}_4\nu_3}$ again represent the unrenormalized 1-loop corrections to the respective decays. Inserting this into the renormalization condition \eqref{eq:rc-alpha_nu4_nu3} leads to
\begin{align}
  \delta \alpha_{\text{OS2}}=\frac{1}{2}c_\alpha s_\alpha\left(\delta Z_{HH}-\delta Z_{hh}\right)+\frac{1}{2}\left(c_\alpha^2\delta Z_{Hh} -s_\alpha^2\delta Z_{hH} \right)+c_\alpha s_\alpha\text{Re}\left[\delta_\text{loop}^{H\bar{\nu}_4\nu_3}-\delta_\text{loop}^{h\bar{\nu}_4\nu_3}\right]
  \label{eq:alpha_RC_Zp}
\end{align}
for the renormalization constant of the Higgs mixing angle.\\
The renormalization conditions \eqref{eq:rc-alpha} and \eqref{eq:rc-alpha_nu4_nu3} have various desirable features. Firstly, they are symmetric in the fields $h$ and $H$. Secondly, the OS renormalization constants $\delta \alpha_{\text{OS}i}$ ($i=1,2$) are gauge independent and numerically stable for degenerated masses $M_\text{h} \sim M_\text{H}$ and have smooth limits for $s_\alpha \rightarrow 0$ and $c_\alpha \rightarrow 0$. In contrast to the OS renormalization of the Higgs mixing angle $\alpha$ of the SM extension by a real Higgs field considered in Ref.~\cite{SDMixingAngles} or an $\MSbar$ renormalization of the DASM, the OS renormalization of $\alpha$ in the DASM is process dependent. This is reflected by the remaining non-vanishing higher-order contributions $\delta_\text{loop}^{h\bar{\nu}_4\nu_j}$ and $\delta_\text{loop}^{H\bar{\nu}_4\nu_j}$, $j=3,4$, in Eqs.~\eqref{eq:alpha-RC} and \eqref{eq:rc-alpha_nu4_nu3}, respectively, and due to the fact that the complex Higgs field $\rho$ of the DASM cannot be coupled to fully gauge-invariant fake fermion fields since $\rho$ is not a singlet under the  $U(1)_\text{d}$ gauge group.\\
As an alternative to OS conditions, the angle $\alpha$ could be renormalized via an $\MSbar$ prescription. The resulting renormalization constant in the $\MSbar$ scheme can be obtained from \eqref{eq:alpha-RC} or  \eqref{eq:alpha_RC_Zp} via
\begin{align}
  \delta\alpha_\MSbar=\delta\alpha_{\text{OS}1}\Bigl|_\text{UV}=\delta\alpha_{\text{OS}2}\Bigl|_\text{UV}.
\label{eq:dalphaMSbOS}
\end{align}
The explicit expression for $\delta\alpha_\MSbar$ in the PRTS, which does not depend on any mass hierarchy of the Higgs bosons, is given in Appendix~\ref{sec:exp_MSbar_CTs}. Note that even though Eq.~\eqref{eq:dalphaMSbOS} is true in any tadpole scheme, the NLO predictions based on the OS renormalization of $\alpha$ are, in contrast to NLO predictions based on $\MSbar$ renormalization, not dependent on the tadpole scheme \mbox{(see Sect~\ref{sec:tadpol})}. The tadpole contributions to $\delta \alpha_\MSbar$ in the three described tadpole schemes explicitly read
\begin{align}
  \delta \alpha_{\MSbar,\text{tad}}^\text{PRTS}&=0,
  \label{eq:dalphaMSbarPRTS}\\
  \delta\alpha_{\MSbar,\text{tad}}^\text{FJTS}&=-2\frac{c_\text{hhH} \Delta v^\text{FJTS}_h+c_\text{hHH}\Delta v_H^\text{FJTS}}{M_\text{H}^2-M_\text{h}^2}\biggr|_\text{UV},
  \label{eq:dalphaMSbarFJTS}\\
  \delta\alpha_{\MSbar,\text{tad}}^\text{GIVS}&=-2\frac{c_\text{hhH} \Delta v^\text{GIVS}_h+c_\text{hHH}\Delta v_H^\text{GIVS}}{M_\text{H}^2-M_\text{h}^2}\biggr|_\text{UV},\label{eq:dalphaMSbarGIVS}
\end{align}
where $c_\text{hhH}$ and $c_\text{HHh}$ are the scalar coupling constants given in Eq.~(2.18) of Ref.~\cite{BogDit}. As described in Sect.~\ref{sec:dgamma}, tadpole terms explicitly appear in the relations between the original bare parameters of the Higgs potential and mixing angles. The bare mixing angles $\alpha_0^\text{PRTS}$ and $\alpha_0^\text{FJTS}$ are related via
\begin{align}
  \alpha_0^\text{PRTS}=\alpha_0^\text{FJTS}-\frac{2}{M_\text{H}^2-M_\text{h}^2}\left(c_\text{hhH} \frac{\delta t^\text{FJTS}_h}{M_\text{h}^2}+c_\text{hHH}\frac{\delta t_H^\text{FJTS}}{M_\text{H}^2}\right).
\end{align}
For the $\MSbar$-renormalized mixing angle we find the (gauge-dependent) shift
\begin{align}
  \alpha_\MSbar^\text{PRTS}-\alpha_\MSbar^\text{FJTS}&=\alpha_0^\text{PRTS}-\alpha_0^\text{FJTS}-\left(\delta\alpha_\MSbar^\text{PRTS}-\delta\alpha_\MSbar^\text{FJTS}\right)\nonumber\\
  &=\frac{2}{M_\text{H}^2-M_\text{h}^2}\left(c_\text{hhH} \frac{ T^h}{M_\text{h}^2}+c_\text{hHH}\frac{ T^H}{M_\text{H}^2}\right)\biggr|_\text{finite}
\end{align}
between the respective PRTS and the FJTS values. Similarly, we find the (gauge-independent) shift
\begin{align}
\alpha_\MSbar^\text{GIVS}-\alpha_\MSbar^\text{FJTS}=\frac{2}{M_\text{H}^2-M_\text{h}^2}\left(c_\text{hhH} \frac{ T^h_\text{nl}}{M_\text{h}^2}+c_\text{hHH}\frac{ T^H_\text{nl}}{M_\text{H}^2}\right)\biggr|_\text{finite}
\end{align}
for the conversion between the GIVS and the FJTS.
 
\subsubsection{Renormalization of the mixing angle {\boldmath$\theta_\text{r}$}}
For the renormalization of the mixing angle $\theta_\text{r}$ in the neutrino sector we consider OS formfactors for appropriate decays. By default, we assume $M_\text{H}> 2 m_{\nu_4}$ for the decays $\text{H}\to \bar{\nu}_4\nu_4/\bar{\nu}_3\nu_4$ to be possible. For $M_\text{H}< 2 m_{\nu_4}$ we sketch an alternative OS renormalization condition for $\theta_\text{r}$ at the end of this section.
The amplitudes for $\text{H}\rightarrow \bar{\nu}_4\nu_4$ and $\text{H}\rightarrow\bar{\nu}_3\nu_4$ read
\begin{align}
\mathcal{M}^{\text{H}\rightarrow \bar{\nu}_4\nu_4}=[\bar{u}_{\nu_4}v_{\nu_4}]_{H}\mathcal{F}^{H\bar{\nu}_4\nu_4},\hspace{0.5cm}\mathcal{M}^{\text{H}\rightarrow\bar{\nu}_3\nu_4}=[\bar{u}_{\nu_4}\omega_\text{R}v_{\nu_3}]_{H}\mathcal{F}^{H\bar{\nu}_4\nu_3},
\end{align}
with the spinors $\bar{u}_{\nu_4}$ and $v_j$, $j=\nu_3,\nu_4$, of the final-state fermions, the right-handed chiral projection operator $\omega_\text{R}=\frac{1}{2}(1+\gamma_5)$ and $[\dots]_H$ denoting the decay kinematics. To fix $\delta \theta_\text{r}$ we demand that the higher-order corrections to the ratio of the LO formfactors $\mathcal{F}_\text{LO}^{H\bar{\nu}_4\nu_4}$ and $\mathcal{F}_\text{LO}^{H\bar{\nu}_4\nu_3}$ vanish up to absorptive parts,
\begin{align}
\frac{\text{Re}\,\mathcal{F}^{H\bar{\nu}_4\nu_4}}{\text{Re}\,\mathcal{F}^{H\bar{\nu}_4\nu_3}}\stackrel{!}{=}\frac{\mathcal{F}_\text{LO}^{H\bar{\nu}_4\nu_4}}{\mathcal{F}_\text{LO}^{H\bar{\nu}_4\nu_3}}=\frac{s_{\theta_\text{r}}}{c_{\theta_\text{r}}}.
\label{eq:rc-thetar}
\end{align}
Combining Eqs.~\eqref{eq:ffhH44} and \eqref{eq:ffhH43} one finds
\begin{align}
\frac{\text{Re}\,\mathcal{F}_\text{NLO}^{H\bar{\nu}_4\nu_4}}{\text{Re}\,\mathcal{F}_\text{NLO}^{H\bar{\nu}_4\nu_3}}={}\frac{s_{\theta_\text{r}}}{c_{\theta_\text{r}}}\biggl\{&1+\frac{\delta s_{\theta_\text{r}}}{s_{\theta_\text{r}}}-\frac{\delta c_{\theta_\text{r}}}{c_{\theta_\text{r}}}+\frac{1}{2}\left[\delta Z^{\nu,\text{R}}_{44}-\delta Z^{\nu,\text{R}}_{33}+\frac{c_{\theta_\text{r}}}{s_{\theta_\text{r}}}\delta Z^{\nu,\text{R}}_{34} -\frac{s_{\theta_\text{r}}}{c_{\theta_\text{r}}}\delta Z^{\nu,\text{R}}_{43}\right] \nonumber\\
&+\text{Re}\left[\delta_\text{loop}^{H\bar{\nu}_4\nu_4}-\delta_\text{loop}^{H\bar{\nu}_4\nu_3}\right]\biggr\}
\end{align}
for the ratio of the real parts of the two formfactors at NLO.
Using Eq.~\eqref{eq:rc-thetar} and $\delta s_{\theta_\text{r}}= c_{\theta_\text{r}} \delta \theta_\text{r}$, we finally find
\begin{align}
\delta \theta^\text{H}_{\text{r,OS}}={}&\frac{1}{2}c_{\theta_\text{r}} s_{\theta_\text{r}}\Bigl(\delta Z^{\nu,\text{R}}_{33}-\delta Z^{\nu,\text{R}}_{44}\Bigr)+\frac{1}{2}\Bigl(s_{\theta_\text{r}}^2\delta Z^{\nu,\text{R}}_{43} -c_{\theta_\text{r}}^2\delta Z^{\nu,\text{R}}_{34} \Bigr)+c_{\theta_\text{r}} s_{\theta_\text{r}}\text{Re}\Bigl[\delta_\text{loop}^{H\bar{\nu}_4\nu_3}-\delta_\text{loop}^{H\bar{\nu}_4\nu_4}\Bigr].
\label{eq:dtheta}
\end{align}
As an alternative for $M_\text{H}<2m_{\nu_4}$, one can formulate an OS renormalization condition using the OS formfactors $\mathcal{F}^{Z'\bar{\nu}^{\text{R}}_{3}\nu^{\text{R}}_{3}}$ and $\mathcal{F}^{Z'\bar{\nu}_4\nu_3}_1$ (defined below) of the $\text{Z}'\to \bar{\nu}^{\text{R}}_{3}\nu^{\text{R}}_{3}$ and \mbox{$\text{Z}'\to \bar{\nu}_3\nu_4$} decays, respectively. Further, in the case $M_{\text{Z}'}<m_{\nu_4}$ one can simply switch from $Z'\to \bar{\nu}_3\nu_4$ to the $\nu_4\to \text{Z}' \nu_{3}$ decay for the formulation of the renormalization condition to cover the whole parameter space. This will not affect the formal result for the renormalization constant $\delta\theta_\text{r,OS}^{\text{Z}'}$ (see Eq.~\eqref{eq:dtheta_alter}). The amplitudes are given by
\begin{align}
  \mathcal{M}^{\text{Z}'\rightarrow \bar{\nu}^{\text{R}}_{3}\nu^{\text{R}}_{3}}&=[\bar{u}_{\nu_3}\slashed{\varepsilon}\omega_\text{R} v_{\nu_3}]_{Z'}\hspace{1pt}\mathcal{F}^{Z' \bar{\nu}^{\text{R}}_{3}\nu^{\text{R}}_{3}},\\
  \mathcal{M}^{\text{Z}'\rightarrow \bar{\nu}_3\nu_4}&=[\bar{u}_{\nu_4}\slashed{\varepsilon}\omega_\text{R}v_{\nu_3}]_{Z'}\hspace{1pt}\mathcal{F}^{Z' \bar{\nu}_4\nu_3}_1+ [\bar{u}_{\nu_4}\omega_\text{R}v_{\nu_3}]_{Z'}\left(\varepsilon_\mu p_{\nu_3}^\mu\right)\mathcal{F}^{Z' \bar{\nu}_4\nu_3}_2,
\end{align}
with the spinors $\bar{u}_{\nu_j}$, $j=3,4$, and $v_{\nu_3}$ of the final-state fermions, the polarization vector $\varepsilon_\mu$ of the $\text{Z}'$ boson, and the right-handed chirality projector $\omega_\text{R}=\frac{1}{2}(1+\gamma_5)$. The momentum of the neutrino $\nu_3$ is given by $p_{\nu_3}$, and $[\dots]_{Z'}$ denotes the decay kinematics. Note that the additional formfactor $\mathcal{F}_2^{ Z' \bar{\nu}_4\nu_3}$ for the $\text{Z}'\to \bar{\nu}_3\nu_4$ decay is loop induced, i.e.~zero at LO. Due to the unique decomposition of the matrix element into different covariants spanning the underlying $\text{Z}'$-truncated Greens function, it is still possible to formulate the renormalization conditions by only using $\mathcal{F}^{Z'\bar{\nu}_4\nu_3}_1$.\\
The OS renormalization condition at NLO then reads
  \begin{align}
    \frac{\text{Re}\,\mathcal{F}_\text{NLO}^{Z'\bar{\nu}^{\text{R}}_{3}\nu^{\text{R}}_{3}}}{\text{Re}\,\mathcal{F}_{1,\text{NLO}}^{Z'\bar{\nu}_4\nu_3}}\stackrel{!}{=}\frac{\mathcal{F}_\text{LO}^{Z'\bar{\nu}^{\text{R}}_{3}\nu^{\text{R}}_{3}}}{\mathcal{F}_{1,\text{LO}}^{Z'\bar{\nu}_4\nu_3}}=-\frac{s_{\theta_\text{r}}}{c_{\theta_\text{r}}}.
    \label{eq:rc-thetar_alt}
  \end{align}
  Following the steps presented above for $\delta\theta_\text{r,OS}^{\text{H}}$ one finds
  \begin{align}
    \delta\theta_{\text{r},\text{OS}}^{\text{Z}'}=\frac{1}{2}s_{\theta_\text{r}}c_{\theta_\text{r}}\left(\delta Z^{\nu,\text{R}}_{44}-\delta Z^{\nu,\text{R}}_{33}\right)+\frac{1}{2}\left(c^2_{\theta_\text{r}}\delta Z^{\nu,\text{R}}_{43}-s_{\theta_\text{r}}^2\delta Z^{\nu,\text{R}}_{34}\right)+s_{\theta_\text{r}}c_{\theta_\text{r}}\text{Re}\left[\delta_\text{loop}^{Z'\nu_4\nu_3}-\delta_\text{loop}^{Z'\nu_{3}\nu_{3}}\right]
    \label{eq:dtheta_alter}
  \end{align}
  at NLO. Here $\delta_\text{loop}^{Z'\nu_i\nu_3}$, $i=3,4$, represent the unrenormalized, relative 1-loop corrections to the respective decays. \\
  Both presented OS renormalization conditions \eqref{eq:rc-thetar} and \eqref{eq:rc-thetar_alt} lead to gauge-independent renormalization constants $\delta\theta_{\text{r},\text{OS}}^{\text{H/Z}'}$, which are well behaved for exceptional values of $\theta_\text{r}$. As already pointed out for the renormalization of the Higgs mixing angle $\alpha$ in the previous section, NLO predictions based on $\theta_\text{r,\text{OS}}$ in an OS scheme are independent of the chosen tadpole scheme, but a process dependence is introduced in $\delta\theta_{\text{r},\text{OS}}^{\text{H/Z}'}$ via the explicit loop corrections appearing in \eqref{eq:dtheta} and \eqref{eq:dtheta_alter}, respectively. Further, \eqref{eq:dtheta} requires the mass hierarchy $M_\text{H}> 2m_{\nu_4}$ in the DASM.\\
As an alternative to OS renormalization, the angle $\theta_\text{r}$ can be renormalized in the $\MSbar$ scheme. The $\MSbar$ version of $\delta\theta_\text{r}$ can be obtained via
\begin{align}
  \delta\theta_{\text{r},\MSbar}=\delta\theta_{\text{r},\text{OS}}^\text{H}\Bigl|_\text{UV}=\delta\theta_{\text{r},\text{OS}}^{\text{Z}'}\Bigl|_\text{UV}.
\end{align}
Again, this relation is true in any tadpole scheme, but NLO predictions based on $\theta_{\text{r},\MSbar}$ are, in general, dependent on the tadpole renormalization scheme (see Sect~\ref{sec:tadpol}). The resulting explicit expression for $\delta\theta_{\text{r},\MSbar}$ in the PRTS can be found in Appendix~\ref{sec:exp_MSbar_CTs}; note that it does not depend on any hierarchy of masses of the Higgs bosons, $\text{Z}'$, or the neutrinos.
The tadpole contributions to $\delta \theta_{\text{r},\MSbar}$ in the tadpole schemes described in Sect.~\ref{sec:tadpol} explicitly read
\begin{align}
  \delta \theta_{\text{r},\MSbar,\text{tad}}^{\text{PRTS}}&=0,\label{eq:thetaRMSbarPRTS}\\
  \delta\theta_{\text{r},\MSbar,\text{tad}}^{\text{FJTS}}&=\frac{1}{m_{\nu_4}}\left(c_{\text{h}\bar{\nu}_4\nu_3}\Delta v_h^\text{FJTS}+c_{\text{H}\bar{\nu}_4\nu_3}\Delta v_H^\text{FJTS}\right)\biggr|_\text{UV},\label{eq:thetaRMSbarFJTS}\\
  \delta\theta_{\text{r},\MSbar,\text{tad}}^{\text{GIVS}}&=\frac{1}{m_{\nu_4}}\left(c_{\text{h}\bar{\nu}_4\nu_3}\Delta v_h^\text{GIVS}+c_{\text{H}\bar{\nu}_4\nu_3}\Delta v_H^\text{GIVS}\right)\biggr|_\text{UV},\label{eq:thetaRMSbarGIVS}
\end{align}
where we have introduced the shorthands
\begin{align}
  c_{\text{h}\bar{\nu}_4\nu_3}=\frac{1}{v_1}s_\alpha s_{\theta_\text{r}}c_{\theta_\text{r}}m_{\nu_4},\qquad
  c_{\text{H}\bar{\nu}_4\nu_3}=-\frac{1}{v_1}c_\alpha s_{\theta_\text{r}}c_{\theta_\text{r}}m_{\nu_4},
\end{align}
for the couplings $c_{\text{h}\bar{\nu}_4\nu_3}h \bar{\nu}_4 \nu_3+c_{\text{H}\bar{\nu}_4\nu_3}H \bar{\nu}_4 \nu_3$ in the Lagrangian.
  As described in the previous sections tadpole terms explicitly appear in the relations between the original bare parameters of the Higgs potential and $\theta_{\text{r},0}$. The bare mixing angles $\theta_{\text{r},0}^\text{PRTS}$ and $\theta_{\text{r},0}^\text{FJTS}$ are related via
\begin{align}
  \theta_{\text{r},0}^\text{PRTS}=\theta_{\text{r},0}^\text{FJTS}+\frac{1}{m_{\nu_4}}\left(c_{\text{h}\bar{\nu}_4\nu_3}\frac{\delta t_h^\text{FJTS}}{M_\text{h}^2}+c_{\text{H}\bar{\nu}_4\nu_3}\frac{\delta t_H^\text{FJTS}}{M_\text{H}^2}\right).
\end{align}
Thus, for the $\MSbar$-renormalized mixing angle we find the (gauge-dependent) shift 
\begin{align}
  \theta_{\text{r},\MSbar}^\text{PRTS}-\theta_{\text{r},\MSbar}^\text{FJTS}&=\theta_{\text{r},0}^\text{PRTS}-\theta_{\text{r},0}^\text{FJTS}-\left(\delta\theta_{\text{r},\MSbar}^\text{PRTS}-\delta\theta_{\text{r},\MSbar}^\text{FJTS}\right)\nonumber\\
  &=-\frac{1}{m_{\nu_4}}\left(c_{\text{h}\bar{\nu}_4\nu_3}\frac{T^h}{M_\text{h}^2}+c_{\text{H}\bar{\nu}_4\nu_3}\frac{T^H}{M_\text{H}^2}\right)\biggr|_\text{finite}
\end{align}
between the respective PRTS and the FJTS values. Similarly, we find the (gauge-independent) shift
\begin{align}
\theta_{\text{r},\MSbar}^\text{GIVS}-\theta_{\text{r},\MSbar}^\text{FJTS}=-\frac{1}{m_{\nu_4}}\left(c_{\text{h}\bar{\nu}_4\nu_3}\frac{T_\text{nl}^h}{M_\text{h}^2}+c_{\text{H}\bar{\nu}_4\nu_3}\frac{T_\text{nl}^H}{M_\text{H}^2}\right)\biggr|_\text{finite}
\end{align}
for the conversion between the GIVS and the FJTS.

\subsubsection{Charge renormalization}
The electric unit charge $e$ is defined as the coupling constant of the fermion--photon interaction with on-shell fermions in the Thomson limit, i.e.\ in the limit of vanishing photon momentum transfer. In the following, we make use of the results of Ref.~\cite{Stefan_Charge_Ren}, where the charge renormalization of the SM was generalized to SM extensions with gauge groups that contain an explicit $U(1)$ factor similar to the broken weak  hypercharge symmetry of the SM. While the treatment of charge renormalization in Ref.~\cite{Stefan_Charge_Ren} is even valid to all orders, here we only consider NLO accuracy as throughout this paper.\\
We briefly sketch the arguments of Ref.~\cite{Stefan_Charge_Ren} for the derivation of the electric charge renormalization constant within the DASM in our notation. The derivation is based on charge universality. Charge universality states that in the Thomson limit higher-order corrections to the coupling strength of the photon to any charged particle do not depend on any specific properties of the charged particle besides its charge, i.e.\ the charge renormalization constant does not depend on the specific charged particle (typically taken as fermion) used to derive it. Charge universality can, e.g., be proven with the background-field method (see e.g.~Refs.~\cite{BDJ,Stefan_Charge_Ren,weinberg_1996,Charge_Uni_Dit}), where the all-order renormalization of electric charge widely works as in pure QED. Exploiting charge universality we introduce a second ``fake fermion'' field\footnote{Note that this ``fake fermion'' field is equivalent to the field $\eta$ in Ref.~\cite{Stefan_Charge_Ren}. We have renamed it here to avoid confusion with the parameter $\eta$ introduced in Sect.~\ref{sec:gauge_part}.} $\kappa$ which only carries infinitesimal weak hypercharge, $Y_{\text{w},\kappa}$, but no other quantum numbers, leading to the infinitesimal electric charge $Q_\kappa=Y_{\text{w},\kappa}/2$. The most general gauge-invariant Lagrangian including the new fermion field that has to be added to the Lagrangian of the DASM is given by
\begin{align}
\mathcal{L}_\kappa={}\bar{\kappa}\biggl({}&\text{i}\slashed{\partial}-\frac{Y_{\text{w},\kappa}}{2}g_{1} \slashed{B}-m_{\kappa}\biggr)\kappa\nonumber\\
={}\bar{\kappa}\biggl\{{}& \text{i}\slashed{\partial}-e Q_\kappa \Bigl[\slashed{A}+\frac{1}{c_{\text{w}}}\Bigl(\left(s_{\text{w}}c_{\gamma} -\eta s_{\gamma}\right)\slashed{Z}-\left(s_{\text{w}}s_{\gamma}+\eta c_{\gamma}\right)\slashed{Z}'\Bigr)\Bigr]-m_{\kappa}\biggr\}\kappa.
\end{align}
The non-chirality of the fermion $\kappa$ allows for a Dirac mass term with mass parameter $m_\kappa$ and ensures that no anomalies are introduced into the theory. Further, taking the limit $Q_\kappa\rightarrow 0$ will decouple the introduced fake fermion field $\kappa$ from any other particles so that the DASM is recovered. Charge universality now implies that one can use this fake fermion to restore the desired physical meaning of the electric charge in higher orders by demanding all relative higher-order corrections to the $A\bar{\kappa}\kappa$ vertex to vanish in the Thomson limit. Therefore, the imposed renormalization condition for the renormalized  $A\bar{\kappa}\kappa$ vertex function is given by
\begin{align}
  \bar{u}(p)\Gamma_{\text{R},\mu}^{A\bar{\kappa}\kappa}(0,-p,p)u(p)\Biggr|_{p^2=m_\kappa^2}=-Q_\kappa e\hspace{1pt} \bar{u}(p)\gamma_\mu u(p),
  \label{eq:charg_ren_cond}
\end{align}
where $m_\kappa$ denotes the renormalized on-shell mass of the fake fermion. At the 1-loop level, the renormalized vertex function is given by
\begin{align}
  \Gamma_{\text{R},\mu}^{A\bar{\kappa}\kappa}(k,\bar{p},p)={}&\left(1+\frac{1}{2} \delta Z_{AA}+\delta Z_{\kappa}+\delta Z_e\right)\Gamma_{\text{LO},\mu}^{A\bar{\kappa}\kappa}(k,\bar{p},p)+Q_\kappa \Lambda_{\mu}^{A\bar{\kappa}\kappa}(k,\bar{p},p)\nonumber\\
  &+\frac{1}{2}\delta Z_{ZA}\Gamma_{\text{LO},\mu}^{Z\bar{\kappa}\kappa}(k,\bar{p},p)+\frac{1}{2}\delta Z_{Z'A}\Gamma_{\text{LO},\mu}^{Z'\bar{\kappa}\kappa}(k,\bar{p},p),
\end{align}
with the LO vertex functions
\begin{align}
 \Gamma_{\text{LO},\mu}^{A\bar{\kappa}\kappa}(k,\bar{p},p)={}&-Q_\kappa e \gamma_\mu,\\
 \Gamma_{\text{LO},\mu}^{Z\bar{\kappa}\kappa}(k,\bar{p},p)={}&-Q_\kappa \frac{e}{c_\text{w}} \left( s_{\text{w}}c_\gamma -\eta s_{\gamma}\right),\\
 \Gamma_{\text{LO},\mu}^{Z'\bar{\kappa}\kappa}(k,\bar{p},p)={}&Q_\kappa \frac{e}{c_\text{w}} \left(s_\text{w} s_{\gamma}+ \eta c_{\gamma} \right),
\end{align}
and the unrenormalized vertex corrections $\Lambda_{\mu}^{A\bar{\kappa}\kappa}(k,\bar{p},p)$.
A crucial point in the derivation is that all possible couplings of the fake fermion to particles within the DASM are proportional to $Q_\kappa$ which implies
\begin{align}
  \delta Z_\kappa={}\mathcal{O}\left(Q_\kappa^2\right),\qquad \Lambda_\mu^{A\bar{\kappa}\kappa}(k,\bar{p},p)={}\mathcal{O}\left(Q_\kappa^2\right).
\end{align}
Taking the limit $Q_\kappa\rightarrow 0$ and keeping only terms linear in $Q_\kappa$, Eq.~\eqref{eq:charg_ren_cond} implies
\begin{align}
  0=\left[\delta Z_e + \frac{1}{2} \delta Z_{AA}+\frac{1}{2c_\text{w}}\Bigl\{\left(s_{\text{w}}c_\gamma -\eta s_{\gamma}\right)\delta Z_{ZA}-\left(s_\text{w} s_{\gamma}+\eta c_{\gamma}\right) \delta Z_{Z'A}\Bigr\}\right]\bar{u}(p)\gamma_\mu u(p),
\end{align}
leading directly to
\begin{align}
 \delta Z_e=-\frac{1}{2}\biggl[\delta Z_{AA}+\frac{s_{\text{w}}c_{\gamma} - \eta s_{\gamma}}{c_\text{w}}\delta Z_{ZA}-\frac{s_{\text{w}}s_{\gamma}+\eta c_{\gamma}}{c_\text{w}}\delta Z_{Z'A}\biggr].
\end{align}
In agreement with charge universality, $\delta Z_e$ is independent of the fermion $\kappa$ used for the formulation of the renormalization condition and is, similarly to the SM, given by a pure combination of gauge-boson self-energies. Obviously, the well-known NLO SM result for the electric charge renormalization constant is obtained by taking the SM limit $\gamma \to 0$.

\section{Prediction for the W-boson mass from muon decay}
\label{se:pheno}
\begin{figure}[b]
\centering
  \setlength{\unitlength}{1pt}
\raisebox{-18pt}{\includegraphics[width=0.2\textwidth]{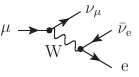}}\qquad\qquad
\raisebox{-18pt}{\includegraphics[width=0.2\textwidth]{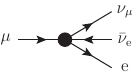}}
\caption{The Feynman diagrams representing muon decay in the DASM (left) and the Fermi theory (right) at LO.}
\label{fig:muon_dec_dia}
\end{figure}

\begin{figure}[b]
\centering
  \setlength{\unitlength}{1pt}
\raisebox{-18pt}{\includegraphics[width=0.2\textwidth]{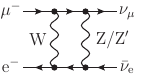}}\quad
\raisebox{-18pt}{\includegraphics[width=0.2\textwidth]{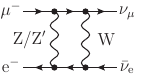}}\quad
\raisebox{-18pt}{\includegraphics[width=0.2\textwidth]{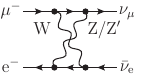}}\quad
\raisebox{-18pt}{\includegraphics[width=0.2\textwidth]{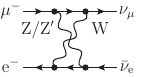}}
\caption{Box diagrams with $\text{Z, Z}'$, and W exchange contributing to muon decay at NLO in the DASM.}
\label{fig:muon_dec_mass_box}
\end{figure}

\begin{figure}
\centering
                \includegraphics[width=16cm]{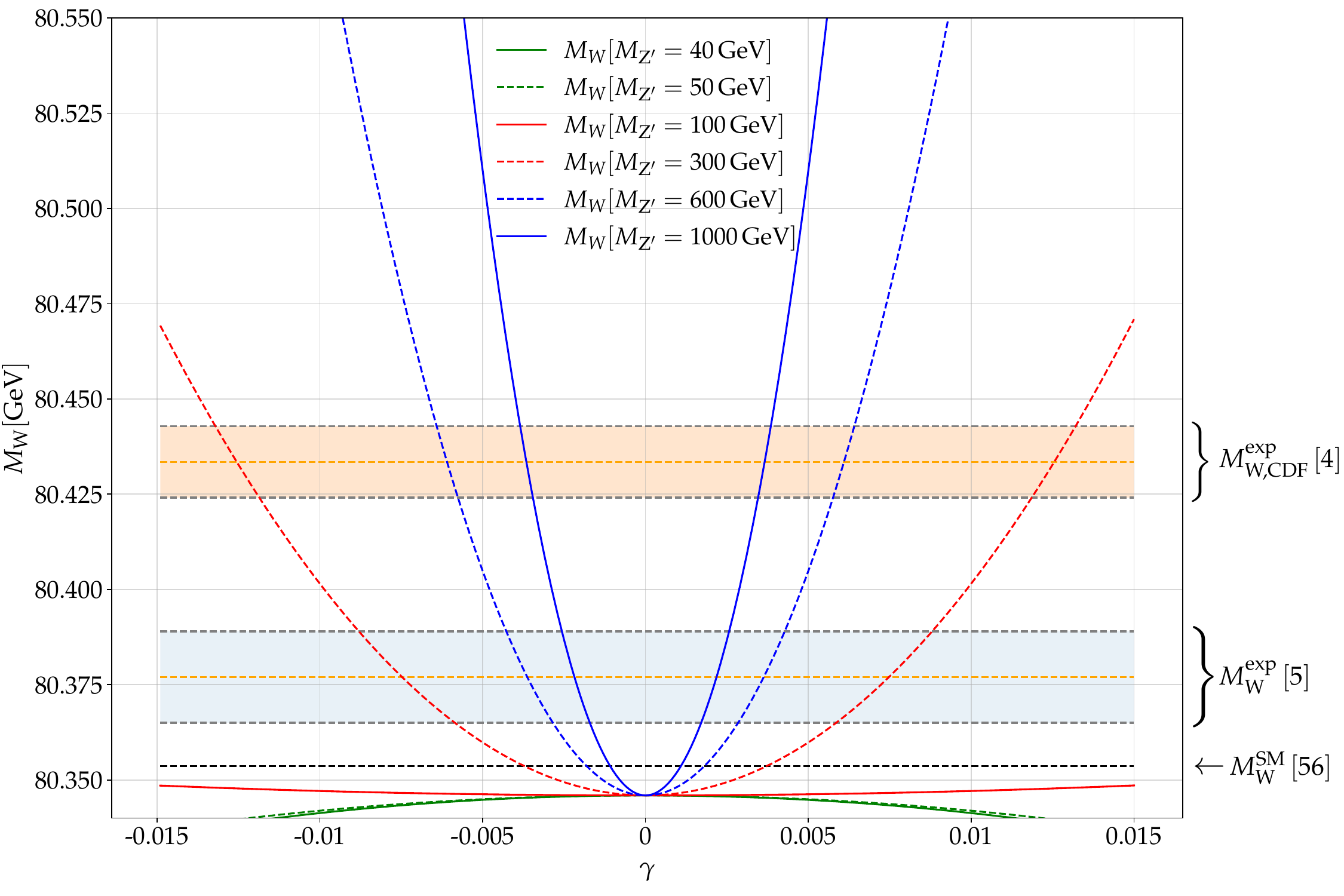}
                \caption{Predictions for $M_\text{W}$ in the DASM for various combinations of $\gamma$ and $M_{\text{Z}'}$. The best SM prediction is given by $M^\text{SM}_\text{W}=80.3536\,\GeV$ \cite{MWmasseSM}, and the measured world average is \mbox{$M^\text{exp}_\text{W}=80.377\pm0.012\,\GeV$ \cite{PDG}.} For completeness we show the result of the CDF experiment $M_\text{W,CDF}^\text{exp}= 80.4335 \pm 0.0094\,\GeV$ \cite{CDFMW}, which is not included in the world average value quoted above. The parameters of the fermion and Higgs sectors are set to the benchmark values given in Eqs.~\eqref{eq:benchmark_dark} and \eqref{eq:Higgs_Bench_Mark}, respectively.}
\label{fig:M_W_NLO}
  \end{figure}

\begin{figure}
\centering
\includegraphics[width=0.66\linewidth]{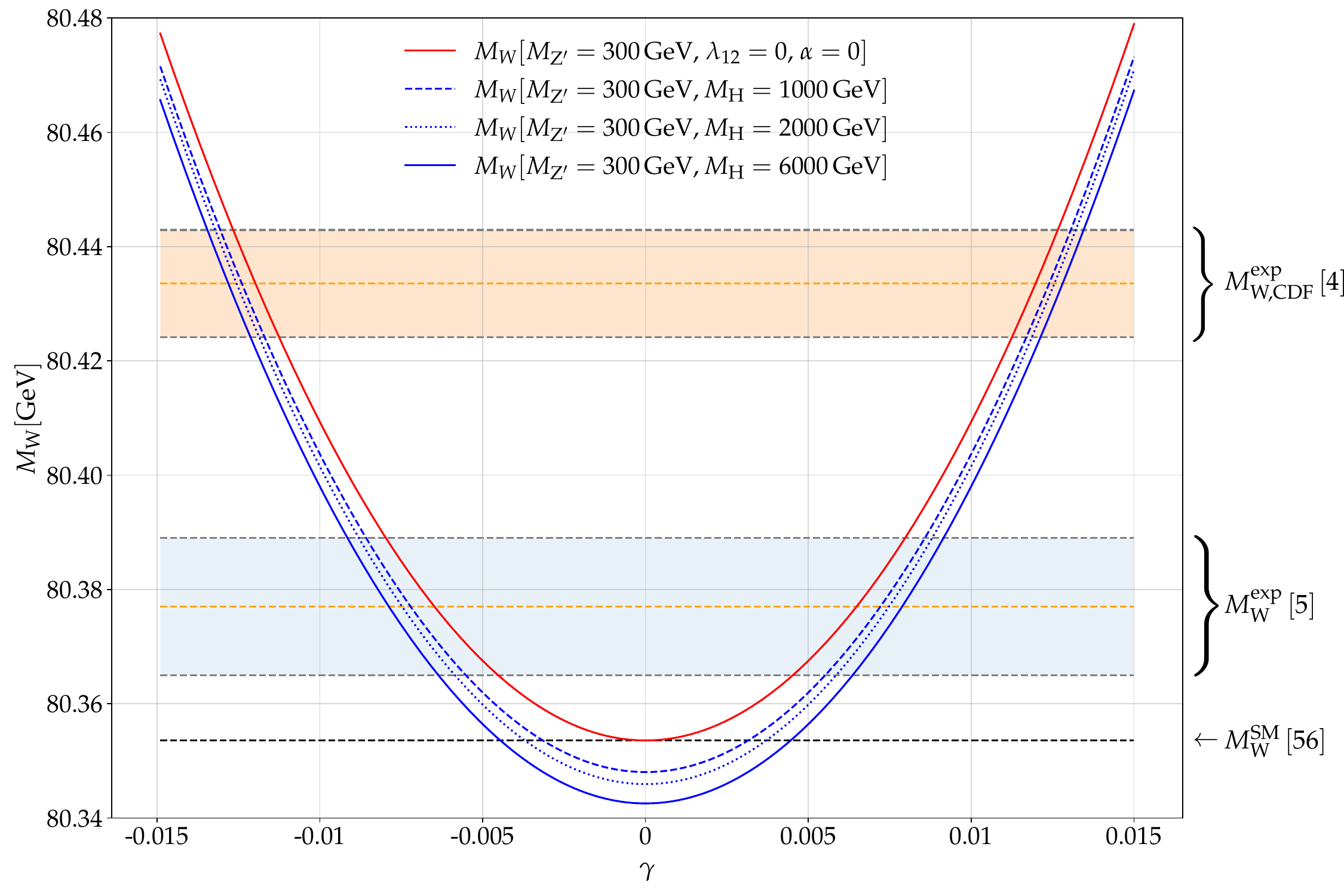}
\includegraphics[width=0.66\linewidth]{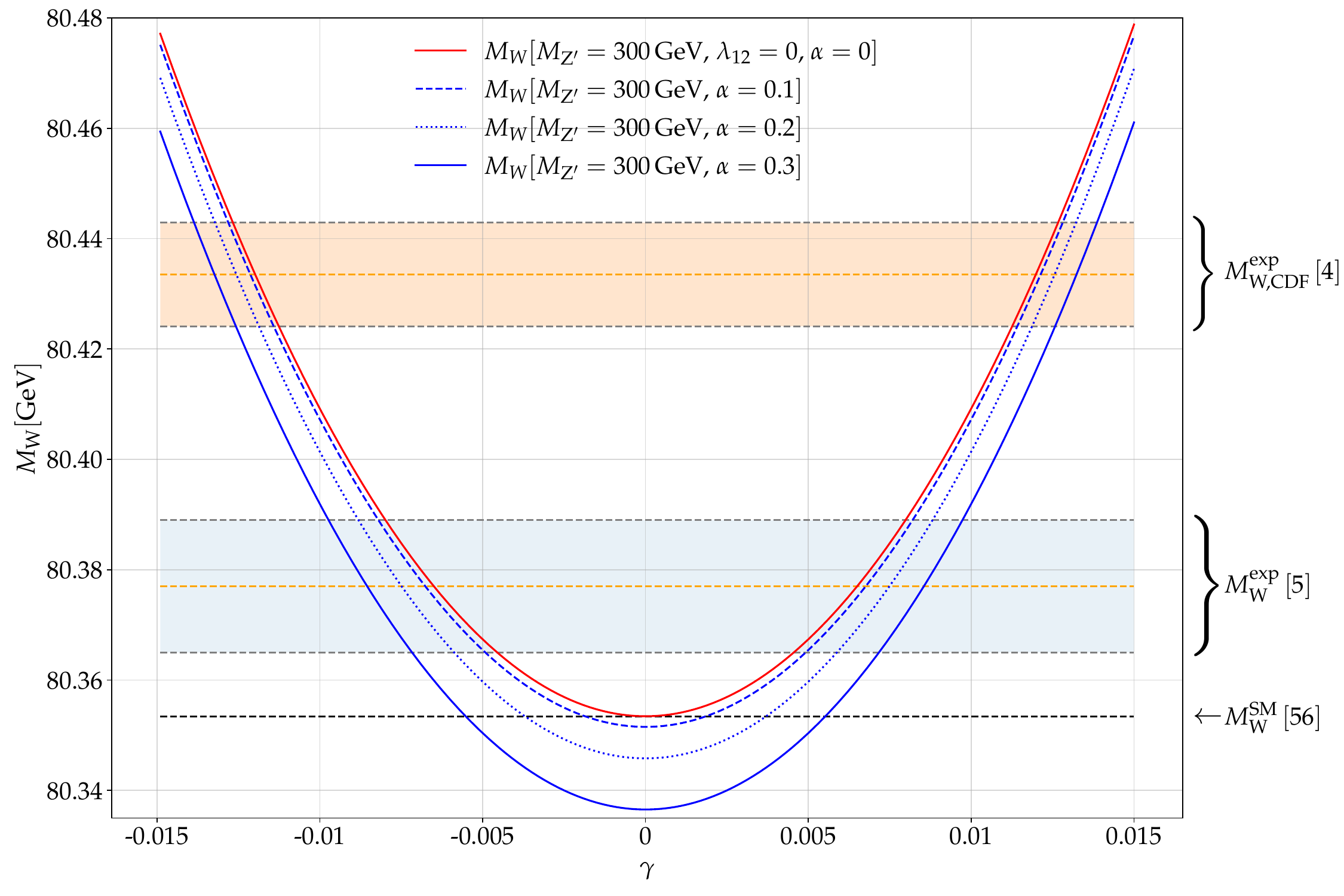}
\includegraphics[width=0.66\linewidth]{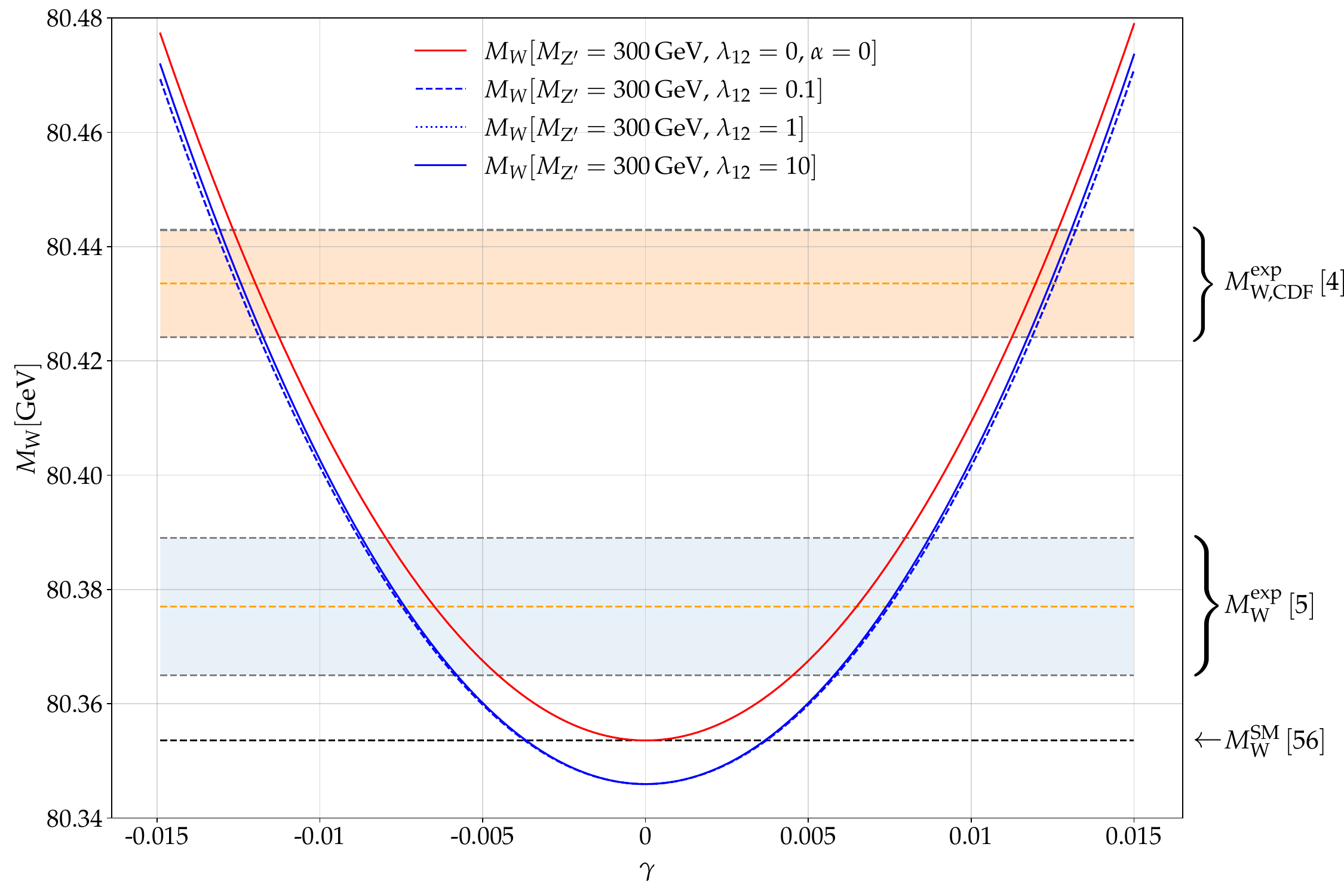}
\caption{The dependence of the DASM prediction for $M_\text{W}$ on $M_\text{H}$ (top), $\alpha$ (middle), and $\lambda_{12}$ (bottom). The red line represents the prediction for a decoupled scalar sector extension. For the blue lines all parameters that are not specified in the plots are set to their values according to Eqs.~\eqref{eq:benchmark_dark} and \eqref{eq:Higgs_Bench_Mark}.}
\label{fig:M_W_H}
\end{figure}

In this section we give a first confrontation of the DASM with a high-precision measurement by investigating the W-boson mass prediction based on muon decay. The relevant LO Feynman diagrams in the DASM (left) and in the Fermi theory (right) are shown in Fig.~\ref{fig:muon_dec_dia}. 
Neglecting terms of order $\mathcal{O}\left(\frac{m^2_\mu}{M_\text{W}^2}\right)$, the comparison between the respective amplitudes leads to the well-known LO relation
\begin{align}
 G_{\text{F}} = \frac{\alpha_\text{em}\pi}{\sqrt{2}s_\text{w}^2 M_\text{W}^2}+\dots,
    \label{eq:mudec-lo}
\end{align}
with the electromagnetic coupling constant $\alpha_\text{em}=\frac{e^2}{4\pi}$, connecting $M_\text{W}$ to the precisely measured Fermi constant $G_\text{F}$. Note that even though this looks similar to the relation valid in the SM, the dependence of $s_\text{w}$ on the gauge-boson masses differs from the corresponding SM relation according to Eq.~\eqref{eq:par-rel-ym} and thus, the relation between $G_\text{F}$ and $M_\text{W}$ in the DASM differs already at LO from the respective SM relation.\\
Higher-order corrections to muon decay are usually quantified in terms of the constant \mbox{$\Delta r$ \cite{Sirlin_d_r}}. The generalization of Eq.~\eqref{eq:mudec-lo} to NLO reads
\begin{align}
  G_\text{F}&=\frac{\alpha_\text{em} \pi}{\sqrt{2}s_\text{w}^2 M_\text{W}^2}\left(1+\Delta r\right)\nonumber\\
  &=\frac{\alpha_\text{em} \pi}{\sqrt{2}s_\text{w}^2 M_\text{W}^2}\left(1+2\delta Z_e-\frac{\delta s_\text{w}^2}{s_\text{w}^2}-\frac{\delta M_\text{W}^2}{M_\text{W}^2}+\frac{\Sigma_\text{T}^{WW}(0)}{M_\text{W}^2}+\delta_\text{vertex+box}\right),
  \label{eq:mudec-nlo}
\end{align}
where
\begin{align}
  \delta_\text{vertex+box}=\delta_\text{vertex}+\delta_\text{box}^\text{massive}+\delta_\text{box}^{\gamma\text{W}}
\end{align}
contains the relative NLO vertex corrections to the $W\mu\bar{\nu}_\mu$ and $W\nu_ee^+$ vertices denoted by $\delta_\text{vertex}$, corrections originating from box diagrams of the muon decay in the DASM as well as the NLO QED corrections of the Fermi theory, and the bremsstrahlung corrections in the two theories.\\
All diagrams needed for the predictions of $\Delta r$ in the DASM were generated using \textsc{FeynArts} \cite{Feynarts} and further evaluated using \textsc{FormCalc} \cite{Formcalc} and \textsc{LoopTools} \cite{Formcalc}. To have an additional cross-check on the DASM implementation we have constructed two completely independent  \textsc{FeynArts} model files.\\
The QED parts in the DASM and the SM are identical. Thus, similar to the SM, for\linebreak \mbox{$m_\text{e}, m_\mu \ll M_\text{W}$} the QED contributions originating from bremsstrahlung and virtual photon exchange between initial- and final-state leptons in the DASM and the Fermi model lead to \cite{hollikmuon}
\begin{align}
\delta_\text{box}^{\gamma\text{W}}={}&\frac{\alpha_\text{em}}{4\pi}\left(\log\frac{M_\text{W}}{m_e}+\log\frac{M_\text{W}}{m_\mu}-2\log\frac{m_e}{\lambda}-2\log\frac{m_\mu}{\lambda}+\frac{9}{2}\right),
\label{eq:QED_muon_dec}
\end{align}
where the photon mass $\lambda$ is used for infrared regularization.
The box diagrams induced by contributions from massive gauge-boson exchange between initial- and final-state particles in the DASM are shown in Fig.~\ref{fig:muon_dec_mass_box}.
They yield the corrections
\begin{align}
\delta_\text{box}^\text{massive}={}&\frac{\alpha_\text{em} M_\text{W}^2}{8\pi c_\text{w}^2s_\text{w}^2}\left[\rule{0cm}{0.75cm}\right.\frac{\log\frac{M_\text{Z}^2}{M_\text{W}^2}}{M_\text{Z}^2-M_\text{W}^2}\left(c_\gamma^2\left(5-10s_\text{w}^2+2s_\text{w}^4\right)+6s_\text{w}^3\eta c_\gamma s_\gamma -3s_\text{w}^2\eta^2s_\gamma^2\right)\nonumber\\
  &+\frac{\log\frac{M_{\text{Z}'}^2}{M_\text{W}^2}}{M_{\text{Z}'}^2-M_\text{W}^2}\left(s_\gamma^2\left(5-10s_\text{w}^2+2s_\text{w}^4\right)-6s_\text{w}^3\eta c_\gamma s_\gamma-3s_\text{w}^2\eta^2c_\gamma^2\right)\left.\rule{0cm}{0.75cm}\right].
\end{align}
Combining $\delta_\text{box}^\text{massive}$ with \eqref{eq:QED_muon_dec} and the vertex corrections one finally finds
\begin{align}
  \delta_\text{vertex+box}={}\frac{\alpha_\text{em}}{16 \pi s_\text{w}^2}\left\{\rule{0cm}{0.7cm}\right.{}&16\left(\Delta_\text{UV}-\log \frac{M_\text{W}^2}{\mu^2}\right)+24-\frac{1}{c_\text{w}^2}\sum_{V=\text{Z},\text{Z}'}\left[\rule{0cm}{0.7cm}\right.\frac{\sigma_V\log\frac{M_V^2}{M_\text{W}^2}}{\left(M_V^2-M_\text{W}^2\right)}\nonumber\\*
    &\times\bigl(\sigma_V [3 M_\text{W}^2[s_\text{w}^2 (2+\eta^2)-1]+10 c_\text{w}^4 M_V^2]+c_{2\gamma}(3 M_\text{W}^2[1+s_\text{w}^2 (\eta^2-2)]\nonumber\\*
    &-10 c_\text{w}^4 M_V^2)+ 6 \eta s_{2\gamma}s_\text{w}^3 M_\text{W}^2\bigr) \left.\rule{0cm}{0.7cm}\right]\left.\rule{0cm}{0.7cm}\right\},\qquad \sigma_{\text{Z/Z}'}=\mp 1,
\end{align}
where $\Delta_\text{UV}$ is the standard 1-loop UV divergence defined in \eqref{eq:stand_UV} and $\mu^2$ denotes the reference scale of dimensional regularization.
There is a strong dependence of $\Delta r$ on the light fermion masses entering through the charge renormalization constant $\delta Z_e$.
The dependence on the light-quark masses results from the non-perturbative effect of the photonic vacuum polarization at low energies induced by hadronic resonances. To absorb this non-perturbative contribution to the $M_\text{W}$ prediction into the input parameters, the running electromagnetic coupling $\alpha_\text{em}\left(M_\text{Z}^2\right)$ is used instead of the fine-structure constant $\alpha_\text{em}(0)$.
Further, the leading SM terms of the top-quark mass dependence, originating from the correction $\Delta \rho$ to the $\rho$-parameter, can be resummed up to $\mathcal{O}\left(\alpha_\text{em}^2\right)$ to further reduce the theoretical uncertainty. This can be summarized by the following replacements \cite{rhoparam,rhoparam2,deltaRcor2,deltaRcor3,deltaRcor5}:
\begin{align}
\alpha_\text{em}\rightarrow\alpha_\text{em} (M_\text{Z}^2)=\frac{\alpha_\text{em}(0)}{1-\Delta \alpha_\text{em}},\hspace{0.3cm} s_\text{w}^2\rightarrow\bar{s}_\text{w}^2=s^2_\text{w}+c_\text{w}^2\Delta \rho,\hspace{0.3cm} \Delta r\rightarrow \Delta r_\text{rem},
\end{align}
with
\begin{align}
  \Delta\alpha_\text{em}={}&\Delta \alpha^{(5)}_\text{had}+\Delta\alpha_\text{lep}=\frac{\partial\Sigma^{AA}_{f\neq t}(k^2)}{\partial k^2}\biggr|_{k^2=0}-\frac{\Sigma^{AA}_{f\neq t}(k^2)}{k^2}\biggr|_{k^2=M_\text{Z}^2},\\
\frac{c_\text{w}^2}{s_\text{w}^2}\Delta\rho={}&\frac{3\alpha_\text{em} m_\text{t}^2 M_\text{Z}^2}{16\pi \bigl(M_{\text{W},1}^2-M_\text{Z}^2\bigr)^2}\nonumber\\
&+\frac{3\alpha_\text{em}s_\gamma^2 m_\text{t}^2\bigl[M_{\text{Z}'}^2\bigl(M_{\text{W},1}^2+M_\text{Z}^2\bigr)-M_\text{Z}^2\bigl(M_{\text{W},1}^2+2M_{\text{W},2}^2+M_\text{Z}^2\bigr)\bigr]}{16 \pi \bigl(M_{\text{W},1}^2-M_\text{Z}^2\bigr)^3} +\mathcal{O}\left(s_\gamma^3\right),\\
\Delta r={}& \Delta \alpha_\text{em}-\frac{c_\text{w}^2}{s_\text{w}^2}\Delta \rho+\Delta r_\text{rem},
\end{align}
where $\Delta\alpha_\text{lep}$ and $\Delta\alpha^{(5)}_\text{had}$ summarize the terms with lepton- and light-quark (all other than the top-quark) mass logarithms introduced by $\delta Z_e$. The leading top-quark mass dependence of $\Delta r$ up to $\mathcal{O}\left(s_\gamma^2\right)$ is absorbed into $\Delta \rho$. The decomposition\footnote{The corresponding expression for $M_\text{W}$ is obtained from solving Eq.~\eqref{eq:mudec-lo} for the W-boson mass.}
\begin{align}
M_\text{W}^2={}&M_{\text{W},1}^2+M_{\text{W},2}^2s_\gamma^2+\mathcal{O}\left(s_\gamma^4\right),
\end{align}
was introduced to keep the expression compact.
At NLO, Eq.~\eqref{eq:mudec-nlo} is then modified to
\begin{align}
  G_\text{F}=\frac{\alpha_\text{em}(M_\text{Z}^2) \pi}{\sqrt{2}\bar{s}_\text{w}^2 M_\text{W}^2}\left(1+\Delta r_\text{rem}\right).
\label{eq:mudecfin}
\end{align}
The DASM admits two SM limits
\begin{alignat}{8}
  &1.&\qquad \gamma\to{}&{}0, \qquad &\alpha\to{}&0, \qquad \lambda_{12}&{}\to 0,\nonumber\\
  &2.&\qquad M_{\text{Z}'}\to{}&{}M_\text{Z}, \qquad &\alpha\to{}& 0, \qquad \lambda_{12}&{}\to 0,\nonumber
\label{eq:decoupling_lim}
\end{alignat}
leading to a complete decoupling of the dark sector from the SM at NLO\footnote{This statement holds as long as the masses $m_{\nu_i}$, $i=1,2,3$, of the SM-like neutrinos are negligible. Taking into account neutrino mass effects, SM limits would require a decoupling of $f'_\text{d}$ from the other three neutrinos. Further, we assume $M_\text{h}^\text{DASM}=M_\text{h}^\text{SM}$ here. For $M^\text{DASM}_\text{H}=M_\text{h}^\text{SM}$, the decoupling limits are given by \eqref{eq:decoupling_lim} with $\alpha\to \pi/2$.}. Therefore, the DASM can provide at least the same level of agreement between theory predictions and measurements as the SM for any precision observable (PO). Equation \eqref{eq:mudecfin} can now be used to obtain the desired prediction for $M_\text{W}$ at NLO. Therefore, we first eliminate $M_\text{W}$ in the radiative corrections $\Delta r_\text{rem}$ and $\Delta\rho$ in favour of $G_\text{F}$ using the LO relation \eqref{eq:mudec-lo} and finally solve Eq.~\eqref{eq:mudecfin} for $M_\text{W}$ after expressing $s_\text{w}$ in terms of $M_\text{W}$, $M_\text{Z}$, $M_{\text{Z}'}$, and $\gamma$ via Eq.~\eqref{eq:par-rel-ym}. Assuming the effects of the BSM sectors on the SM predictions to be small, as clearly favoured by experimental data, we will take the best SM predictions of $M^\text{SM}_\text{W}$ \cite{MWmasseSM} and add the difference between the NLO DASM and SM predictions according to
\begin{align}
M_\text{W}^\text{DASM}=M_\text{W}^\text{SM}+\Delta M_\text{W},\qquad \Delta M_\text{W}=M_\text{W,NLO}^\text{DASM}-M_\text{W,NLO}^\text{SM},
\end{align}
where the best SM prediction is given by $M^\text{SM}_\text{W}=80.3536\,\GeV$ \cite{MWmasseSM}, to obtain the best predictions within the DASM, which includes SM corrections beyond NLO. For the explicit values of the SM-like input parameters we closely follow Ref.~\cite{PDG} and take
 \begin{alignat}{4}
   \alpha_\text{em}(0)&=\frac{1}{137.035999180},&\qquad G_\text{F}&={}{}1.1663788\cdot10^{-5}\,\GeV^{-2},\nonumber\\
\Delta\alpha^{(5)}_\text{had}&=0.02768, &\qquad  \alpha_\text{s}(M^2_\text{Z})&={}{}0.1179,\nonumber\\
      \Delta\alpha_\text{lep}&=0.0314977, &\qquad  M_\text{Z}&={}{}91.1876\,\GeV,\nonumber\\
    M_\text{h}&=125.25\,\GeV, &\qquad m_\text{e}&={}{}0.51099895\cdot10^{-3}\,\GeV,\nonumber\\
     m_\mu&=0.1056583755\,\GeV,&\qquad m_\tau&={}{}1.77686\GeV,\nonumber\\
     m_\text{u}&=0.1\,\GeV,&\qquad m_\text{d}&={}{}0.1\GeV,\nonumber\\
     m_\text{s}&=0.1\,\GeV,&\qquad m_\text{c}&={}{}1.27\,\GeV,\nonumber\\
     m_\text{b}&=4.18\GeV,&\qquad m_\text{t}&={}{}172.5\GeV.
 \end{alignat}
Note that the values for $\Delta\alpha_\text{lep}$ \cite{deltaalphalep} and $\alpha_\text{s}(M^2_\text{Z})$ are only used in the determination of the best SM value $M^\text{SM}_\text{W}$. Further, we use $\alpha_\text{em}(M_\text{Z}^2)$ throughout the whole calculation and $\alpha_\text{em}(0)$ is only needed for the determination of $\alpha_\text{em}(M_\text{Z}^2)$. Owing to the use of $\alpha_\text{em}(M_\text{Z}^2)$, the light-fermion masses hardly play a role. In our calculation we assume $M_\text{H}>M_\text{h}$ and identify the $\text{h}$ boson with the SM-like Higgs boson, i.e.~$M_\text{h}^\text{DASM}=M_\text{h}^\text{SM}$.\\
   In Eq.~\eqref{eq:mudec-nlo} the newly introduced parameters from the fermion sector $m_{\nu_4}$ and $\theta_\text{r}$ only enter via $\delta s_\text{w}^2$ and are of order $\mathcal{O}\left(s_\gamma^2\right)$. Therefore, we find their influence to be negligible for the shown predictions for $M_\text{W}$ and choose the benchmark scenario
\begin{align}
  m_{\nu_4}=10\,\GeV,\qquad \theta_\text{r}=0.2.
  \label{eq:benchmark_dark}
\end{align}
Figure~\ref{fig:M_W_NLO} shows the dependence of the DASM predictions for $M_\text{W}$ as a function of the mixing angle $\gamma$ for various $\text{Z}'$ masses $M_{\text{Z}'}$.
Due to the appearance of renormalization constants in  Eq.~\eqref{eq:mudecfin} all additional parameters of the theory implicitly appear in the prediction of $M_\text{W}$. However, the prediction for $M_\text{W}$ is most sensitive to $\gamma$ and $M_{\text{Z}'}$, because these parameters already enter the LO prediction for $M_\text{W}$ via $s_\text{w}$. The additional parameters introduced in the Higgs sector lead to a small shift of the W-boson mass prediction towards smaller values for $M_\text{H}>M_\text{h}$. This shift manifests itself in the results shown in Fig.~\ref{fig:M_W_NLO} as the deviation from the SM result (black dashed line) at $\gamma=0$. To further illustrate the effects of the Higgs sector extension we give more results for various values of $M_\text{H}$, $\alpha$, and $\lambda_{12}$ in Fig.~\ref{fig:M_W_H}.
\noindent
As expected, the dependence of the prediction on the parameters from the Higgs sector extension compared to the $M_{\text{Z}'}$ and $\gamma$ dependence, already present at LO, is small. Therefore, we choose the benchmark scenario
 \begin{align}
   M_\text{H}=2000\,\GeV,\qquad \alpha=0.2,\qquad \lambda_{12}=0.2,
   \label{eq:Higgs_Bench_Mark}
 \end{align}
 for the further investigation of the influence of the gauge-sector extension on the W-boson mass prediction in the DASM. Figure \ref{fig:M_W_NLO} shows that for $M_{\text{Z}'}>M_\text{Z}$ the DASM prediction for $M_\text{W}$ rises and intersects the $1\,\sigma$ uncertainty bands for various combinations of $M_{\text{Z}'}$ and $\gamma$.
  The larger the difference between the Z- and the Z$'$-boson masses the steeper this rise becomes. For $M_{\text{Z}'}<M_\text{Z}$ the predicted values for $M_\text{W}$ decreases in the DASM, so that the DASM shows in these regions of the parameter space a worse compatibility with measurements than the SM. As a first rough estimate for the parameter regions of the DASM that can explain the measured value of $M_\text{W}$ we give the intervals for the $\gamma$ values where the theory predictions intersect with the $1\,\sigma$ uncertainty band of the measured values in Tab.~\ref{tab:M_W_limits}.
 \begin{table}
   \centering
\begin{tabular}{|c||c|c|c|c|c|c|}
\hline
$M_{\text{Z}'}\,[\GeV]$& $40$ & $50$ & $100$ & $300$ & $600$ & $1000$ \\
\hline
$|\gamma|/ 10^{-2}$ & -- & -- & $4.06-6.10$ & $0.59-0.88$ & $0.28-0.43$ & $0.17-0.26$ \\
\hline
$|\gamma|_\text{CDF}/ 10^{-2}$& -- & -- & $8.22-9.16$ & $1.19-1.32$ &$0.58-0.64 $  & $0.35-0.38$ \\
\hline
 \end{tabular}
\caption{The intervals of the values of $|\gamma|$ for several exemplary values of $M_{\text{Z}'}$ that show agreement between the theoretical prediction of $M_\text{W}$ and the $1\,\sigma$ uncertainty band of the measurements in the benchmark scenarios \eqref{eq:benchmark_dark} and \eqref{eq:Higgs_Bench_Mark} for the fermion and Higgs sector, respectively. Here we distinguish between the values $|\gamma|$ that lead to agreement with the experimental world average $M_\text{W}^\text{exp}$ and the values $|\gamma|_\text{CDF}$ that lead to agreement with the measured value of the CDF collaboration $M_\text{W,CDF}^\text{exp}$.} 
\label{tab:M_W_limits}
\end{table}
Generally the confrontation of the DASM with a measured value of $M_\text{W}$ only constrains $|\gamma|$, and the sign of $\gamma$ has to be determined by including additional observables in a fit of the DASM to data.

\section{Conclusions}
\label{se:conclusions}
In the absence of any spectacular direct signal of physics beyond the SM, ``discovery via precision'' seems to be the potential path to BSM physics, so that the need for precise theoretical predictions is greater than ever, not only in the SM but also for BSM physics. In order to achieve these highly accurate predictions, higher-order corrections, at least to NLO, need to be taken into account. Moreover, to parameterize the theory by an intuitive and phenomenologically sound set of input parameters, a judicious choice of the renormalization scheme is important in the SM and its extensions. While the renormalization of the SM is well understood since about 30 years, most of the extensions of scalar, gauge-boson, and/or fermion sectors introduce further subtleties in the renormalization procedure, especially by introducing mixing angles. \\
In this work we have given a full theoretical setup of the DASM in $R_\xi$ gauge and defined a particularly intuitive and experimentally easy accessible set of input parameters. The DASM extends the SM by a widely generic dark abelian sector containing a spontaneously broken $U(1)_\text{d}$ gauge group. This $U(1)_\text{d}$ gauge group is broken by a Higgs field that carries only dark charge and, thus, is a singlet under the SM gauge group. Further, introducing right-handed SM-like neutrinos as well as a fermion to the dark sector allows for three portals from the SM to this possible dark sector. The influence of these portal terms should be quite generic, so that the DASM provides a quite generic extension of the SM to study the sensitivity of EW precision data to a broad range of BSM models.\\
The DASM inherits mixing of fields in the scalar, fermion, and gauge sectors, giving rise to mixing angles in each of the three sectors mentioned above. We have performed a full renormalization of the model at NLO and derived explicit results for renormalization constants in OS and $\MSbar$ renormalization schemes. The proposed OS renormalization schemes for mixing angles have several desirable properties:
\begin{itemize}
\item All renormalization conditions (besides the one for the scalar self-coupling $\lambda_{12}$) are based on S-matrix elements. Thus, all NLO predictions based on these OS parameters are gauge independent.
\item The complete OS renormalization of all parameters related to masses leads to a systematic cancellation between tadpole corrections in mass counterterms and self-energies in the calculation of predictions for observables, making the presented OS prescription independent of the chosen tadpole scheme. 
\item The proposed OS renormalization conditions render predictions for observables perturbatively stable for degenerate masses of the particles involved in the mixing process, respectively.
\item All mixing-angle renormalization constants have smooth limits for exceptional values of the respective mixing angles. 
\end{itemize}
On the other hand, $\MSbar$ renormalization schemes have the benefit of being very simplistic and symmetric in the fields that mix. Further, the variation of the renormalization scale offers a simple way to estimate the perturbative stability of predictions. Nevertheless, $\MSbar$ renormalization of mixing angles suffers from severe downsides, such as issues with perturbative stability in certain parameter regions, for instance for degenerate masses of the particles corresponding to the mixing fields or for extreme values of the respective mixing angles. In addition, $\MSbar$ renormalization schemes for mixing angles are prone to introduce gauge dependences in the parameterization of predictions for observables. Having both OS and $\MSbar$ renormalization prescriptions available is an ideal situation to asses perturbative uncertainties from missing higher-order corrections by studying renormalization scale and scheme uncertainties. The formulation of the renormalization done in this work may serve as a proposal for the renormalization of models including mixing angles, e.g., due to kinetic mixing in the gauge-boson sector or similar phenomena.\\
As a first application of our OS scheme we have presented NLO predictions for $M_\text{W}$ in the DASM. Assuming the influence of new physics on the prediction of $M_\text{W}$ to be small, we include state-of-the-art corrections in the SM limit to further improve the precision of our prediction. We find a large part of the parameter space of the DASM that is capable of describing the experimental results better than the SM does. The DASM prediction for $M_\text{W}$ is independent of the sign of $\gamma$ and large $\text{Z}'$ masses $M_{\text{Z}'}>M_\text{Z}$ are preferred for all values of $\gamma$. For several values of $M_{\text{Z}'}$ and $\gamma$ the full region between the experimental world average $M_\text{W}^\text{exp}$ and the measured value of the CDF collaboration $M_\text{W,CDF}^\text{exp}$ can be covered by the DASM prediction. For $M_\text{Z}<M_{\text{Z}'}<1\,\TeV$, $|\gamma|$ ranges from $\sim 10^{-1}$ to $\sim 10^{-3}$, where $M_\text{W,CDF}^\text{exp}$ needs larger $|\gamma|$ than $M_\text{W}^\text{exp}$. Thus, the DASM, as generic extension of the SM, remains a promising candidate for the search of possible BSM physics. \\
With the complete NLO setup for the DASM provided in this work the next logical step is the confrontation of NLO predictions within the DASM with EW precision data. These precision tests can then be used to resolve the correlation between $\gamma$ and $M_{\text{Z}'}$ and  might further clarify whether the generic extensions introduced in the DASM are capable of significantly loosening the tensions between some SM predictions and measurements, e.g., for $(g-2)_\mu$, while keeping the good agreement of the SM between most predictions and data, and, thus, whether these types of SM extensions remain promising candidates in the search for BSM physics.
\section*{Acknowledgments}
S.D. and J.R. thank the Research Training Group GRK2044 of the German Research Foundation (DFG) for financial support. H.R.'s research is funded by  the Deutsche Forschungsgemeinschaft (DFG, German Research Foundation)---project no.\ 442089526; 442089660.
\appendix

\section*{Appendix}

\section{Explicit form of the ghost Lagrangian}
\label{app:ghost-part}
In this appendix we list the infinitesimal gauge transformations of the fields in the DASM as well as the explicit expression of the ghost Lagrangian $\mathcal{L}_\text{FP}$, adopting the conventions of Ref.~\cite{DEDI20201} for the field-theoretical SM quantities.\\
The infinitesimal gauge transformations of the gauge fields $W_\mu^a$, $B_\mu$, and $C_\mu$ are given by
\begin{align}
\delta W^a_\mu= \partial_\mu\delta\theta^a+g_2 f^{abc}W^b_\mu\delta\theta^c,\qquad\delta B_\mu= \partial_\mu\delta\theta^\text{Y},\qquad\delta C_\mu= \partial_\mu\delta\theta^\text{C},
\end{align}
where $\delta\theta^a$, $\delta\theta^\text{Y}$, and $\delta\theta^\text{C}$ are the gauge group parameters of the $SU(2)_{\text{W}}$, $U(1)_\text{Y}$, and $U(1)_{\text{d}}$ gauge groups, respectively.
For the Higgs doublet and singlet we have
\begin{align}
\delta \Phi= \left(-\frac{\text{i} g_1}{2}\delta\theta^\text{Y}+\frac{\text{i} g_2\tau^a}{2}\delta\theta^a\right)\Phi,\qquad \delta \rho= \left(-\text{i}e_\text{d}\delta\theta^\text{C}\right)\rho.
\end{align}
For the fields corresponding to the gauge and scalar bosons we find
\begin{align}
\delta W^\pm={} &\partial_\mu \delta\theta^\pm\pm\frac{ie}{s_\text{w}}\left[W^\pm_\mu\left[c_\text{w}(c_\gamma\delta\theta^\text{Z}-s_\gamma\delta\theta^{\text{Z}'})-s_\text{w}\delta\theta^\text{A}\right]\right.\\
& \left.+\left[s_\text{w} A_\mu-c_\text{w} (c_\gamma Z_\mu-s_\gamma Z'_\mu)\right]\delta\theta^\pm\right],\\ 
\delta A_\mu={}& \partial_\mu\delta\theta^\text{A}+\text{i}e\left(W^+_\mu\delta\theta^--W_\mu^-\delta\theta^+\right),\\
\delta Z_\mu={}& \partial_\mu\delta\theta^\text{Z}-\text{i}ec_\gamma\frac{c_\text{w}}{s_\text{w}}\left(W_\mu^+\delta\theta^--W^-_\mu\delta\theta^+\right),\\
\delta Z'_\mu={}&\partial\delta\theta^{\text{Z}'}+\text{i}es_\gamma\frac{c_\text{w}}{s_\text{w}}\left(W^+_\mu\delta\theta^--W^-_\mu\delta\theta^+\right)\\
\delta \phi^\pm={}&\mp\text{i}e\phi^\pm\left[\delta\theta^\text{A}+\delta\theta^\text{Z}\left[-c_\text{w} s_\gamma\eta+\frac{s_\text{w}^2-c_\text{w}^2}{2s_\text{w} c_\text{w}}\left(c_\gamma-s_\text{w}s_\gamma\eta\right)\right]\right.\nonumber\\
&\left.-\delta\theta^{\text{Z}'}\left[c_\text{w} c_\gamma\eta+\frac{s_\text{w}^2-c_\text{w}^2}{2s_\text{w} c_\text{w}}\left(s_\gamma+s_\text{w}c_\gamma\eta\right)\right]\right]\nonumber\\
&\pm\frac{\text{i}e}{2s_\text{w}}\left[v_2+c_\alpha h+s_\alpha H\pm\text{i}\left(c_x\chi-s_x\chi'\right)\right]\delta\theta^\pm,\\
\delta h={}&-\tilde{e}s_\alpha\left(c_\gamma\delta\theta^{\text{Z}'}+s_\gamma\delta\theta^\text{Z}\right)\left(c_x\chi'+s_x\chi\right)+\frac{ec_\alpha}{2s_\text{w}c_\text{w}}\left(c_x\chi-s_x\chi'\right)\nonumber\\
&\times\left[\delta\theta^\text{Z}\left(c_\gamma-s_\text{w}\eta s_\gamma\right)-\delta\theta^{\text{Z}'}\left(s_\gamma+s_\text{w}\eta c_\gamma\right)\right]+\frac{\text{i}ec_\alpha}{2s_\text{w}}\left(\phi^+\delta\theta^--\phi^-\delta\theta^+\right),\\
\delta H={}&\tilde{e}c_\alpha\left(c_\gamma\delta\theta^{\text{Z}'}+s_\gamma\delta\theta^\text{Z}\right)\left(c_x\chi+s_x\chi'\right)+\frac{s_\alpha e}{2s_\text{w}c_\text{w}}\left(c_x\chi-s_x\chi'\right)\nonumber\\
&\times\left[\delta\theta^\text{Z}\left(c_\gamma-s_\text{w}\eta s_\gamma\right)-\delta\theta^{\text{Z}'}\left(s_\gamma+s_\text{w}\eta c_\gamma\right)\right]+\frac{\text{i}es_\alpha}{2s_\text{w}}\left(\phi^+\delta\theta^--\phi^-\delta\theta^+\right),\\
\delta \chi={}&\delta\theta^\text{Z}\left[-\tilde{e}s_\gamma s_x\left(c_\alpha H-s_\alpha h+v_1\right)-\frac{ec_x}{2s_\text{w}c_\text{w}}\left(v_2+c_\alpha h+s_\alpha H\right)\left(c_\gamma-s_\text{w}\eta s_\gamma\right)\right]\nonumber\\
&+\delta\theta^{\text{Z}'}\left[\frac{ec_x}{2s_\text{w}c_\text{w}}\left(v_2+c_\alpha h+s_\alpha H\right)\left(s_\gamma+s_\text{w} \eta c_\gamma\right)-\tilde{e}c_\gamma s_x \left(c_\alpha H-s_\alpha h+v_1\right)\right]\nonumber\\
&+\frac{ec_x}{2s_\text{w}}\left(\phi^+\delta\theta^-+\phi^-\delta\theta^+\right),\\
\delta \chi'={}&\delta\theta^{\text{Z}}\left[\frac{es_x}{2s_\text{w}c_\text{w}}\left(v_2+c_\alpha h+s_\alpha H\right)\left(c_\gamma-s_\text{w}\eta s_\gamma\right)-\tilde{e}s_\gamma c_x\left(c_\alpha H-s_\alpha h+v_1\right)\right)\nonumber\\
&-\delta\theta^{\text{Z}'}\left[\frac{es_x}{2s_\text{w} c_\text{w}}\left(s_\gamma+s_\text{w}\eta c_\gamma\right)\left(v_2+c_\alpha h+s_\alpha H\right)+\tilde{e}c_\gamma c_x\left(c_\alpha H-s_\alpha h+v_1\right)\right]\nonumber\\
&-\frac{es_x}{2s_\text{w}}\left(\phi^+\delta\theta^-+\phi^-\delta\theta^+\right),
\end{align}
where the variations of the gauge group parameters 
\begin{align}
&\delta\theta^\pm=\frac{\delta\theta^1\mp i\delta\theta^2}{\sqrt{2}},\\
&\delta\theta^\text{A}=a c_\text{w} \delta\theta^\text{C}+c_\text{w}\delta\theta^\text{Y}-s_\text{w}\delta\theta^3,\\
&\delta\theta^\text{Z}=\left(a c_\gamma s_\text{w} +s_\gamma \sqrt{1-a^2}\right)\delta\theta^\text{C}+c_\gamma s_\text{w} \delta\theta^\text{Y}+c_\gamma c_\text{w}\delta\theta^3,\\
&\delta\theta^{\text{Z}'}=\left(\sqrt{1-a^2}c_\gamma-a s_\gamma s_\text{w}\right) \delta\theta^\text{C}-s_\gamma s_\text{w}\delta\theta^\text{Y}-s_\gamma c_\text{w}\delta\theta^3
\end{align}
have been used.\\
With the help of this transformation behaviour of the fields as well as Eqs.~\eqref{eq:gauge-fixing-funct-FtH} and \eqref{eq:lag-fp} one finds the following Faddeev--Popov ghost Lagrangian,
\begin{align}
\mathcal{L}_\text{FP}=&-\bar{u}^A&&\hspace{-12pt}\partial^\mu\partial_\mu u^A+\text{i}e(\partial^\mu \bar{u}^A)\left(W^+_\mu u^--W^-_\mu u^+\right)\nonumber\\
&-\bar{u}^Z&&\hspace{-15pt}\left(\partial_\mu\partial^\mu+\xi_\text{V}M_\text{Z}^2\right)u^Z-\frac{\text{i}ec_\gamma c_\text{w}}{s_\text{w}}\left(\partial^\mu\bar{u}^Z\right)\left(W^+_\mu u^--W^-_\mu u^+\right)\nonumber\\
&+&&\hspace{-28pt}\frac{c_xe}{2s_\text{w}}\xi_\text{V}M_\text{Z}\bar{u}^Z\left(\phi^+u^-+\phi^-u^+\right)\nonumber\\
&-\bar{u}^Z&&\hspace{-13pt} \xi_\text{V}M_\text{Z}\hspace{-2pt}\left\{\left[\frac{ec_xc_\alpha}{2s_\text{w}c_\text{w}}\left(c_\gamma-s_\text{w}\eta s_\gamma\right)-\tilde{e}s_\gamma s_x s_\alpha\right]\hspace{-2pt}h\hspace{-2pt}+\hspace{-2pt}\left[\tilde{e}s_\gamma s_x c_\alpha+\frac{ec_xs_\alpha}{2s_\text{w}c_\text{w}}\left(c_\gamma-s_\text{w}\eta s_\gamma\right)\right]\hspace{-2pt}H\hspace{-1pt}\right\}u^Z\nonumber\\
&+\bar{u}^Z&&\hspace{-13pt}\xi_\text{V} M_\text{Z}\hspace{-2pt}\left\{\left[\tilde{e}c_\gamma s_x s_\alpha+\frac{ec_xc_\alpha}{2s_\text{w}c_\text{w}}\left(s_\gamma+s_\text{w}\eta c_\gamma\right)\right]\hspace{-2pt}h\hspace{-2pt}-\hspace{-2pt}\left[\tilde{e}c_\gamma s_xc_\alpha-\frac{ec_xs_\alpha}{2s_\text{w}c_\text{w}}\left(s_\gamma+s_\text{w}\eta c_\gamma\right)\right]\hspace{-2pt}H\right\}\hspace{-1pt}u^{Z'}\nonumber\\
&-\bar{u}^{Z'}&&\hspace{-13pt}\left(\partial_\mu\partial^\mu+\xi_\text{V}M_{\text{Z}'}^2\right)u^{Z'}+\frac{\text{i}es_\gamma c_\text{w}}{s_\text{w}}\left(\partial^\mu\bar{u}^{Z'}\right)\left(W^+_\mu u^--W^-_\mu u^+\right)\nonumber\\
&-\bar{u}^{Z'}&&\hspace{-11pt}\frac{s_xe}{2s_\text{w}} \xi_\text{V}M_{\text{Z}'}\left(\phi^+u^-+\phi^-u^+\right)\nonumber\\
&-\bar{u}^{Z'}&&\hspace{-11pt} \xi_\text{V}M_{\text{Z}'}\hspace{-2pt}\left\{\left[\frac{es_xc_\alpha}{2s_\text{w}c_\text{w}}\left(s_\gamma+s_\text{w}\eta c_\gamma\right)-\tilde{e}c_\gamma c_x s_\alpha\right]\hspace{-2pt}h\hspace{-2pt}+\hspace{-2pt}\left[\frac{es_xs_\alpha}{2s_\text{w}c_\text{w}}\left(s_\gamma+s_\text{w}\eta c_\gamma\right)+\tilde{e}c_\gamma c_xc_\alpha\right]\hspace{-2pt}H\hspace{-2pt}\right\}\hspace{-1pt}u^{Z'}\nonumber\\
&+\bar{u}^{Z'}&&\hspace{-11pt}\xi_\text{V}M_{\text{Z}'}\hspace{-2pt}\left\{\left[\frac{es_xc_\alpha}{2s_\text{w}c_\text{w}}\left(c_\gamma-s_\text{w}\eta s_\gamma\right)+\tilde{e}s_\gamma s_\alpha c_x\right]\hspace{-2pt}h\hspace{-2pt}+\hspace{-2pt}\left[\frac{es_xs_\alpha}{2s_\text{w}c_\text{w}}\left(c_\gamma-s_\text{w}\eta s_\gamma\right)-\tilde{e}s_\gamma c_x c_\alpha\right]\hspace{-2pt}H\hspace{-2pt}\right\}u^Z\nonumber\\
&+\biggl\{&&\hspace{-18pt}-\bar{u}^+(\partial_\mu\partial^\mu+\xi_\text{W}M^2_\text{W})u^++\text{i}e(\partial^\mu\bar{u}^+)\left[A_\mu-\frac{c_\text{w}}{s_\text{w}}(c_\gamma Z_\mu-s_\gamma Z'_\mu)\right]u^+\nonumber\\
&&&\hspace{-18pt}\left.-\bar{u}^+\frac{e}{2s_\text{w}}\xi_\text{W}M_\text{W}\left[c_\alpha h+s_\alpha H+\text{i}(c_x \chi-s_x\chi')\right]u^+-\text{i}e\left(\partial^\mu\bar{u}^+\right)W^+_\mu u^A\right.\nonumber\\
&&&\hspace{-18pt}\left.+\bar{u}^+\xi_\text{W}M_\text{W}e\phi^+u^A+\frac{\text{i}ec_\text{w}}{s_\text{w}}\left(\partial^\mu\bar{u}^+\right)W^+_\mu\left(c_\gamma u^Z-s_\gamma u^{Z'}\right)\right. \nonumber \\
&&&\hspace{-18pt} +\bar{u}^+ \xi_\text{W}M_\text{W}e\phi^+\left.\left[\left(\frac{c_\gamma\left(s_\text{w}^2-c_\text{w}^2\right)}{2c_\text{w}s_\text{w}}-\frac{s_\gamma\eta}{2c_\text{w}}\right)u^Z-\left(\frac{s_\gamma\left(s_\text{w}^2-c_\text{w}^2\right)}{2s_\text{w}c_\text{w}}+\frac{c_\gamma\eta}{2c_\text{w}}\right)u^{Z'}\right]\right.\nonumber\\
&&&\hspace{-18pt}\left.+(u^+\rightarrow u^-,\bar{u}^+\rightarrow\bar{u}^-,W^+\rightarrow W^-,\phi^+\rightarrow\phi^-,\text{i}\rightarrow -\text{i}) \right.\biggr\}.
\end{align}

\section{The non-linear Higgs representation of the DASM}
\label{app:non_lin_Higgs}
Here we give a brief account of the chosen non-linear Higgs representation of the DASM used in the calculation of the tadpole renormalization constants of the GIVS. For the SM-like Higgs doublet $\Phi$ we closely follow Refs.~\cite{https://doi.org/10.48550/arxiv.2203.07236,SDTad2,non_lin1,non_lin2}. Therefore, we introduce the $ 2\times 2$ matrix notation for the linearly represented Higgs doublet
\begin{align}
\mathbf{\Phi}\equiv \left(\Phi^\text{C},\Phi\right)= \frac{1}{\sqrt{2}}\left[\left(v_2+h_{2}\right)\mathbf{\mathbb{1}}+2 \text{i} \mathbf{\phi}\right],\qquad \mathbf{\phi}\equiv \frac{\phi_j \mathbf{\sigma}_j}{2}.
\end{align}
In our notation we use bold symbols to denote matrices. The quantities $\mathbf{\sigma}_i$, $i=1,2,3$, are the Pauli matrices and $\phi_i$ the three real would-be Goldstone-boson fields. They are related to the would-be Goldstone-boson fields of Eq.~\eqref{eq:higgs-fields} according to
\begin{align}
  \phi^\pm=\frac{1}{\sqrt{2}}\left(\phi_2\pm \text{i}\phi_1\right),\qquad \chi_2=-\phi_3.
\end{align}
In the matrix representation the gauge-invariant mass operator of the Higgs doublet $\Phi^\dagger \Phi$, appearing in the Higgs potential, is simply given by
\begin{align}
\Phi^\dagger \Phi=\frac{1}{2}\tr\left[\mathbf{\Phi^\dagger \Phi}\right], 
\end{align}
and, similarly, the kinetic terms for the Higgs doublet are given by the trace
\begin{align}
\mathcal{L}_{\Phi,\text{kin}}=\frac{1}{2}\tr\left[\left(D_\mu \mathbf{\Phi}\right)^\dagger\left(D^\mu \mathbf{\Phi}\right)\right].
\end{align}
A matrix formulation for the Higgs field $\rho$ is not needed since it is only charged under a $U(1)_\text{d}$ gauge group. Now we can easily switch to the non-linear representations of the Higgs fields
\begin{align}
\mathbf{\Phi}=\frac{1}{\sqrt{2}}\left(h_2^\text{nl}+v_2\right)\exp\left(\frac{ \text{i} \zeta_j \mathbf{\sigma}_j}{ v_2}\right),\qquad \rho= \frac{1}{\sqrt{2}}\left(h_1^\text{nl}+v_1\right)\exp\left(\frac{\text{i} \chi_1^\text{nl}}{v_1}\right),
\end{align}
where $h^\text{nl}_1,h^\text{nl}_2$ are the physical, gauge-invariant Higgs fields of the non-linear representation and $\zeta_i, \chi_1^\text{nl}$, $i=1,2,3$ represent real would-be Goldstone-boson fields.
With this choice of non-linear representation the component fields of the linear and non-linear representations are connected via
\begin{alignat}{3}
 h_1&=\left(h_1^\text{nl}+v_1\right)\cos\left(\frac{\chi_1^\text{nl}}{v_1}\right)-v_1,&\qquad& h_2&=&\left(h_2^\text{nl}+v_2\right)\cos\left(\frac{|\vec{\zeta}|}{v_2}\right)-v_2,\\
  \chi_1&=\left(h_1^\text{nl}+v_1\right)\sin\left(\frac{\chi_1^\text{nl}}{v_1}\right),&\qquad& \phi_i&=&\left(h_2^\text{nl}+v_2\right)\sin\left(\frac{|\vec{\zeta}|}{v_2}\right)\frac{\zeta_i}{|\vec{\zeta}|}.
\end{alignat}
Note that the respective fields of the linear and non-linear representations agree to linear order in the Goldstone-boson fields\footnote{Therefore, the respective interaction terms of the Lagrangian containing at most one Goldstone field are identical in the linear and non-linear representation.}.

\section{Explicit expressions for {\boldmath $\MSbar$} renormalization constants of mixing angles}
\label{sec:exp_MSbar_CTs}
Here we list the explicit expressions for the $\MSbar$ renormalization constants of the mixing angles $\gamma$, $\alpha$, and $\theta_\text{r}$ in the 't Hooft--Feynman gauge. They are obtained by keeping only terms proportional to the standard 1-loop UV divergence
\begin{align}
  \Delta_\text{UV}=\frac{2}{4-D}-\gamma_\text{E}+\log 4\pi,
\label{eq:stand_UV}
\end{align}
where $\gamma_\text{E}$ is the Euler--Mascheroni constant and $D=4-2 \epsilon$ is the number of space-time dimensions used to calculate loop integrals in dimensional regularization, from the respective OS renormalization constants. In the PRTS the renormalization constants are given by
\begin{align}
\delta\gamma_\MSbar^\text{PRTS}=\Delta_\text{UV}\Biggl\{&\frac{\alpha_\text{em} \left[s_{2 \gamma} \left(1-s_\text{w}^2 \eta^2  \right) +2 s_\text{w} \eta c_{2 \gamma}\right]}{16 \pi c_\text{w}^2 s_\text{w}^2  M^2_{\text{ZZ}'-}} \sum_{l,u,d}  \left[m_l^2+3 (m_u^2+ m_d^2)\right]\nonumber\\
&-\frac{c_\text{w}^2 s_\text{w}^2 s_{2\gamma} s^2_{\theta_\text{r}}  \lambda_{12}^2 m_{\nu_4}^2 M_\text{Z}^2 M_{\text{Z}'}^2}{\alpha_\text{em} \pi ^3 s_{2\alpha}^2  M^2_{\text{ZZ}'-} M^4_{\text{Hh}-}}+\frac{1}{768 \pi ^3 \alpha_\text{em} s_\text{w}^2c_\text{w}^4 M_\text{W}^2 }\biggl\{\pi ^2 \alpha_\text{em}^2 \biggl\{\frac{2}{ M^2_{\text{ZZ}'-}}\nonumber\\
&\times\Bigl[c_{2 \gamma} \Bigl(3 c_{2 \alpha}c_\text{w}^2  M^2_{\text{Hh}-} \left[4 s_\text{w} \eta M_\text{W}^2-2 s_\text{w} c_\text{w}^2 \eta  M^2_{\text{ZZ}'+}+s_{2 \gamma}c_\text{w}^2  M^2_{\text{ZZ}'-} \left(s_\text{w}^2 \eta^2 -1\right) \right]\nonumber\\
&-2 s_\text{w} \eta \bigl[3 c_\text{w}^2  M^2_{\text{Hh}+}  (2 M_\text{W}^2-c_\text{w}^2  M^2_{\text{ZZ}'+} )+2 c_\text{w}^4 \left[M^4_{\text{ZZ}'+}-2 M_\text{Z}^2M_{\text{Z}'}^2\right]
\nonumber\\
&+2c_\text{w}^2   M^2_{\text{ZZ}'+} M_\text{W}^2 (82 s_\text{w}^2-1)+48 M_\text{W}^4 \left(2 c_\text{w}^2 s_\text{w}^2+s_\text{w}^2 \eta^2 +1\right)\bigr]\Bigr)\nonumber\\
&+s_{2 \gamma} \Bigl(3 c_{2 \alpha}c_\text{w}^2  M^2_{\text{Hh}-} \left(s_\text{w}^2 \eta^2 -1\right) (c_\text{w}^2  M^2_{\text{ZZ}'+} -2 M_\text{W}^2)\nonumber\\
&-3 c_\text{w}^2   M^2_{\text{Hh}+} \left(s_\text{w}^2 \eta^2 -1\right) (c_\text{w}^2  M^2_{\text{ZZ}'+} -2 M_\text{W}^2)+c_\text{w}^4 M^4_{\text{ZZ}'+} \left(s_\text{w}^2 \eta^2 -1\right)\nonumber\\
&+2c_\text{w}^2   M^2_{\text{ZZ}'+} M_\text{W}^2 \left[81 s_\text{w}^2 \eta^2 +4 s_\text{w}^2 (5-23 s_\text{w}^2)-9\right]\nonumber\\
&+48 M_\text{W}^4 \left[2 c_\text{w}^2 \left(s_\text{w}^2 \eta^2 -2 s_\text{w}^2+1\right)+s_\text{w}^4 \eta^4-1\right]\Bigr)\Bigr]+2 s_\text{w} c_\text{w}^2 \eta \Bigl(3 c_\text{w}^2 (  M^2_{\text{Hh}+}-  M^2_{\text{ZZ}'+})\nonumber\\
&-6c_{2 \alpha} c^2_{2 \gamma} c_\text{w}^2  M^2_{\text{Hh}-} +c_{4 \gamma}c_\text{w}^2 (3   M^2_{\text{Hh}+}-  M^2_{\text{ZZ}'+}) \nonumber\\
&+4 M_\text{W}^2 (7 c_\text{w}^2-75 s_\text{w}^2-6)\Bigr)-\left[s_{4 \gamma}c_\text{w}^4 \left(s_\text{w}^2 \eta^2 -1\right) (3   M^2_{\text{Hh}+}-  M^2_{\text{ZZ}'+}) \right]\biggr\}
\nonumber\\
&+\frac{8s_{2 \gamma}s_\text{w}^4 c_\text{w}^6 \lambda_{12}^2 M_\text{Z}^2M_{\text{Z}'}^2}{c_\alpha^2s_\alpha^2  M^2_{\text{ZZ}'-} M^4_{\text{Hh}-}}\Bigl[-3c_{2 \alpha}  M^2_{\text{Hh}-}(c_\text{w}^2  M^2_{\text{ZZ}'+}+c_{2 \gamma}c_\text{w}^2  M^2_{\text{ZZ}'-}\nonumber\\
  &-2 M_\text{W}^2)+c_{2 \gamma}c_\text{w}^2  M^2_{\text{ZZ}'-} (  M^2_{\text{ZZ}'+}-3   M^2_{\text{Hh}+}) -3c_\text{w}^2   M^2_{\text{Hh}+}   M^2_{\text{ZZ}'+}\nonumber\\
  &+6   M^2_{\text{Hh}+} M_\text{W}^2+c_\text{w}^2M^4_{\text{ZZ}'+} +8   M^2_{\text{ZZ}'+} M_\text{W}^2+48 c_\text{w}^2 M_\text{Z}^2M_{\text{Z}'}^2\Bigr]\biggr\}
\Biggr\}.
\label{eq:gammaMSbarexpli}
\end{align}
Here we introduced the shorthands
\begin{align}
M^2_{ij\pm}\equiv M_i^2\pm M_j^2,\qquad M^4_{ij\pm}\equiv\left(M_{ij\pm}^2\right)^2.
\label{eq:MSb_shorthands}
\end{align}
Note that the explicit expressions for $\delta \gamma_{\MSbar}$ in the FJTS and the GIVS are easily obtained from \eqref{eq:gammaMSbarexpli} using Eqs.~\eqref{eq:gammaMSbarPRTS}, \eqref{eq:gammaMSbarFJTS}, and \eqref{eq:gammaMSbarGIVS}.
For the $\MSbar$-renormalization constant of the Higgs mixing angle we find
\begin{align}
\delta\alpha_\MSbar^\text{PRTS}={}&\Delta_\text{UV}\Biggl\{ \frac{\alpha_\text{em} s_{2 \alpha} \Lambda_{\text{ZZ}'}^2}{32 \pi M_\text{W}^2 M^2_{\text{Hh}-}( \Lambda_{\text{ZZ}'}^2-2 M_\text{W}^2)} \sum_{f=l,u,d}N_{C,f}\left[ m_f^2 (  M^2_{\text{Hh}+}-8 m_f^2) \right]\nonumber\\
&
+\frac{s^2_{\theta_\text{r}}\lambda_{12}^2 m_{\nu_4}^2 M_\text{W}^2  \left(2 M_\text{W}^2- \Lambda_{\text{ZZ}'}^2\right)  \left(2 c_{2 \theta_\text{r}}m_{\nu_4}^2 -6 m_{\nu_4}^2+  M^2_{\text{Hh}+}\right)}{4 \alpha_\text{em}\pi ^3 s_\alpha c_\alpha  \left(M^2_{\text{Hh}-}\right)^3 \Lambda_{\text{ZZ}'}^2}
\nonumber\\
&+\frac{1}{512 \pi ^3 M_\text{W}^4}\biggl\{ \frac{\alpha_\text{em} \pi ^2M_\text{W}^2}{s_\text{w}^2 \Lambda_{\text{ZZ}'}^4 } \biggl(\frac{2 s_{\alpha} c_{\alpha}}{M^2_{\text{Hh}-}} \biggl[2   c_{4 \alpha} M^4_{\text{Hh}-} \Lambda_{\text{ZZ}'}^4\nonumber\\
&-24 c_{2 \alpha}  M^2_{\text{Hh}+}M^2_{\text{Hh}-}  \Lambda_{\text{ZZ}'}^4+c_{4 \gamma} M^4_{\text{ZZ}'-} \Bigl[11  M^4_{\text{Hh}+}+ 48 M^4_{\text{ZZ}'+}\nonumber\\
    &-16 M_\text{h}^2  M^2_{\text{ZZ}'+} +4M_\text{H}^2 M_\text{h}^2-32 M_\text{h}^2 M_\text{W}^2+96 M_\text{W}^4\Bigr]+4 c_{2 \gamma} \Bigl[11   M^2_{\text{ZZ}'+} M^2_{\text{ZZ}'-}M^4_{\text{Hh}+}\nonumber \\
    &+4   M^2_{\text{ZZ}'+} M^2_{\text{ZZ}'-} \left[M_\text{H}^2 M_\text{h}^2+8 M_\text{W}^2 \left(3 M_\text{W}^2-M_\text{h}^2\right)\right]\nonumber\\
&+16 \left[M_\text{h}^2\left(M_{\text{Z}'}^6- M_\text{Z}^6\right)+3\left( M_\text{Z}^8- M_{\text{Z}'}^8\right)\right]\Bigr]+\left(11 M^4_{\text{Hh}+}+4 M_\text{H}^2 M_\text{h}^2\right) \nonumber\\
&\times\left(2M_\text{Z}^2  M^2_{\text{ZZ}'+} +M_\text{Z}^4+3 M_{\text{Z}'}^4\right)+16 M_\text{h}^2 \Bigl[8 M_\text{Z}^2 M_{\text{Z}'}^2 \left(  M^2_{\text{ZZ}'+}+M_\text{W}^2\right)\nonumber\\
    &-3 M^4_{\text{ZZ}'+} \left(  M^2_{\text{ZZ}'+}+2 M_\text{W}^2\right)\Bigr]-2 s_\text{w}^2 c_\text{w}^2\Bigl[21 \left(  M^2_{\text{ZZ}'+}\right)^4-136 M_\text{Z}^2 M_{\text{Z}'}^2  M^4_{\text{ZZ}'+}\nonumber\\
    &+80 M_\text{Z}^4 M_{\text{Z}'}^4\Bigr]+s_\text{w}^4 \Bigl[136 M_\text{Z}^2 M_{\text{Z}'}^2 M^4_{\text{ZZ}'+}-21 \left(  M^2_{\text{ZZ}'+}\right)^4-80 M_\text{Z}^4 M_{\text{Z}'}^4\Bigr]\nonumber\\
  &+96 M_\text{W}^4 \left(2 M_\text{Z}^2  M^2_{\text{ZZ}'+}+M_\text{Z}^4+3 M_{\text{Z}'}^4\right)+3 \left(55-7 c_\text{w}^4\right) \left(M_\text{Z}^8+M_{\text{Z}'}^8\right)\nonumber\\
  &+6 M_\text{Z}^4 M_{\text{Z}'}^4\left(5+11 c_\text{w}^4\right)-52 s_\text{w}^2M_\text{Z}^2 M_{\text{Z}'}^2(1+c_\text{w}^2)  \left(M_\text{Z}^4+M_{\text{Z}'}^4\right)\biggr] \nonumber\\
&-16 s_{2 \alpha} \Lambda_{\text{ZZ}'}^2 \Bigl[c_{2 \gamma}M^2_{\text{ZZ}'-} \left(  M^2_{\text{ZZ}'+}+2 M_\text{W}^2\right)+2 M_\text{W}^2  M^2_{\text{ZZ}'+}+M_\text{Z}^4+M_{\text{Z}'}^4\Bigr]\biggr)\nonumber\\
&+\frac{16 \lambda_{12}^2 M_\text{W}^2 s_\text{w}^2 }{\alpha_\text{em} s_\alpha c_\alpha  \left(M^2_{\text{Hh}-}\right)^3}\biggl[M_\text{W}^4 \Bigl[4 M_\text{H}^2 M_\text{h}^2-11 M^4_{\text{Hh}+}\Bigr]-M_\text{W}^4M^2_{\text{Hh}-} \nonumber\\
  &\times\left[12 c_{2 \alpha}   M^2_{\text{Hh}+}+c_{4 \alpha}M^2_{\text{Hh}-}\right]+4 c_{2 \gamma} c_\text{w}^4 M_\text{Z}^2 M_{\text{Z}'}^2 M^2_{\text{ZZ}'-}   M^2_{\text{Hh}+}\nonumber\\
  &+4 c_\text{w}^4 M_\text{Z}^4 M_{\text{Z}'}^2  M^2_{\text{Hh}+}+4 c_\text{w}^4 M_\text{Z}^2 M_{\text{Z}'}^4  \left(  M^2_{\text{Hh}+}-12 M_\text{Z}^2\right)\biggr]\nonumber\\
&+\frac{2 \pi  \lambda_{12}}{ \Lambda_{\text{ZZ}'}^4  M^4_{\text{Hh}-}}\biggl[\frac{1}{c_\alpha s_\alpha} \biggl[-2 c_{6 \alpha} M_\text{W}^4  M^4_{\text{Hh}-} \Lambda_{\text{ZZ}'}^4\nonumber\\
    &+c_{2 \alpha} \Bigl[M^2_{\text{ZZ}'-} \Bigl(M^2_{\text{ZZ}'-} \Bigl[c_{4 \gamma} \Bigl(M_\text{W}^4 M^4_{\text{Hh}-}-16 M_\text{Z}^2 M_{\text{Z}'}^2 \Bigl[c_\text{w}^4M^4_{\text{ZZ}'+} \nonumber\\
          &-c_\text{w}^2 M_\text{W}^2  M^2_{\text{ZZ}'+}+4 M_\text{W}^4\Bigr]\Bigr)+8 c_{6 \gamma}c_\text{w}^2 M_\text{Z}^2 M_{\text{Z}'}^2 M^2_{\text{ZZ}'-} \left(M_\text{W}^2-2 c_\text{w}^2  M^2_{\text{ZZ}'+}\right)\Bigr]\nonumber\\
      &+4 c_{2 \gamma} \Bigl[M_\text{W}^4  M^2_{\text{ZZ}'+}  M^4_{\text{Hh}-} -2 c_\text{w}^2  M_\text{Z}^2 M_{\text{Z}'}^2 M^4_{\text{ZZ}'-} \left(M_\text{W}^2-2 c_\text{w}^2  M^2_{\text{ZZ}'+}\right)\Bigr]\Bigr)\nonumber\\
      &+2 M_\text{W}^2 M_\text{Z}^2   M^2_{\text{ZZ}'+}\left[M_\text{W}^2 M^4_{\text{Hh}-} -8 c_\text{w}^2 M_{\text{Z}'}^2 M^4_{\text{ZZ}'-}\right]\nonumber\\
&+4 M_\text{Z}^2 M_{\text{Z}'}^2  M^4_{\text{ZZ}'-}\left[c_\text{w}^4 \left(5  M^4_{\text{ZZ}'+}-4 M_\text{Z}^2 M_{\text{Z}'}^2 \right)+16 M_\text{W}^4\right]\nonumber\\
      &+M_\text{W}^4  M^4_{\text{Hh}-}\left(M_\text{Z}^4+3 M_{\text{Z}'}^4\right)\Bigr]+4 M_\text{Z}^2 M_{\text{Z}'}^2  M^4_{\text{ZZ}'-}\Bigl(c_\text{w}^4 \Bigl[5 M^4_{\text{ZZ}'+}-4 M_\text{Z}^2 M_{\text{Z}'}^2\Bigr]\nonumber\\
    &-4 c_\text{w}^4 \left(  M^2_{\text{ZZ}'+} \left[c_{4 \gamma}  M^2_{\text{ZZ}'+}+c_{6 \gamma}M^2_{\text{ZZ}'-}\right]-c_{2 \gamma}   M^2_{\text{ZZ}'+}M^2_{\text{ZZ}'-}\right)\nonumber\\
&+16 c_{4 \gamma}M_\text{W}^4-16 M_\text{W}^4\Bigr)\biggr]-8 \frac{c_\alpha}{s_\alpha} c_{8 \gamma} c_\text{w}^4 M_\text{Z}^2 M_{\text{Z}'}^2 \left( M^2_{\text{ZZ}'-}\right)^4\biggr]\biggr\}\Biggr\},
\label{eq:alphaMSbarexpli}
\end{align}
where $N_{C,f}$ is the respective colour factor of the fermions. In addition to Eq.~\eqref{eq:MSb_shorthands}, we have introduced the abbreviations 
\begin{align}
 \Lambda_{\text{ZZ}'}^2=  M^2_{\text{ZZ}'+}+ M^2_{\text{ZZ}'-} c_{2\gamma},\qquad \Lambda_{\text{ZZ}'}^4\equiv \left(\Lambda_{\text{ZZ}'}^2\right)^2,
\end{align}
to further compactify the result.
The explicit expressions for $\delta \alpha_{\MSbar}$ in the FJTS and the GIVS are easily obtained from \eqref{eq:alphaMSbarexpli} using Eqs.~\eqref{eq:dalphaMSbarPRTS}, \eqref{eq:dalphaMSbarFJTS}, and \eqref{eq:dalphaMSbarGIVS}.
The $\MSbar$ renormalization constant for the mixing angle in the fermion sector is given by
\begin{align}
 \delta\theta_{\text{r},\MSbar}^\text{PRTS}={}&\frac{\Delta_\text{UV} \lambda_{12}^2 s^2_\text{w} s_{\theta_\text{r}} c_{\theta_\text{r}}}{64 \pi ^3 \alpha_\text{em} c_\alpha^2 s_\alpha^2(M_\text{H}^2-M_\text{h}^2)^2}\biggl\{56 c^2_\text{w} M_\text{Z}^2 M_{\text{Z}'}^2\nonumber\\& +2 m_{\nu_4}^2 s^2_{\theta_\text{r}} \bigl[3 c^2_\text{w} c_{2 \gamma} (M_{\text{Z}'}^2-M_\text{Z}^2)-3 c^2_\text{w} (M_\text{Z}^2+M_{\text{Z}'}^2)+10 M_\text{W}^2\bigr]\biggr\}.
\label{eq:thetaRMSbarexpli}
\end{align}
Again, the explicit expressions for $\delta \theta_{\text{r},\MSbar}$ in the FJTS and the GIVS are easily obtained from \eqref{eq:thetaRMSbarexpli} using Eqs.~\eqref{eq:thetaRMSbarPRTS}, \eqref{eq:thetaRMSbarFJTS}, and \eqref{eq:thetaRMSbarGIVS}.

\clearpage

\bibliographystyle{JHEPmod}
\bibliography{DASM}

\end{document}